\documentclass[aps]{revtex4}
\usepackage{amsmath,amssymb}
\usepackage{color,graphicx}
\usepackage{slashbox}
\usepackage{multirow}

\newcommand{\be}{\begin{equation}}
\newcommand{\ee}{\end{equation}}
\begin{document}
\title{Nonlocally-induced (quasirelativistic) bound states: Harmonic confinement  and
the 
finite  well}
\author{Piotr Garbaczewski  and   Mariusz  \.{Z}aba}
\affiliation{Institute of Physics, University of Opole, 45-052
Opole, Poland}
\date{\today }
\begin{abstract}
Nonlocal Hamiltonian-type   operators, like e.g.   fractional and
quasirelativistic, seem to be instrumental for a conceptual
broadening of current quantum paradigms. However physically relevant
properties of   related quantum systems have not yet  received due
(and scientifically undisputable) coverage in the literature. That
extends to peculiarities of their nonlocally-induced dynamics and
 painfully  lacking explicit  insight  into  energy spectra under confining conditions.
In the present paper we address Schr\"{o}dinger-type eigenvalue
problems for $H=T+V$, where a kinetic term  $T=T_m$  is    a
quasirelativistic energy operator $T_m = \sqrt{-\hbar ^2c^2 \Delta + m^2c^4} - mc^2$
of  mass $m\in (0,\infty )$  particle. A potential
$V$ we assume to refer to the harmonic confinement   or finite well
of an  arbitrary depth.
 We analyze  spectral solutions of the  pertinent  nonlocal   quantum systems
 with a focus on  their   $m$-dependence.   Extremal  mass  $m$ regimes  for  eigenvalues and eigenfunctions of
  $H$  are investigated: (i) $m\ll 1$ spectral  affinity ("closeness") with the
Cauchy-eigenvalue problem   ($T_m \sim T_0=\hbar c |\nabla |$)  and    (ii)  $m \gg 1$ spectral affinity
with  the    nonrelativistic eigenvalue problem ($T_m \sim -\hbar ^2 \Delta /2m $).
 To this  end  we  generalize  to nonlocal  operators   an efficient
computer-assisted method to solve   Schr\"{o}dinger  eigenvalue
problems, widely used in  quantum physics  and quantum chemistry. A resultant  spectrum-generating  algorithm
  allows to carry out all computations directly in the  configuration space of the nonlocal   quantum system.
 This  allows  for  a proper assessment  of  the  spatial  nonlocality  impact on
 simulation outcomes. Although the   nonlocality of $H$  might seem to    stay   in    conflict with  various
  numerics-enforced  cutoffs,  this  potentially serious obstacle   is   kept under control  and  effectively tamed.
   \end{abstract}
\maketitle

\section{Motivation}

The   standard  unitary  quantum  dynamics   $\exp(-iHt/\hbar )$
and the  Schr\"{o}dinger semigroup-driven random motion  $\exp
(-tH/\hbar )$  are examples of dual evolution scenarios, connected
 by means of an    analytic continuation in time (here e.g. $it \rightarrow t$ for times $t\geq 0$).
 Both  types of motion   share in  common a local  Hamiltonian operator $H$.
 Its  spectral resolution is known to determine simultaneously transition
 amplitudes of the Schr\"{o}dinger picture  quantum
motion in $L^2(R^n)$  and   transition probability densities of  a
space-time homogeneous  diffusion  process in $R^n$, with $n\geq 1$.
The considered Hamiltonians  have  the form $H=T+V$, where
the energy operator $T$ derives from the Laplacian and $V$ is a
locally defined  confining  potential.

Within the general theory of so-called infinitely divisible
probability laws  the familiar Laplacian (probabilistically interpreted as the Wiener noise
 or Brownian motion generator) is  merely one isolated member of a  rich family
of non-Gaussian  L\'evy noise generators. They  stem from the
fundamental L\'{e}vy-Khintchine formula, provided we restrict considerations to   symmetric (typically
heavy-tailed)     probability distributions of spatial jumps and
resultant jump-type Markov processes, c.f.  \cite{GS}.

  The emergent
L\'{e}vy generators  are manifestly nonlocal (pseudo-differential)
operators  that   give rise to L\'{e}vy-Schr\"{o}dinger semigroups and
 to  nonlocally-induced  random  dynamics.  The   dual (Euclidean)
image  of the latter,  comprises unitary dynamics scenarios that exemplify   an
inherently   nonlocal  quantum behavior.

 The  canonical quantization  concept we   introduce indirectly,  by  choosing   the Hilbert space $L^2(R^n)$ as an arena for
  investigations. From the start we have the Fourier transformation realized as a unitary operation in this space, a pre-quantum  version
of uncertainty relations (due to G. H. Hardy, c.f.  \cite{GS}) and
standard (i.e. pedestrian  nonrelativistic quantum mechanics)  notions of  position and momentum operators
as a consequence.     Our further discussion will be restricted to one
spatial dimension ($1D$), which is  not a must but
 a pragmatic simplification of otherwise  potentially clumsy reasoning, see  however  Ref. \cite{GS} for  less restrictive considerations.

The   L\'{e}vy-Khintchine formula, while  tailored   for our
purposes,  derives from a Fourier transform of a symmetric
probability density function. A variety of such  probability laws
for random noise is classified by means of a characteristic function
which, while restricted to $1D$,  is an exponent $\eta (p)$ of the
$(2\pi )^{1/2}$-multiplied Fourier transform of that probability
density function (pdf), $\int dx\, \rho (x) \exp(\pm ipx)= \exp[\eta
(p)]$.  The canonical quantization recipe  $p  \rightarrow \hat{p} =
-i\nabla $ (natural units  $\hbar =1= c$ being implicit),
 while  executed  upon  the characteristic function,   induces
  L\'{e}vy-Schr\"{o}dinger semigroups $ \exp[ t\eta (\hat{p})]$ that drive   random  jump-type processes.

As mentioned before,  dual partners  of semigroup operators  are unitary evolution
operators $ \exp[it \eta (\hat{p})]$  which  here by  set a  broadened  (e.g.
going beyond the local paradigm)  quantum   mechanical context.

By  redefining the characteristic exponent as $\eta (p) = - F(p)$
and subsequently $F(\hat{p})= T$, we can
 classify  probability laws   of interest and next  the  emergent energy operators,    in  a bit  more physical vein.
 Those are: (i) symmetric stable laws that correspond to $F^{\mu }(p) = |p|^{\mu } $, with
 $\mu \in (0,2)$ and give rise to   so-called fractional energy operators  $T^{\mu }= (-\Delta )^{\mu /2} \doteq  |\nabla |^{\mu }$,
  (ii) quasirelativistic probability law   inferred from   $F_m(p) =
 \sqrt{p^2+ m^2} - m$, $m>0$ which  is  a rescaled  version (no $c$)
  of a classical relativistic  Hamiltonian  $\sqrt{  m^2c^4 + c^2p^2} - mc^2$.  Accordingly, in natural units  $\hbar = c =1$,
    $T_m= \sqrt{-\Delta + m^2} -m$
  stands for the quasirelativistic energy operator.
  We note that $F^{1}(p) = |p|$  determines  the Cauchy probability law and gives rise to  Cauchy
   operator, here   denoted  $T^1= (-\Delta )^{1/2} \doteq |\nabla |$. Clearly $T^1=T_0$.

We  are interested in solving   Schr\"{o}dinger-type  eigenvalue
problems for  Hamiltonians of the form $H=T+V$, where $T$  may be  a
nonlocal energy operator, while   $V$ is   a locally defined
confining  potential. The latter we  specify   to be either harmonic
or   refer to  a  finite well of arbitrary depth.  Under these
confining conditions, the  Cauchy  oscillator and  Cauchy   (in)finite
well were investigated Ref. \cite{GZ}.

 In the present paper we
shall consider  quasirelativistic Hamiltonians   $T_m$  as energy
operators of interest and subsequently   compute  a number of their
nonlocally-induced bound states in  harmonic and finite well
regimes. Recently reported    approximate  quasirelativistic infinite well spectral solution ($m>0$  "particle in the box" problem
\cite{KKM}), together with that for the Cauchy infinite well \cite{K,GZ}  and known   spectral solution for the Cauchy (massless) harmonic oscillator \cite{SG,LM},
 provide  verification tools for our quasirelativistic spectral  results, once we turn over to the  $m\ll 1$ regime  of the corresponding  quasirelativistic spectral
 problems.  In the $m\gg1$ extreme  a  direct comparison will prove possible with the  standard   nonrelativistic spectral data.  We shall give  more explicit
 meaning to   those   "small" versus  "large"  mass regimes in below.

   If an   analytic  solution of  the "normal"  Laplacian-based
  Schr\"{o}dinger eigenvalue problem  is not in the reach,  a recourse   to the
    imaginary time propagation technique  (to evolve the system in "imaginary time", to employ "diffusion algorithms") is a standard routine
 \cite{BBC}-\cite{Chin}.
  There exist a plethora of methods (mostly computer-assisted, on varied levels of sophistication and approximation finesse) to address
the spectral solution of  local   1D-3D  Schr\"{o}dinger operators
in various areas of quantum physics  and quantum chemistry.
Special emphasis is paid  there  to  low-lying  bound states, were
"low-lying" actually means that  even few hundred of them are
computable.

The major  goal of  the present  paper  is  to  generalize  the
above  mentioned  "diffusion  algorithms"  so that the resultant
"jump-type algorithms"
 would   provide   {\it reliable}   high accuracy  approximations to {\it true}  spectral solutions  for
  the   quasirelativistic Hamiltonian  in the   wide  mass parameter range $m\in (0,\infty )$.
All computations are  carried out in  configuration space, thus
deliberately  avoiding a customary  usage of Fourier transforms
which  blur an  inherent spatial nonlocality of the problem. We
keep  under control   the balance between   the nonlocality impact
and   various (lower and upper)  bounds upon the integration volume
and  the space-time  intervals  partitioning  finesse,  that are unavoidable in  numerical procedures.

We are very detailed about the (bottom)  part of the spectrum, somewhat disregarding higher eigenvalues
(except for a number of approximate formulas). Some steps (like e.g. the choice of
the Gram-Schmidt orhonormalization procedure) of the spectrum generating
 algorithm were tailored specifically to this end.

Compared with nonlocal spectral problems considered in the literature so far, even though our  computations are carried out
for rescaled versions of original  models (thus devoid of  explicit  physical dimensions),  we  have kept intact the
 mass (for all models) and  the  well width and depth    dependence.
 Moreover, albeit with dimensionless   computation  outcomes  in hands,   we  can   fully  recover all  physically relevant
 characteristics of considered models. An extended Appendix C gives details about how to  eliminate and reintroduce  physical
 (dimensional)  constants, plus an assessment of involved  length and energy scales.

\section{Spectrum-generating algorithm: An outline.}

To deduce a spectral resolution  (e.g. find  eigenvalues and
eigenfunctions) of   a self-adjoint   non-negative  operator
   $H$, it is   the "imaginary time propagation"  i.e.
    the semigroup dynamics $\exp (-tH)$ with  $t\geq 0$ which is particularly well suited to this end, \cite{BBC,AK}.
    That, in view of  obvious domain and  convergence/regularization  properties which   are implicit in the
Euclidean (or statistical like e.g. the partition function
evaluation) framework.

Let  us consider the  eigenvalue  problem for   a self-adjoint
operator  $H$  of the form  $H = T + V$, assuming that (at least a
part of)  the spectrum is strictly positive,  discrete  and
non-degenerate $0<E_1 < E_2 < E_3 < \ldots $ (the latter restriction
may be lifted,
  since it is known how to handle degenerate spectral problems, \cite{BBC,AK}):
 \be H\,\psi_i(x) =
E_i\psi_i(x),\qquad i=1,2,\ldots, \ee
 where  $T$   is \it  not \rm
 necessarily a local differential operator (like the negative of the
Laplacian), but a  nonlocal (pseudo-differential) operator.

In below we shall mostly refer to   nonlocal operators  $T$ defined
through their action on  suitable $L^2(R)$  functions in the domain
of $H$:
 \be
  T\,\psi(x)  = p.v. \int [\psi(x)-\psi(x+z)]\, \nu (dz),
\ee where $\nu (dz)=\nu(z) dz$ stands for so-called L\'{e}vy measure and
generically  the  $1D$  integral  in Eq. (2)  is interpreted in
terms of its  Cauchy principal value: $p.v. \int f(z)\nu (dz) =
\lim\limits_{\varepsilon\to
0}\int_{R\backslash(-\varepsilon,\varepsilon)} f(z) \nu (dz)$.

The choice of  $\nu (z)= 1/(\pi z^2 )$   identifies the  Cauchy
operator   $T = (-\Delta )^{1/2} \doteq |\nabla |$, while that of
\be \nu_m(z)=\frac{m}{\pi}\frac{K_1(m|z|)}{|z|},\label{l3} \ee where
$K_1$  is a modified Bessel function of the third kind, defines the
quasirelativistic operator $T_m = \sqrt{-  \Delta + m^2} -m$.

To define the spectrum-generating algorithm, we first need  to
introduce an  approximation of the original semigroup dynamics $\exp
(-tH) \psi $, of  a suitable  initial data vector
 $\psi $ for arbitrary $t>0$,  by a composition of  a large number of consecutive
  small time  "shifts".  To this end a recourse
 to Trotter-type formulas is necessary and the  Strang  splitting  method produces
  a number of their approximations  of varied orders.

In the present paper we shall focus on  the  simplest  second order
Strang   approximation of  the semigroup  operator $\exp (- H \Delta
t)$, where $H= T+V$  and $\Delta t \ll 1$,  that has been widely
used \cite{Auer} in quantum physics and quantum chemistry contexts.
The splitting identity
 \be
e^{-H\, \Delta t}  \approx e^{-\frac{\Delta t}{2}V}e^{-\Delta
t\,T}e^{-\frac{\Delta t}{2}V} \label{l1}
 \ee
holds   true   up to terms of order  $ \mathcal{O}((\Delta t)^3)$.
Like in the standard  quantum mechanical  perturbation theory, the
interpretation of  the $\mathcal{O}(t^3)$
 term as "sufficiently  small" remains somewhat obscure, unless
specified with reference to its action  on functions  in the
domain of $H$.

  A   preferably long  sequence     of consecutive small time  $\Delta t\doteq h$   "shifts"    of
    an initially given function $\psi(x,0) \rightarrow \psi (x,kh)$ with  $k= 1, 2, ...$,
    mimics  the  actual continuous evolution
    of $\psi (x,t)$ in  the time interval $[0, kh]$.
For   sufficiently small times $\Delta t \doteq  h$ we may take one
more approximations step (keeping  e.g. second and higher
   order terms of the Taylor series would improve
an  approximation  accuracy):
  \be e^{-h H}\approx
e^{-\frac{h}{2}V}\left(1-h T\right)e^{-\frac{h}{2}V}  \doteq
\mathcal{S}(h).\label{shift} \ee

 The induced approximation error
depends on the time step  $h$ value. If $h$ is small,  the error is
small as well but
  the number of iterations towards  first  convergence symptoms is becoming large. Thus a proper balance between
  the two goals, e.g. the  accuracy level and the  optimal convergence performance, need to be established.
  (One more source of inaccuracies   is rooted in the nonlocality of involved operators and
  spatial cutoffs needed to evaluate integrals. This issue we shall discuss later.)

We note that an optimal  value of a "small" time shift   $\Delta t
=h$,
 appears to be model-dependent.   Subsequently, we shall   refer to   $h=0.001$.

  An outline of the  algorithm that is appropriate for a numerical implementation
   and ultimately  is capable of   generating   approximate  eigenvalues and eigenfunctions
    of $H$, reads as follows:

  (i) We  choose a finite number  $n$
  of    trial  state  vectors (preferably linearly independent) $\{ \Phi_i^{(0)}, \, 1\leq i\leq n\} $, where
     $n$ is correlated with an ultimate number
   of eigenvectors of $H$ to be obtained in the numerical procedure;
      at the moment we disregard an issue of their optimal (purpose-dependent) choice.

   (ii)  For all trial functions   the time  evolution beginning    at $t=0$ and
terminating at  $t= h$, for all $1\leq i \leq n$  is  mimicked by
the time shift  operator $S(h)$  of Eq. (\ref{shift})
 \be \Psi_i^{(1)}(x)=S(h)\Phi_i^{(0)}(x). \ee

 (iii) The obtained set of  linearly independent  vectors
$\{\Psi_i^{(1)}\}$ should be made orthogonal  (we shall use the
familiar Gram-Schmidt  procedure, although there are many others,
\cite{AK})  and normalized. The
 outcome  constitutes a {\it new} set   of trial states
 $\{ \Phi_i^{(1)}, i=1,2,\ldots, n\}$.

(iv) Steps (ii)  and (iii) are next repeated  consecutively, giving
rise to a temporally ordered sequence of    $n$-element
orthonormal   sets  $\{ \Phi_i^{(k)}(x), i=1,2,\ldots, n\}$ and the
resultant  set of linearly independent vectors \be
\Psi_i^{(k+1)}(x)=S(h)\Phi_i^{(k)}(x),\qquad i=1,2,\ldots,n,
\nonumber \ee
 at time $t_{k+1}= (k+1) \cdot h$.  We main abstain from its orthonormalization  and stop the iteration procedure, if
 definite symptoms of convergence  are detected.
 A discussion of   operational convergence criterions can be found e.g.  in Ref. \cite{Chin}.

(v) The   temporally ordered  sequence of $\Phi_i^{(k)}(x)$,  $k\geq
1$
  for  sufficiently  large  $k$  is   expected to converge to an eigenvector  of $S(h)$, according to:
\be S(h)\Phi_i^{(k)}(x)= e^{- h E_i^{(k)}}\Phi_i^{(k)}(x) \approx
e^{-  h E_i}\psi_i(x),  \label{E} \ee
 where
$\psi_i$ actually   stands for an eigenvector of  $H$  corresponding
to the eigenvalue $E_i$. Here: \be
 E_i^{(k)}(h) = -\frac{1}{h}\ln(\mathcal{E}_i^{k}(h)),  \label{l3}
 \ee
where \be
\mathcal{E}_i^{k}(h)=<\Phi_i^{(k)}|\Psi_i^{(k+1)}>=<\Phi_i^{(k)}|S(h)\Phi_i^{(k)}>,\nonumber
\ee is  an expectation value of   $S(h)$ in the $i$-th state
$\Phi_i^{(k)}$.

 It is   the evaluation of  $\Phi_i^{(k)}(x)$ and    $E_i^{(k)}(h)$ that is amenable to
  computing routines  and yields approximate eigenfunctions and   eigenvalues of $H$. The degree of
 approximation   accuracy is   set by the terminal  time  value $t_k=kh$,  at which
  earlier  detected symptoms of convergence ultimately   stabilize, so that the iteration (i)-(v) can be   stopped.\\

{\bf Remark:} Even in the high-fidelity  computation
regime (c.f. \cite{BBC}-\cite{Chin}), we  never arrive at  {\it
exact} eigenfunctions and eigenvalues, but   at  their  more or less
accurate approximations.  Therefore we should  properly identify and
keep under control various computation inaccuracies,  coming from
different sources. A   model-independent inaccuracy source lies in
our choice   $h=0.001$  of the  "elementary" time shift
  (actually, a partition unit for any time interval). It is a matter of a preparatory numerical "experimentation"
 whether  the  $h$   choice  needs to be finer or not (e.g. $10^{-4}$ or $10^{-5}$).  The price paid is a significant
  computing  time increase.
Besides    a  low (second)   order of the   Strang splitting of the
semigroup operator,  other inaccuracies of numerical procedures
are model-dependent and come from the  spatial
 nonlocality of involved operators  (2)  that stays  in conflict with  cutoffs needed
  to evaluate the integrals. In $1D$,  we  need  a priori  to declare that  $x\in[-a,a]$, $a>0$.
 How wide the spatial interval should be  to yield  reliable simulation outcomes, especially  for eigenvalues
 (the eigenfunction computation is less sensitive to the choice of $a\geq 50$),
 is again a  matter of a  numerical   experimentation.  We  set the   spatial partition unit   $\Delta x=0.001$.
In view of pre-selected  $[-a,a]$  integration  boundary limits,
irrespective of the initial data choice  $\{ \Phi_i^{(0)} \in
L^2(R)\}$, the simulation  outcome     is automatically placed in
$L^2([-a,a])$. For  the  quasirelativistic  and   Cauchy
oscillators, {\it true} eigenfunctions extend over the whole real line.
 Therefore, a computer-assisted spectral solution effectively provides  an approximation of  {\it true} eigenfunctions
  by suitable  approximating  functions with a support  in  $[-a,a]$.
 Clearly, the value of  $a$ cannot be too small. We have found a   threshold  value  $a=50$   to be
  an optimal  choice (accuracy versus computation time, see also \cite{GZ}). This pertains as well to
  the  computationally "dangerous" regime of  small masses $m\in (0,1]$.  Then e.g.  the  eigenfunctions falloff at infinity
  becomes close to inverse polynomial ($\leq 1/|x|^4$ in the Cauchy case).
   We note that one  can improve an accuracy of computations in the small mass regime.
    To this  end  a partitioning of the integration interval
   should be make finer than  the adopted one   $\Delta x=0.001$  (like e.g. $0.0001$).

\section{Quasirelativistic harmonic oscillator.}

In Ref. \cite{GZ} we have  tested  a predictive power of the just
outlined  computer-assisted  method of solution of the
Schr\"{o}dinger-type  spectral problem for a non-local operator $H$,
through a comparison with  an  available   analytic solution of the
$1D$   Cauchy oscillator problem  \cite{SG,LM}. That  was
subsequently followed by  an analysis of  to the   Cauchy  finite
well problem   and an in-depth  analysis of various inadequacies of
hitherto
 proposed  (would-be)  spectral solutions of  the Cauchy  infinite well problem.

\begin{figure}[h]
\begin{center}
\centering
\includegraphics[width=100mm,height=100mm]{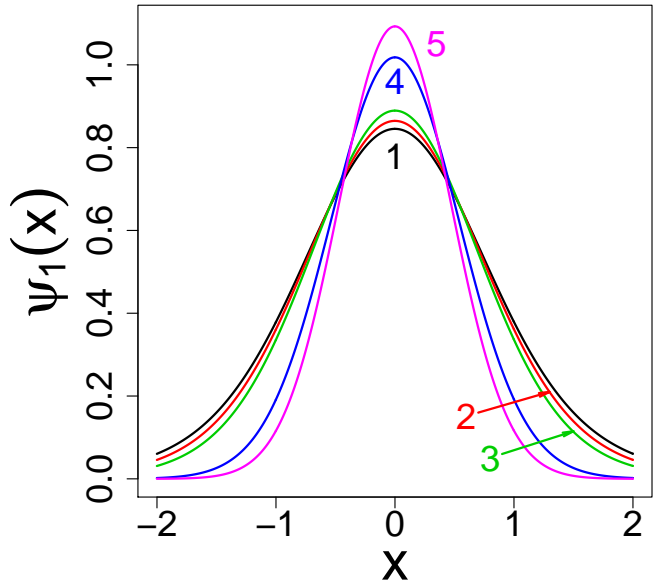}
\caption{Quasirelativistic oscillator ground state  (n=1) is  depicted for masses  $m=0.01,\,0.5,\,1,\,5,\,10$,  labeled respectively by  $1,\, 2,\, 3,\, 4,\, 5$.  A  clear
distinction  is seen  between     tentative
"small" mass $m\leq 1$ and "large"  mass $m\geq 5$ regimes. The $m=0.01$ curve is {\it fapp} identical with the ground state of the Cauchy oscillator,
 whose decay is known to be  inverse polynomial   $\sim C/x^4$, \cite{SG,GZ}.}
\end{center}
\end{figure}

In contrast to  the  $m=0$ regime, spectral data for $m>0$  quasirelativistic  harmonic oscillator (in $1D$-$3D$) are  scarce and  not available in a closed analytic form.
That enforces a computer-assisted approach, where the $m$-dependence needs to be optimally  accounted for, in the whole range $m\in (0, \infty )$.
 As far as we know the literature on the subject, neither the quasirelativistic oscillator nor the quasirelativistic finite  well problems were  ever addressed
  on a similar to  \cite{BBC}-\cite{Chin}  level of  computational accuracy.
   In fact, we can safely
 conjecture  that the spectral solution in $1D$ and $2D$ is  non-existent in the literature, while the available  $3D$ data  are rather  limited, \cite{Hall,Lucha,Remb}.

  We are aware of a long-term research on  quasirelativistic bound states  (primarily in $3D$) for various confining potentials, including
 that of the harmonic oscillator  \cite{Hall,Lucha} and the radial version of the $3D$ Cauchy oscillator \cite{Remb}.
  Interestingly, the high-fidelity computer algorithm we advocate,  has never been employed  nor mentioned  in those  contexts.
Moreover, we quite  intentionally carry   out  spatial computations only, while computations   of Refs. \cite{Lucha,Remb} were   performed
   directly in the Fourier (momentum) representation, thus  with no access  to  nonlocality-sensitive  spatial  diagnostics.

We are interested in  spectral properties (eigenvalues and eigenfunctions) of the quasirelativistic harmonic oscillator
$H= T_m  +V  = \sqrt{-\hbar ^2c^2 \Delta + m^2c^4} - mc^2   + kx^2/2$.  For computational simplicity and comparison with a  number of related
references, we shall work with a rescaled form  of that  Hamiltonian where, except for $m$, other dimensional parameters (or constants)
 are  eliminated:
\be
H= T_m  +V =  [\sqrt{-  \Delta + m^2} -m]   + x^2
\ee
The  traditional coefficient $k/2$ in $V(x) =kx^2/2$ has been  scaled away and  the
 natural system of units  $\hbar =1 =c$ is implicit.  How to eliminate or  reintroduce  dimensional constants and  infer
   typical  energy scales  c.f. the Appendix.

The major preparatory guess, for an execution of  the spectrum-generating algorithm,
 amounts to pre-selecting a suitable set  (comprising one, two or more  elements, see e.g. \cite{GZ}  for more detailed discussion)
of linearly independent trial functions.   There is a large freedom for  that choice in $L^2(R)$  and in  Ref. \cite{GZ} the nonrelativistic harmonic oscillator basis (hermite functions) has
been employed.

We are  motivated by the fact that  whatever this trial set is  and whatever is its support ( $R$ or  $[-1,1]\subset R$),
   in view of the integration volume restriction to $[-a,a]$,     simulation
 outcomes are  unavoidably placed  in $L^2([-a,a])$ and  $a=50$ is used   throughout the paper.
  A computationally convenient choice    of trial functions appears to be the standard
 nonrelativistic  infinite well ("Laplacian in the interval")  eigenbasis   for  $[-1,1] \subset R$  which can be  trivially extended  to
  orthonormal   $L^2(R)$ functions  as follows:
\be
\Phi_{n=2l-1}^{(0)}(x)=\left\{
                    \begin{array}{ll}
                      A\cos\left(\frac{n\pi x}{2}\right), & \hbox{$|x|<1$,} \\
                      0, & \hbox{$|x|\geqslant 1$}
                    \end{array}
                  \right.\qquad
\Phi_{n=2l}^{(0)}(x)=\left\{
                    \begin{array}{ll}
                      A\sin\left(\frac{n\pi x}{2}\right), & \hbox{$|x|<1$,} \\
                      0, & \hbox{$|x|\geqslant 1$}
                    \end{array}
                  \right.\qquad
l=1,2,\ldots\nonumber
\ee
Here $A=\pm 1$.

Anticipating further discussion, we need to mention that numerical
outcomes for simulated eigenvalues  are $a$-sensitive in the small
mass regime $m\ll 1$. Here small means e.g. $m=0.001,\,  0.01$, albeit our subsequent discussion will validate $m=0.5$
or even $m=1$  to be "sufficiently"  small.  However one  needs to know that  for $m=1$ the choice of $a=20$ gives
practically the same outcomes as those  for $a=50$  or $a=100, \, 200$.
(Our previous Cauchy oscillator discussion, \cite{GZ}  (see e.g. Figs. 1, 3 and 6),  proved that appreciable (detectable)  differences between computed
   lowest  eigenvalues  decrease, but still  persist,  while $a$ increases from $a=50$ through $a=100$, up to  $a=500$.)

 To the contrary, approximate  low energy  eigenfunctions can be satisfactorily reproduced within relatively small spatial  interval  like e.g.  $[-3,3]$ or
 $[-5,5]$, beyond which  these functions  quickly decay. Their shape dependence on the  integration bound  $a \geq 50$ is residual and  for all practical purposes
 ({\it fapp})   can be  neglected.

  Our   numerical experimentation has  shown  definite  stabilization/convergence  symptoms
    after about $1.500 - 2000$ small  $h$-time shifts (5)-(8), when
 computed eigenvalues (and shapes of  eigenfunctions)  effectively   stop to change    within  the adopted error limits (that pertains to the eigenvalues evaluation up to four decimal digits). \
 We have found  $k=2500$  to set  an optimal  {\it  terminal}  stabilization
  "time" $t_k= k\, h$ at which our spectrum-generating algorithm can be  stopped and data stored.   To get more accurate data   (up to the seven or eight  decimal digits),
  the stabilization time   should  be increased (to $4.000$ or   more $h$-time steps).

\begin{figure}[h]
\begin{center}
\centering
\includegraphics[width=70mm,height=70mm]{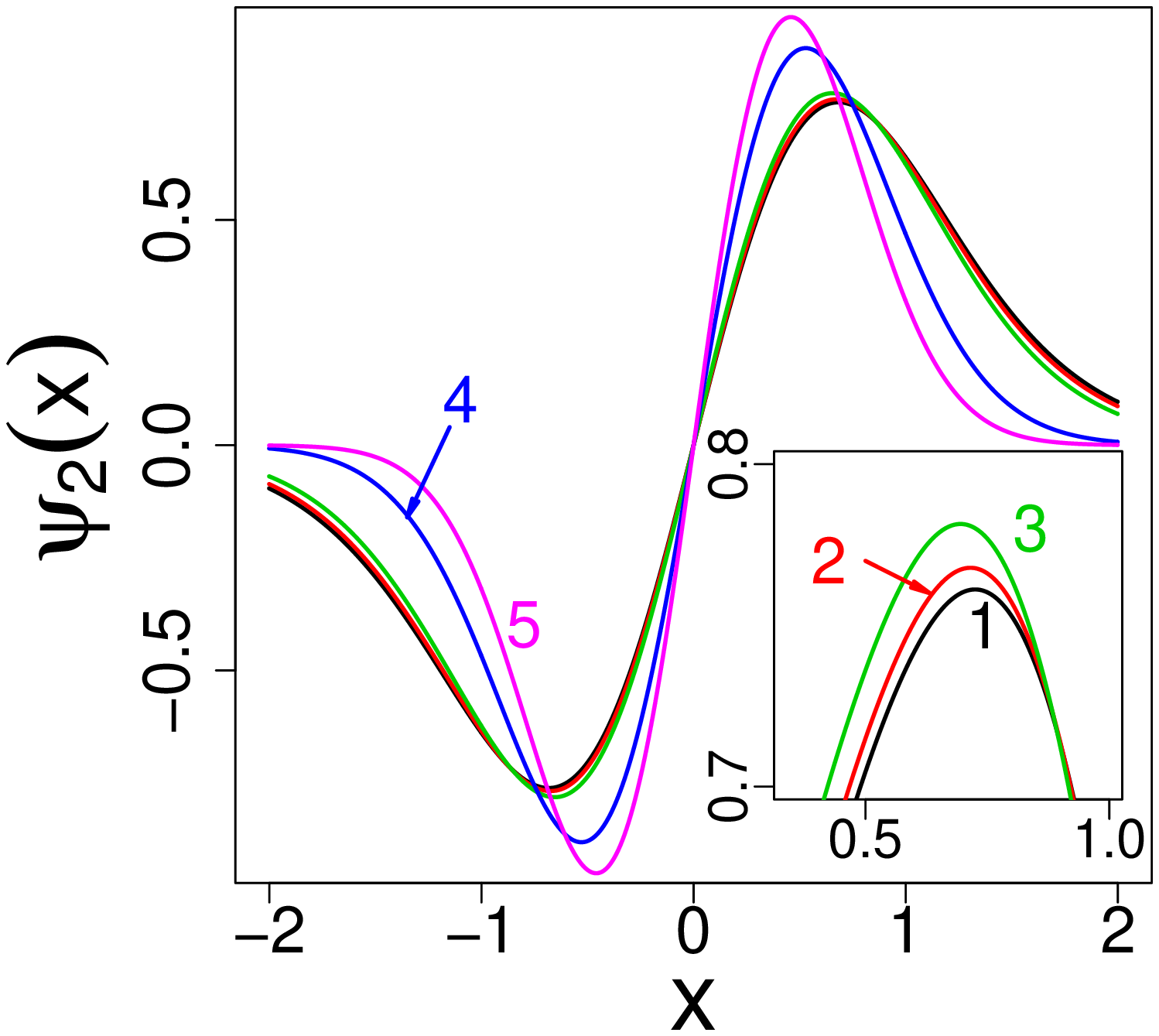}
\includegraphics[width=70mm,height=70mm]{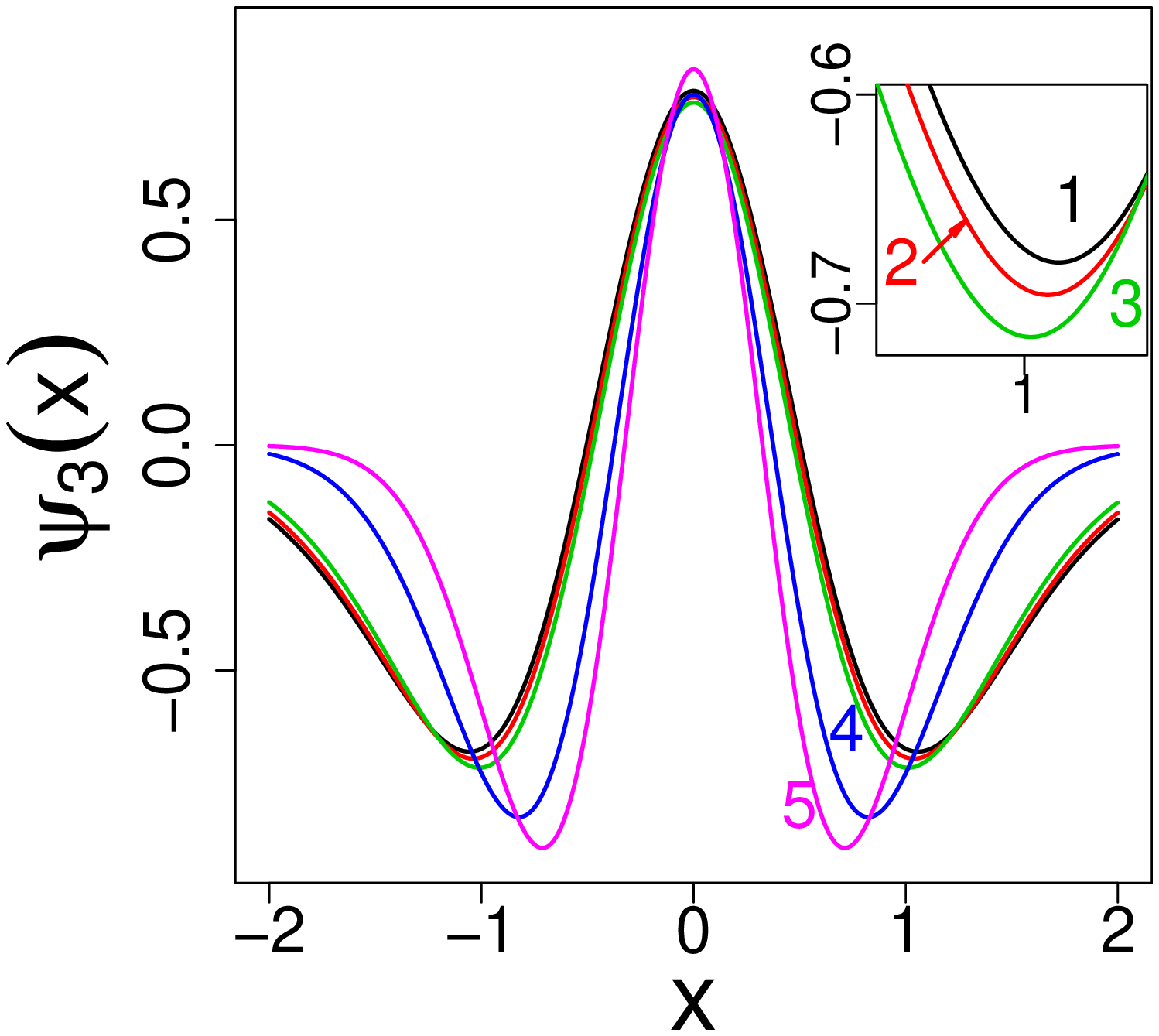}\\
\includegraphics[width=70mm,height=70mm]{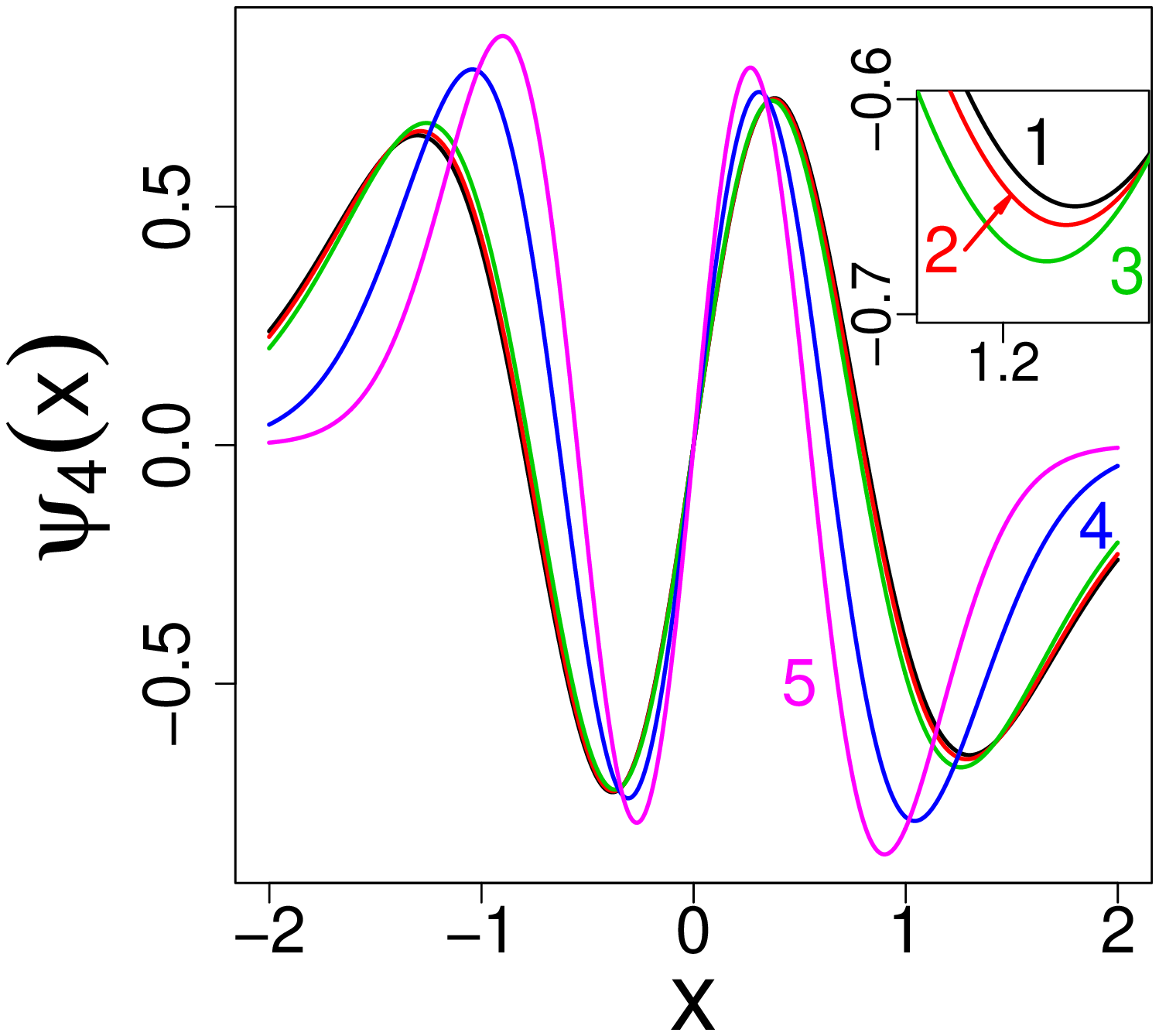}
\includegraphics[width=70mm,height=70mm]{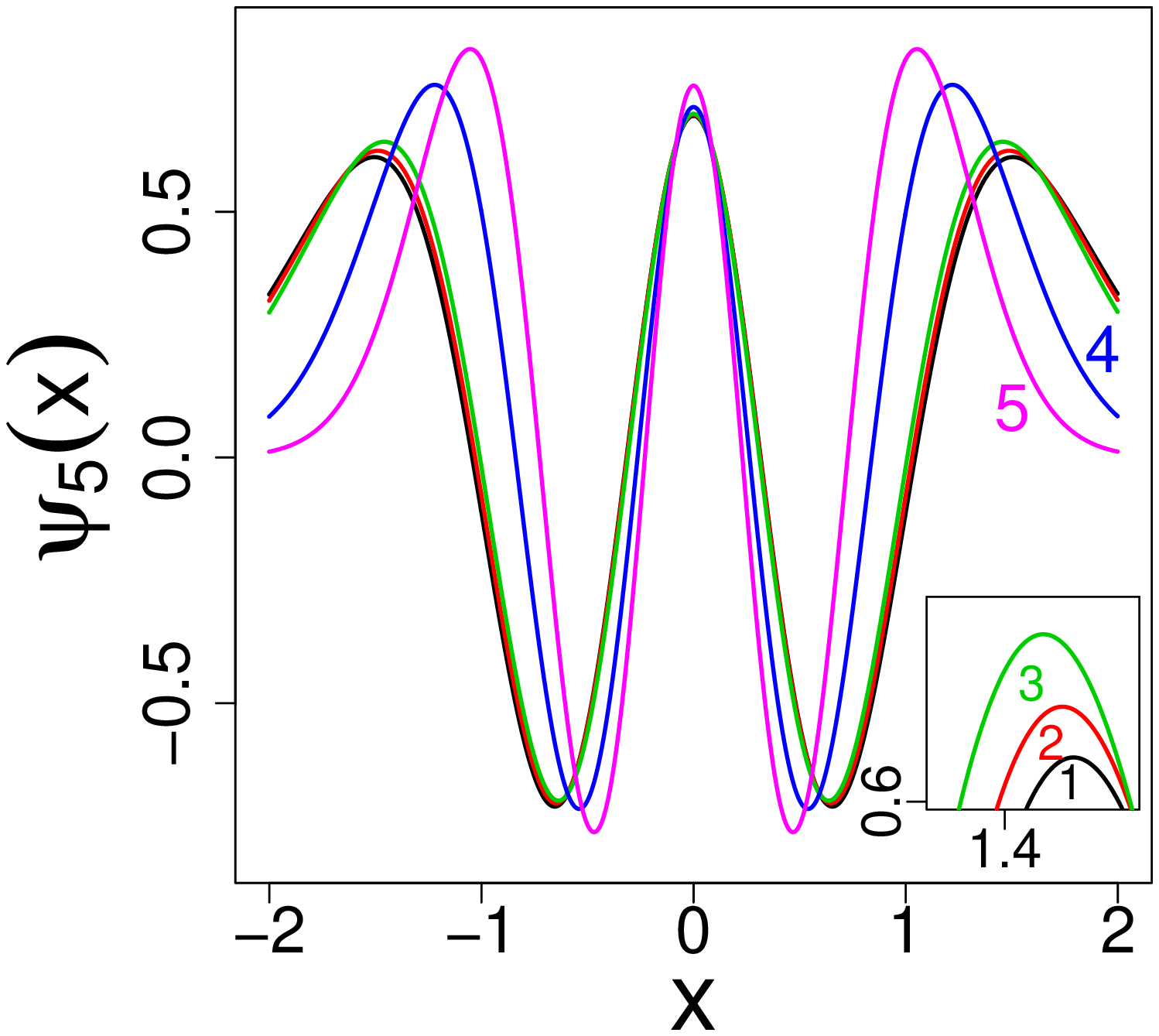}
\caption{Quasirelativistic oscillator excited states   ($n=2, 3, 4, 5$)  for  $m=0.01,\,0.5,\,1,\,5,\,10$, labeled respectively by  $1,2,3,4,5$, (parametr $a=50$). We note a clustering of
curves in the "low" mass regime. Insets depict an  enlarged  vicinity of the local  minimum/maximum  for
 curves $1,2,3$,  identifiable by  respective  $(x,\psi (x))$  coordinates.}
\end{center}
\end{figure}

\begin{figure}[h]
\begin{center}
\centering
\includegraphics[width=80mm,height=80mm]{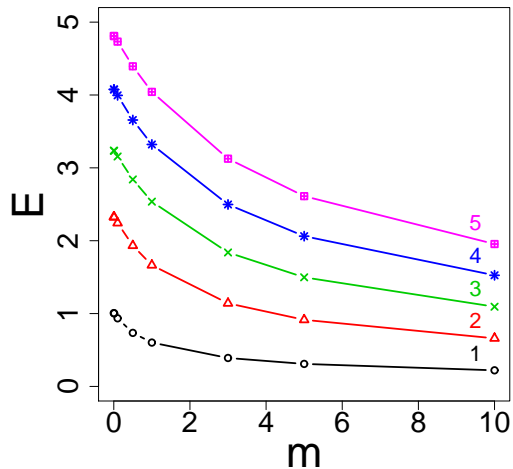}
\caption{The $m$ dependence of the quasirelativistic oscillator eigenvalues with $n=1,\, 2,\, 3,\, 4,\, 5$.
 Employed  $m>0$  values read:  $0.001,\, 0.01,\, 0.1, \,  0.5,\, 1,\, 3,\, 5,\, 10$. The $m=0$ energy values have
  been  directly  imported  from the spectral solution of the Cauchy oscillator \cite{SG,LM}  and cannot be
   graphically distinguished from those for
  $m=0.001$.}
\end{center}
\end{figure}

 Since,  for the quasirelativistic oscillator   we are interested in the $m$-variability   of eigenvalues and eigenfunctions of
 $H$, albeit  unfortunately  with no analytic formulas at hand,  spectral data  need to be computed for a number of  explicit  representative
 values of $m\in (0, \infty )$.    We  systematically refer  to $m= 0.01,\, 0.5,\, 1,\, 3,\,  5,\, 10$,  with brief appearances of  $m=0.001$  and  $m=20,\, 50,\, 100,\, 200$, if
 a deeper insight into $m\ll 1$ or $m\gg 1$ regimes  is necessary.

 Let us add  that for low-lying part of the spectrum, the decay properties  of involved Bessel functions  (3) get  amplified  by the mass parameter  increase.
 Thus e.g.  in case of  $m=1$, for   $|x|> 20=a$,  tails of  Bessel functions are bounded from above by  $10^{-21}$.  For $m=5$, the integration
 bound $a=15$ or $a=20$ would give as good approximate results as that of $a=50$.
Even for  a relatively small mass  $m=0.5$, the integration bound $a=40$ would suit pragmatically oriented scholars (e.g. accepting some degree of robustness
 in numerical calculations and the above mentioned  {\it fapp } criterion).

In Fig. 1, the  ground state function of  $H= H_m= \sqrt{-  \Delta + m^2} -m  + x^2$  is depicted  for mass parameter values
 $m=0.01,\,0.5,\,1$  (regarded as "small"; notice a conspicuous clustering of pertinent curves) and $5,\,10$
 (tentatively regarded as "large").
For small $m$  values  curves stay in a close vicinity of the  Cauchy  oscillator Hamiltonian  (an ultrarelativistic $m=0$ limit  of $H=H_m$).
In case of $m=0.01$, within adopted graphical accuracy limits,   the corresponding curve  $1$ cannot be distinguished from the Cauchy oscillator ground state
(c.f. Fig. 1 in Ref. \cite{GZ}).

Lowest  excited states ($n=2, 3, 4, 5$)   are depicted in Fig. 2, for the same masses and  $a$   as in Fig. 1.
 "Small" $m$  curves  $1,\, 2,\, 3\, $ cluster   in a close vicinity of the  Cauchy oscillator excited states.  Those labeled by $1$ are {\it fapp} identical with their
Cauchy relatives, see \cite{GZ}.

As  mentioned before, for the quasirelativistic oscilator,   an accuracy  with which   the eigenvalues in the  "small" mass regime
 are computable, is  $a$ sensitive. This issue we shall discuss in the
next subsection.

 Interestingly, beginning from  $m \geq 1$  this  $a$-sensitivity practically disappears, and our choice of $a=50$ is definitely  oversized.
 Since the computing time drops down considerably for smaller   values of $a$, we  have positively tested  an adequacy of $a<50$ integration bounds.
  Below we list  first five   numerically obtained eigenvalues,   where  for   $m=1, 3, 5$   integrations we use  $a=20$ ,
  for  $m=10, 20$  we have found  $a=10$ to be reliable,   while  for  $m\geqslant 50$,  the  bound   $a=5$ proved to be sufficient.

\begin{table}[t]
\begin{tabular}{|c||c|c|c|c|c|c|c|}
  \hline
     $V(x)=x^2$  & m=1   & $m=3$ & $m=5$ & $m=10$ & $m=20$ & $m=50$ & $m=100$ \\
        \hline\hline
  $E_1$ & 0.6020   & 0.39043 & 0.30891 & 0.22112 & 0.15669 & 0.09936 & 0.06865 \\
  $E_2$ & 1.6638  & 1.1408  & 0.91436 & 0.65998 & 0.46904 & 0.29639 & 0.20562 \\
  $E_3$ & 2.5362  & 1.8385  & 1.4974  & 1.0939  & 0.77957 & 0.49125 & 0.34230 \\
  $E_4$ & 3.3210  & 2.4971  & 2.0620  & 1.5252  & 1.0886  & 0.68591 & 0.47874 \\
  $E_5$ & 4.0426  & 3.1253  & 2.6111  & 1.9540  & 1.3962  & 0.88136 & 0.61508 \\
  \hline
\end{tabular}
\caption{Quasirelativistic oscillator: $m$-dependence of lowest five eigenvalues.}
\end{table}

In Fig. 3 we display the $m$-dependence ($m \in [0.001,10]$) of first five computed quasirelativistic oscillator eigenvalues, where the  small mass behavior clearly
indicates a convergence towards the Cauchy oscillator spectrum. On the other hand,  the large mass  extreme (here  reaching merely $m=10$) allows us to anticipate
 an affinity with the spectral solution of the  nonrelativistic harmonic oscillator, to be analyzed subsequently.

In Table I, for  reference,   the $m$-dependence of  five  lowest eigenvalues is  presented in the mass range $[1,100]$.
The detailed analysis of the small mass regime  we have relegated to the  separate subsection.

\subsection{$m\ll 1$ regime}

\subsubsection{Low mass eigenvalues}

Small mass spectrum of the quasirelativistic oscillator, like that in the Cauchy case \cite{GZ}, needs the integration interval  bound $a$
 not to be small.  Actually, in the Cauchy case we have found $a=500$ to be reliable for lowest eigenvalues, while $a=50$ is
predominantly  employed in the present paper.  Therefore it essential to investigate the $a$-dependence of computed eigenvalues
for small mass values.

\begin{figure}[h]
\begin{center}
\centering
\includegraphics[width=50mm,height=50mm]{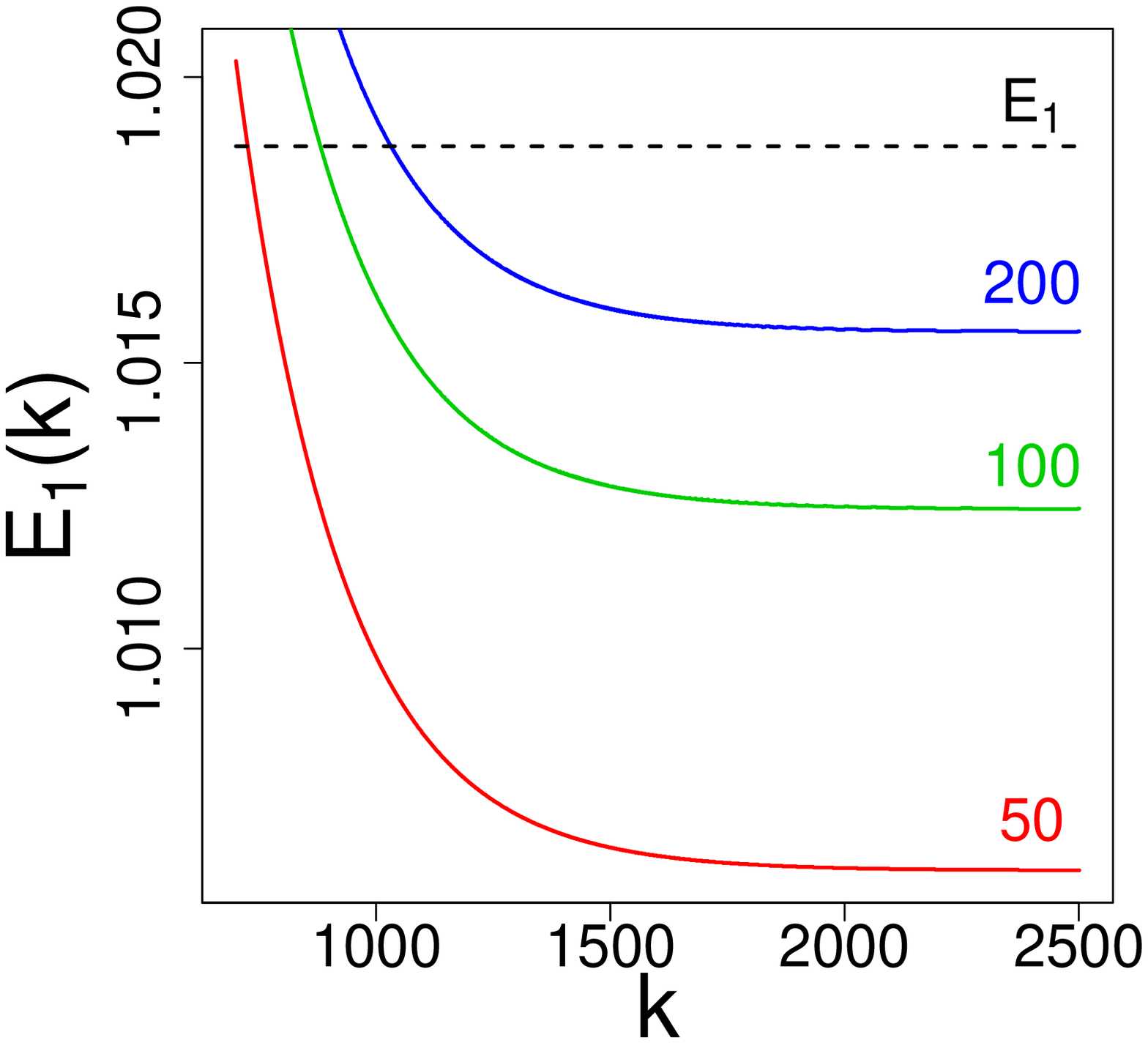}
\includegraphics[width=50mm,height=50mm]{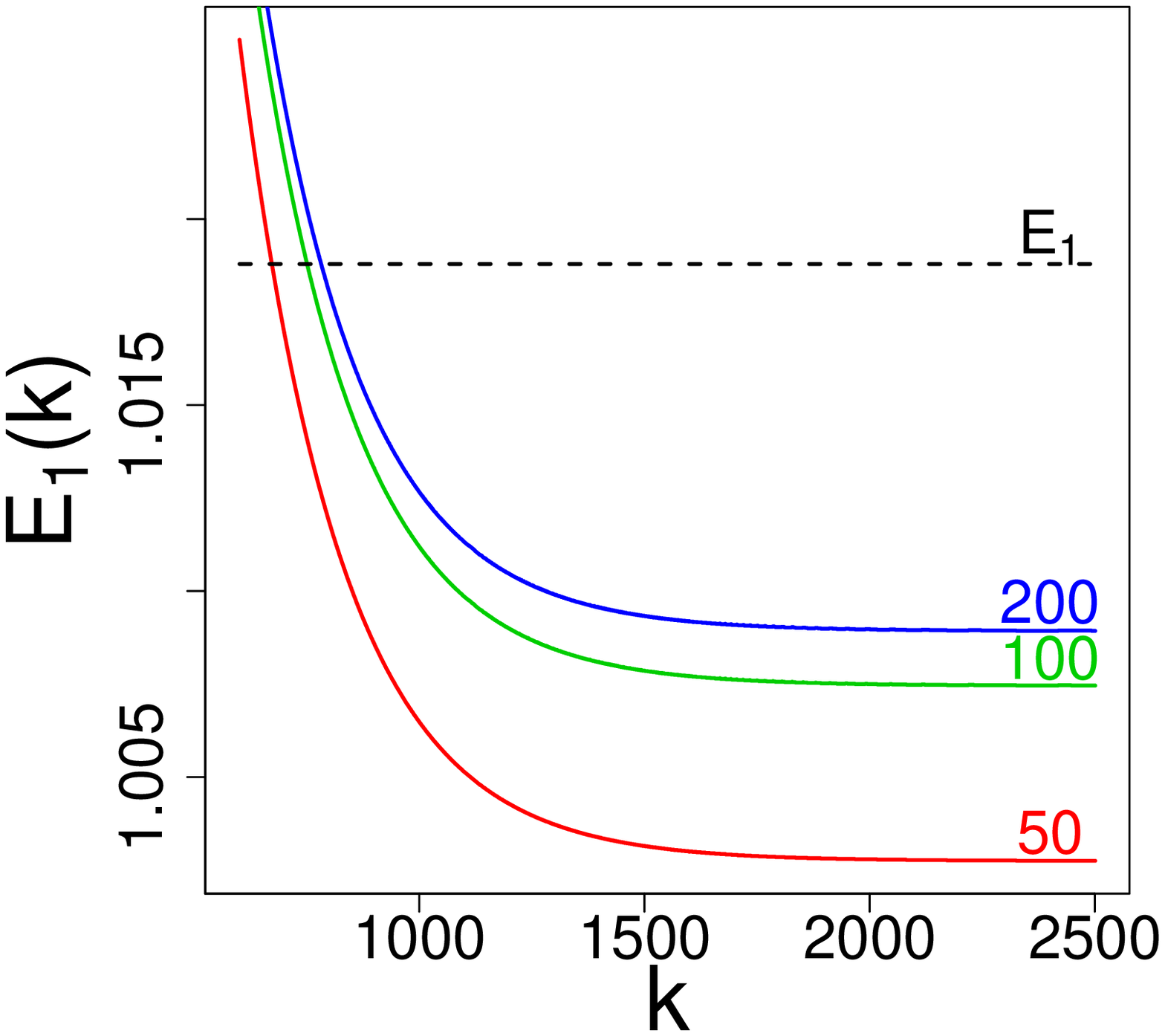}
\includegraphics[width=50mm,height=50mm]{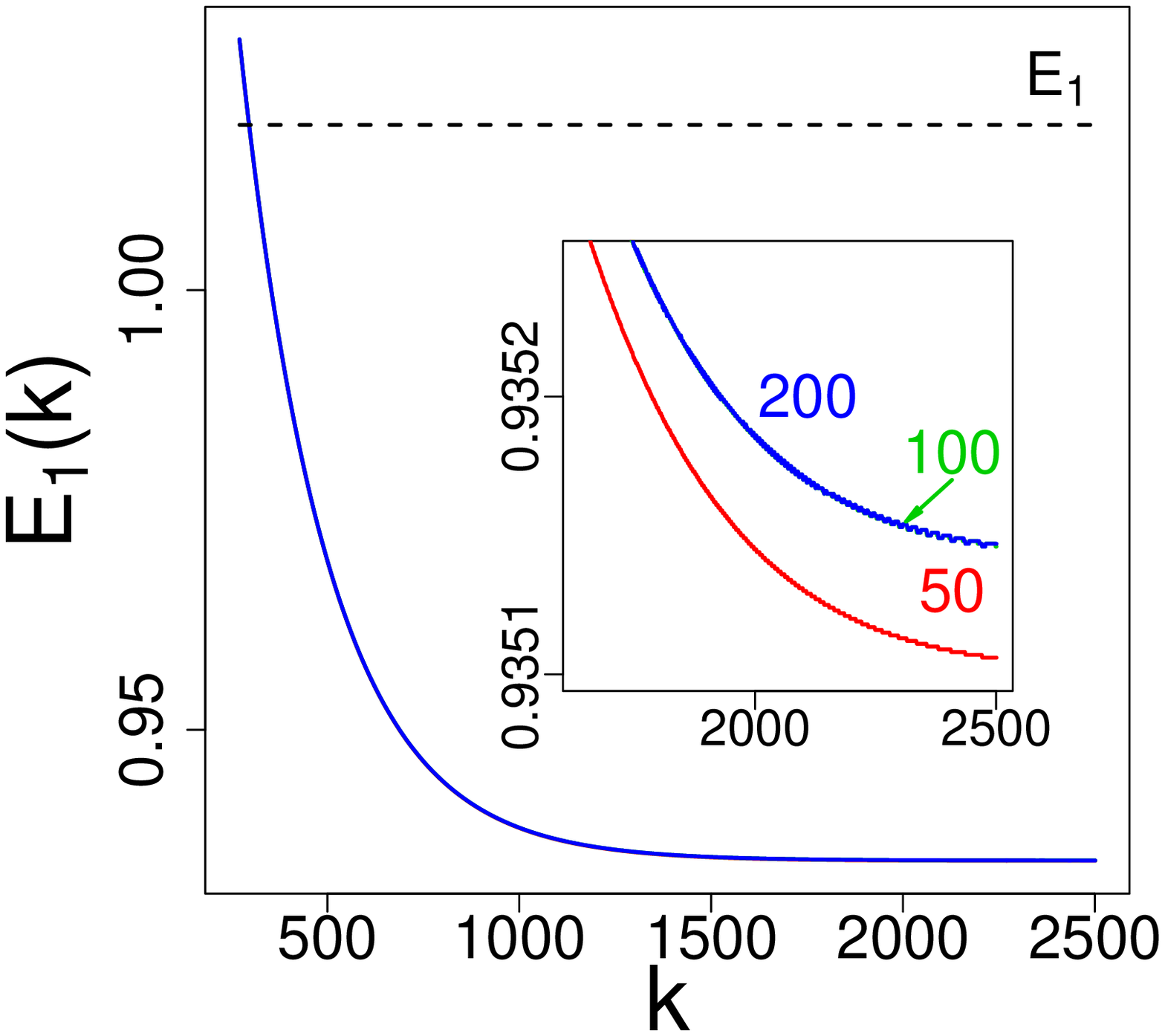}
\caption{(k)-time evolution  of $E_1^{(k)}(h) =
-\frac{1}{h}\ln(\mathcal{E}_1^{k}(h))$ (8)    and  the stabilization
symptoms  in the computation of the  ground state value:  $m=0.001$
(left panel), $m=0.01$ (middle panel) and
 $m=0.1$ (right  panel), for  $a= 50,\, 100,\, 200$.  For reference we have depicted the
 energy level  $E_1 = 1.018792$ which is set   by the Cauchy oscillator bottom eigenvalue.}
\end{center}
\end{figure}

\begin{figure}[h]
\begin{center}
\centering
\includegraphics[width=50mm,height=50mm]{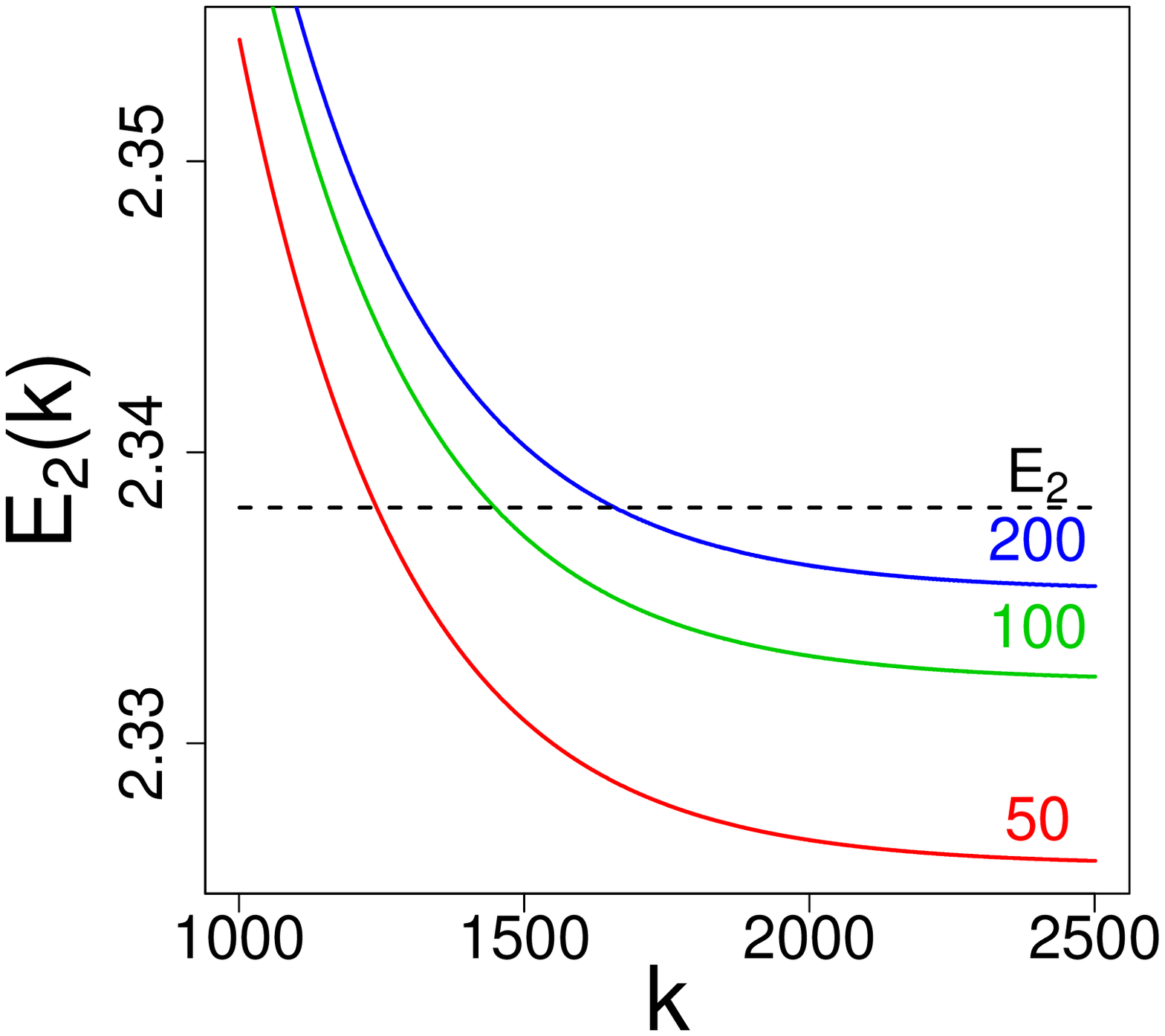}
\includegraphics[width=50mm,height=50mm]{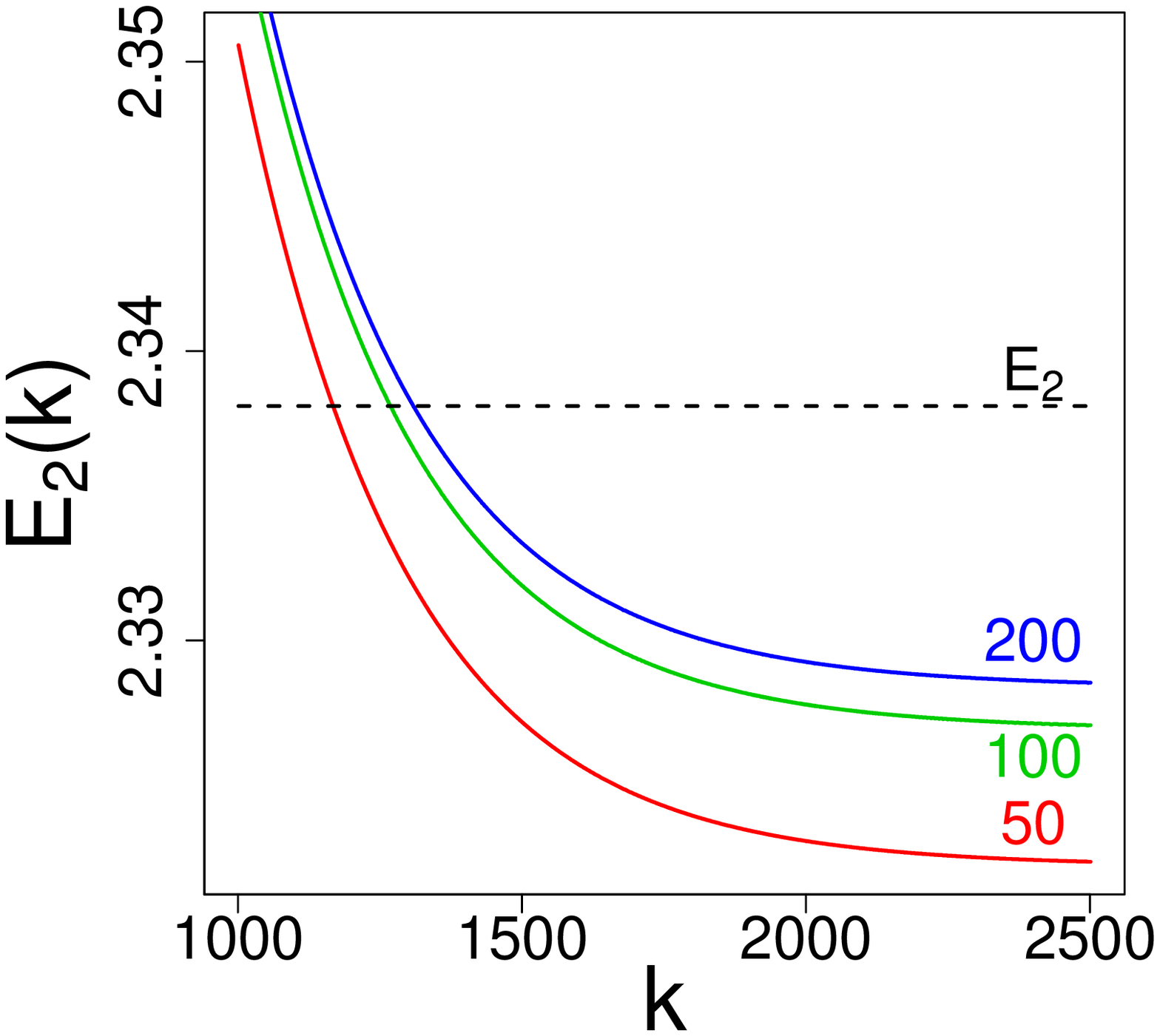}
\includegraphics[width=50mm,height=50mm]{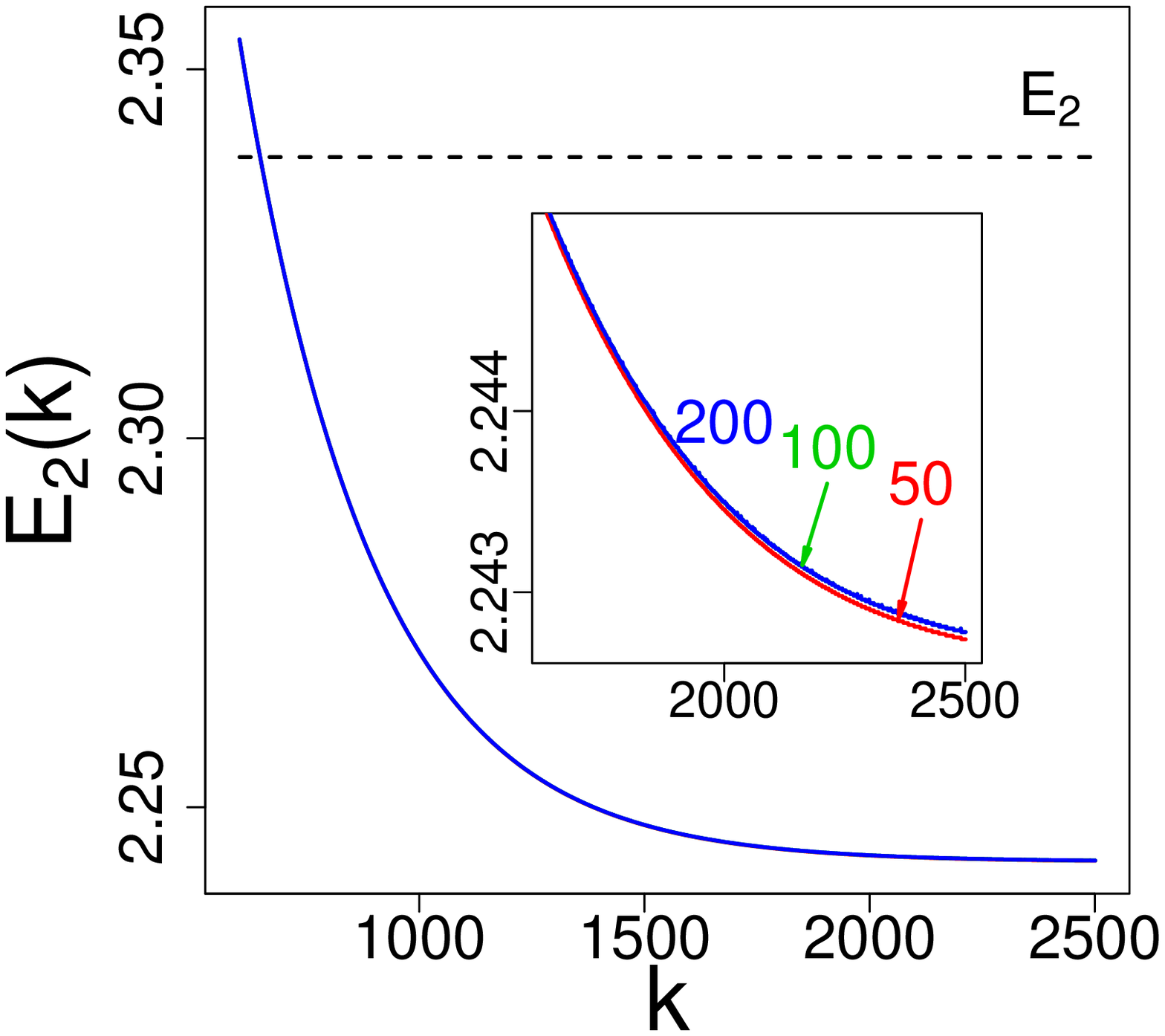}
\caption{ (k)-time evolution  of $E_2^{(k)}(h) =
-\frac{1}{h}\ln(\mathcal{E}_2^{k}(h))$ (8).  Computation of the
first excited eigenvalue for $m=0.001$ (left panel), $m=0.01$
(middle panel), $m=0.1$ (right panel),  for  $a= 50,\, 100,\, 200$.
$E_2=2.338107$ is the first excited Cauchy oscillator eigenvalue. }
\end{center}
\end{figure}

Our results are displayed  In Figs. 4 and 5 for  $m=0.001, 0.01,
0.1$, where a stabilization of the (k)-evolution (8)  is clearly
seen.   A comparison of Fig. 4 with Fig.1 of Ref. \cite{GZ} proves
that the computed $m=0.001$ ground state
 eigenvalue for $a=200$ is extremely close to that
obtained in the Cauchy case proper ($m=0$).  With the growth of $m$
the bottom spectral value drops down. Moreover, the $a$-sensitivity
quickly deteriorates. For $m=0.1$, $a=100$ and $a=200$  computation
outcomes cannot be graphically distinguished in the scale  employed.
 Albeit our primary bound  $a=50$  still can be   (residually) distinguished under
an amplified resolution, as seen in the inset of   the  Fig. 4 right
panel.
 In case of $m=1$ (not displayed), there would be no graphical
differentiation  at all  between $a=50$, $a=100$ and $a=200$
computation outcomes.

An analogous stabilization behavior can be seen in the (k)-evolution
(8) towards the first excited eigenvalue. The deterioration of
$a$-sensitivity with the growth of $m$ is perfectly seen in middle
and right panels (see the inset  for details) of Fig. 5.

\subsubsection{Spectral convergence to the Cauchy  oscillator.}

 For reference, we first display five lowest eigenvalues of the
Cauchy oscillator up to the sixth decimal digit, \cite{SG,LM}. Altogether
 $18$ lowest eigenvalues are listed  in the Appendix.
 One should be aware  that the    finesse of
  explicit    expressions  for    Cauchy oscillator  eigenvalues varies
  in the literature and happens to  extend  to  $14$ or more decimal digits. \\
\begin{table}[h]
\begin{tabular}{|c||c|c|c|c|c|}
  \hline
  m=0 & $E_1$ & $E_2$ & $E_3$ & $E_4$ & $E_5$ \\
  \hline\hline
  \cite{SG,LM} & 1.018792 & 2.338107 & 3.248197 & 4.087949 & 4.820099 \\
  \hline
\end{tabular}
\caption{Cauchy oscillator lowest eigenvalues.}
\end{table}
\begin{table}[h]
\begin{tabular}{|c||c|c|c|c|c|}
  \hline
  m=0.001 & $E_1$ & $E_2$ & $E_3$ & $E_4$ & $E_5$ \\
  \hline\hline
  a=50 & 1.00612 & 2.32596 & 3.23723 & 4.07956 & 4.81614 \\
  a=100 & 1.01245 & 2.33229 & 3.24356 & 4.08590 & 4.82248 \\
  a=200 & 1.01555 & 2.33540 & 3.24667 & 4.08901 & 4.82560 \\
  \hline
\end{tabular}
\caption{Quasirelativistic oscillator: $a$-dependence of lowest eigenvalues for  $m=0.001$. }
\end{table}
\begin{table}[h]
\vspace{0.5cm}
\begin{tabular}{|c||c|c|c|c|c|}
  \hline
  m=0.01 & $E_1$ & $E_2$ & $E_3$ & $E_4$ & $E_5$ \\
  \hline\hline
  a=50 & 1.00275 & 2.32235 & 3.23367 & 4.07593 & 4.81255 \\
  a=100 & 1.00746 & 2.32707 & 3.23839 & 4.08066 & 4.81728 \\
  a=200 & 1.00893 & 2.32854 & 3.23987 & 4.08213 & 4.81876 \\
  \hline
\end{tabular}
\caption{$a$-dependence for m=$0.01$.}
\end{table}
\begin{table}[h]
\begin{tabular}{|c||c|c|c|c|c|}
  \hline
  m=0.1 & $E_1$ & $E_2$ & $E_3$ & $E_4$ & $E_5$ \\
  \hline\hline
  a=50 & 0.935106 & 2.24274 & 3.15568 & 3.99499 & 4.73274 \\
  a=100 & 0.935146 & 2.24278 & 3.15573 & 3.99503 & 4.73278 \\
  a=200 & 0.935147 & 2.24278 & 3.15573 & 3.99503 & 4.73278 \\
  \hline
\end{tabular}
\caption{$a$-dependence for $m=0.1$.}
\end{table}

A comparison   (Tables II to V) of  quasirelativistic oscillator eigenvalues,  in the descending mass order $m=  0.1,\, 0.01, \, 0.001$,   with those for the Cauchy oscillator  clearly indicates
the {\it  spectral   convergence}  of the quasirelativistic  oscillator to the Cauchy one as $m$ approaches $0$.

The   clustering of "small" mass curves in Figs. 1 and 3, corresponding to  $m\in (0,1]$, gives  support  to the  statement about
the   convergence of    quasirelativistic  spectral data to   ultrarelativistic ones  as $m$ drops down to $0$.
   In the Appendix  we give additional analytic hints to
 this   conclusion.

 Accordingly, for  small masses,   the Cauchy oscillator provides  a reliable
  spectral  approximation  of the quasirelativistic one   in the whole spectral range  (i.e. for arbitrarily large  $n$).
Thence, it is of interest to recall  asymptotic ("large" $n$)  regularities of Cauchy oscillator  eigenvalues.
Those may   be adopted to approximate  higher eigenvalues
 of the small mass quasirelativistic  system.
  These regularities  are quantified   by means of   handy  analytic    formulas \cite{SG,LM}.
For odd labels $n$ we have:
\be
E_{n=2k-1} \sim  \left(\frac{3\pi}{2}
\right)^{2/3} \left( {n + \frac{3}{4}} \right)^{2/3} \label{odd}
\end{equation}
while for even $n$ there holds
\be
E_{n=2k}\sim  \left(\frac{3\pi}{2}
\right)^{2/3} \left( {n + \frac{1}{4}} \right)^{2/3} \label{even}
\ee
with  $k=1,2,3,...$.   Concerning an  approximation accuracy, we must  decide  how large the label  $n$ needs to be.
 The  approximation   finesse clearly  depends on  the  a priori  chosen   robustness level and can be fine-tuned.
   In the present discussion we have found formulas (\ref{odd}) and (\ref{even}) to give  reliable  approximations  for
{\it  relatively low} labels   $n \geq 6$,  see the   Appendix   for detailed data.

\subsection{$m\gg 1$ regime}

 The $m$-dependence of quasirelativistic oscillator eigenvalues  for $m\in (0,10]$  depicted  in Fig. 3,
  clearly indicates    symptoms of   $m\gg 1$  spectral regularities which need  to be verified more convincingly.
  See e.g. the Appendix for analytic hints to this end.
 Clearly,  mass values  should be  picked out  well beyond the interval $(0,10]$.
 In Fig. 6  a sequence of  eight  consecutive  (lowest)   eigenvalues  is depicted  separately for
each mass parameter   $m=10,20,50,100$  separately.
The dependence of   $E_n(m)$  on  $n$ indicates  approximately  equal  spacings between consecutive energy levels.

\begin{figure}[h]
\begin{center}
\centering
\includegraphics[width=75mm,height=75mm]{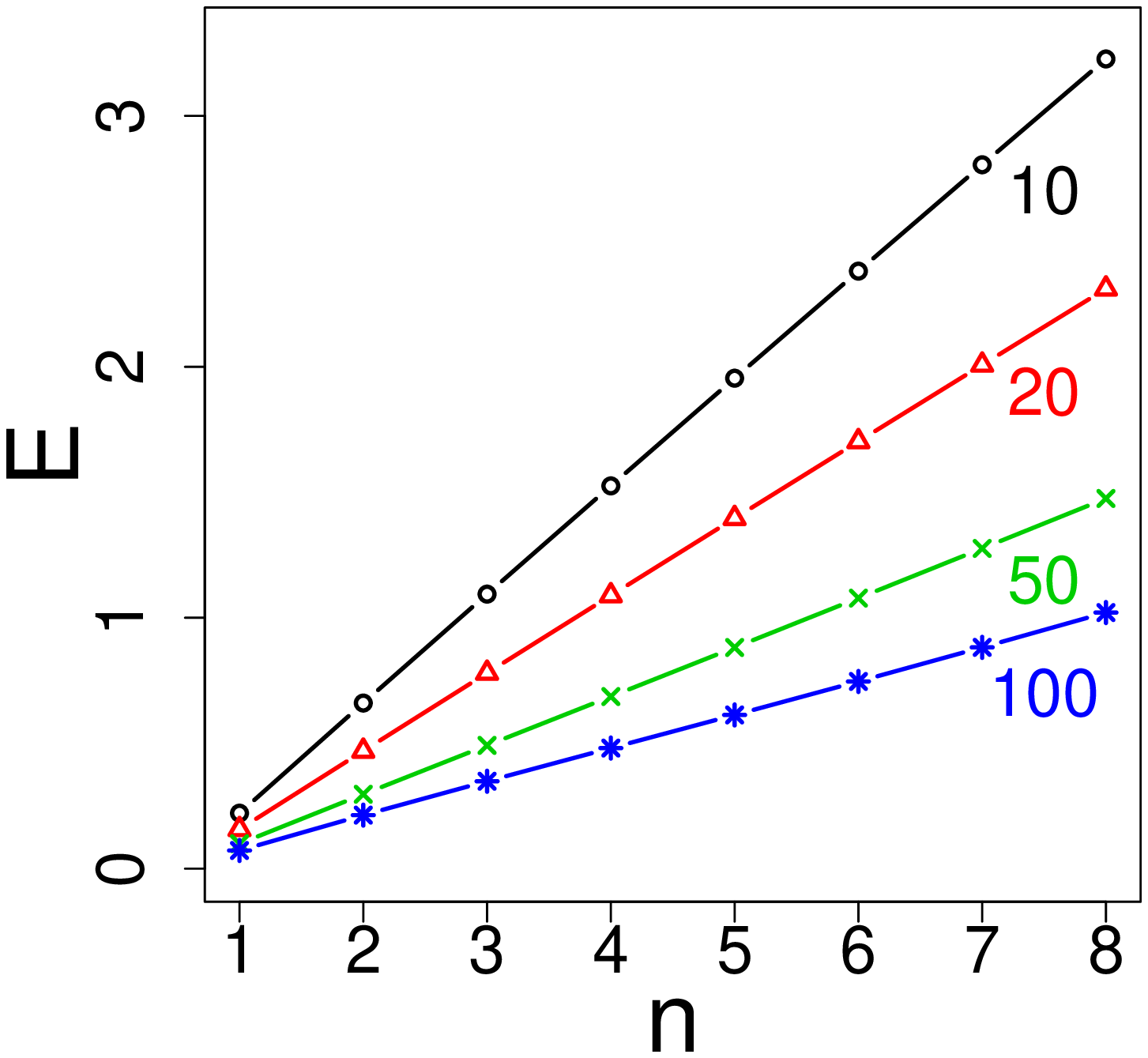}
\includegraphics[width=75mm,height=75mm]{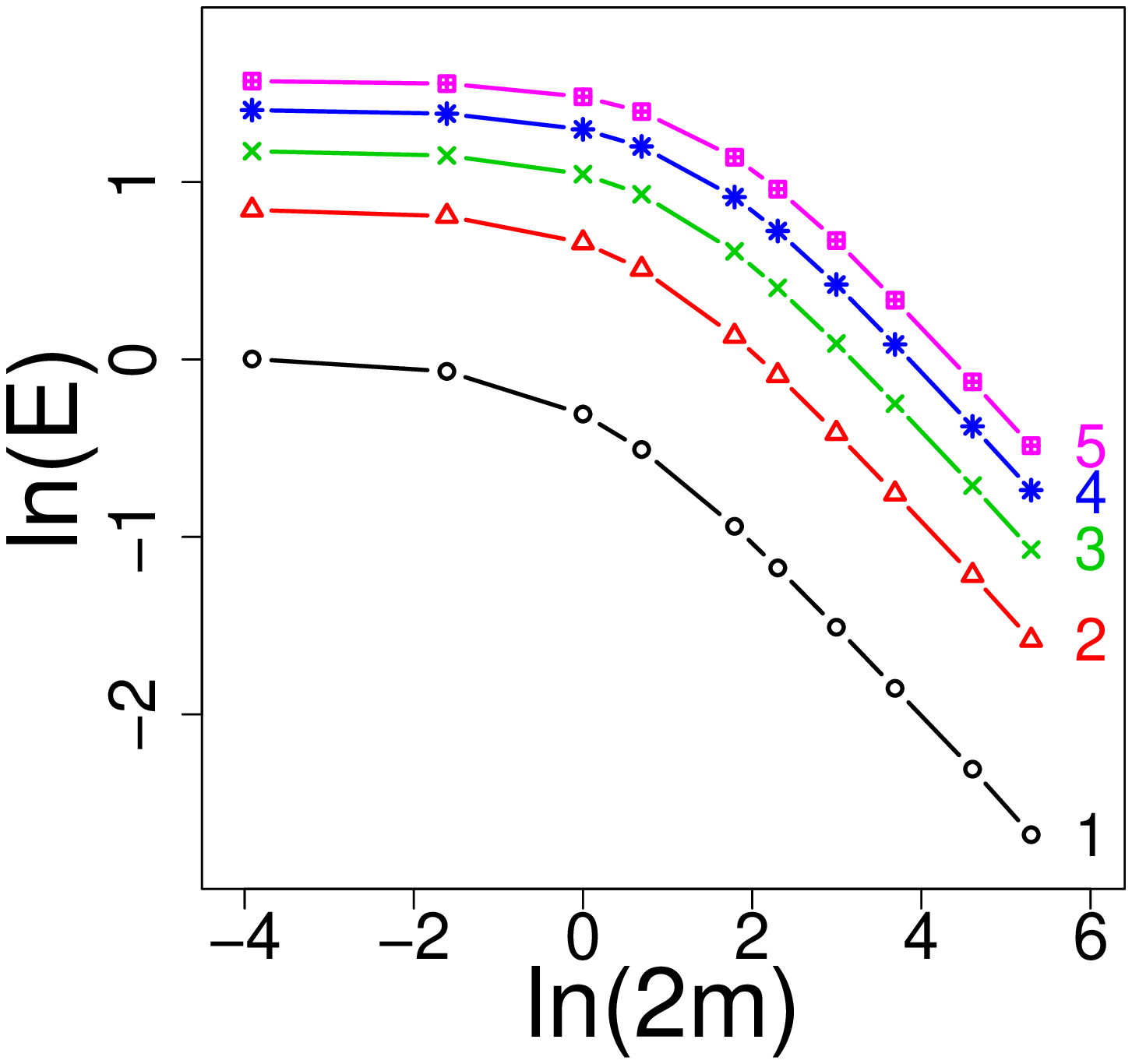}
\caption{Quasirelativistic $m\gg 1$ regime. Left  panel: eight consecutive   eigenvalues  $E_n(m)$, for
 masses  $m= 10,\, 20,\, 50,\, 100$, build  an approximate a straight line
  $E_n(m)= {\frac{1}{\sqrt{2m}}} (2n-1)$, $n\geq 1$. The  best result  is   obtained  if fitting
  employs   $m\geq 10$.  Right panel:  doubly logarithmic scale gives access to  a wider  mass range:  $m=0.01, 0.1,  0.5,
1, 3, 5, 10,  20, 50, 100$. Note  that for $m>3.7$ i.e. $\ln (2m)>2$,   straight line segments  are mimicked by
 $\ln(E_n(m))=-\frac{1}{2}\ln(2m)+\ln(2n-1),\, n=1,2,3,4,5$, thus  reproducing the  nonrelativistic oscillator spectral pattern.  }
\end{center}
\end{figure}

 In the right panel of Fig. 6,   the  $E_n(m)$ data have been displayed (in a doubly logarithmic scale)
against $2m$,  for each fixed $n$   separately. That clearly identifies  the $m$-dependence  of the  $n$-th eigenvalue ($n=1,...,8$)  in a relatively wide mass range $m\in (0,100]$.
The equal spacing conjecture receives   even stronger support by fitting the  numerically computed data   to  {\it  approximating} straight lines
(that in Fig. 4) of the form
\be
E_n(m)= \frac{1}{\sqrt{2m}}(2n-1),\qquad n=1,2,\ldots\qquad m \gg 1,\label{r1}
\ee
or  equivalently (that in Fig. 5)
\be
\ln[E_n(m)]=-\frac{1}{2}\ln(2m)+\ln(2n-1),\qquad n=1,2,\ldots\qquad m\gg 1.\label{r2}
\ee
These formulas  are approximately valid  for sufficiently large $m$ and  the $E_n(m)$ dependence on $n$  definitely   appears to follow the nonrelativistic
harmonic oscillator spectral pattern.  In fact
 $E_n= \hbar \omega (n+{\frac{1}{2}}), \, n=0,1,..., $ where $\omega =\sqrt{k/m}$ derives from
  $H= -(\hbar ^2/2m)\Delta + k x^2/2$.  By scaling away $k$ (set formally $k=2$) and eliminating  $\hbar =1$,
  we are left with   $H= -(1/2m) \Delta + x^2$ whose spectral solution reads
    $E_n= \sqrt{\frac{2}{m}}\, (n+{\frac{1}{2}}), \, n\geq 0$.  By relabeling that spectrum  according to $n \rightarrow n-1$,
     where the former $n=0$ is replaced by the new $n=1$, we  ultimately
    arrive at the formula $E_n= \sqrt{\frac{2}{m}}\, (n -{\frac{1}{2}})   = \frac{1}{\sqrt{2m}} (2n-1), \, n\geq 1$,
     i.e.  (\ref{r1}).

Concerning the fitting procedure, let us point out that in Fig. 6  we encounter   functions of the form   $\ln(E(m,n)=f[\ln(2m)]$.  For mass values obeying $\ln(2m)> 2$ (e.g. $m>3.7$)
 we can approximate   the resultant curves by  straight line segments  of  the form $\ln(E(m,n))=a \ln(2m)  + b$.
 There, "ideally" we should  have $a=-1/2$ and $b=  \ln (2n-1)$. Although  an
  "ideal" outcome is  never the case,  approximate values for $a$ and $b$  (retrieved form computed data) quite well fit to the nonrelativistic oscillator picture.

 For concreteness we  reproduce  approximate values for  $a$ and $b$  that determine  straight line segment fits  in Fig. 7,  for first five eigenvalues.
  Error bounds were evaluated by means  of the  least square  deviation method for computed spectral data.
The  fitting of straight  lines  has been  actually  started from $m=5$ for $n=1, 2$ and $m=10$  for $n>2$.
 \begin{align*}
n&=1, & (-0.501&\pm 0.005)\ln(2m)+(-0.012\pm 0.018), & &n=(0.994\pm 0.009).\\
n&=2, & (-0.497&\pm 0.006)\ln(2m)+(1.069\pm 0.021), & &n=(1.96\pm 0.03).\\
n&=3, & (-0.504&\pm 0.005)\ln(2m)+(1.606\pm 0.019), & &n=(2.99\pm 0.05).\\
n&=4, & (-0.503&\pm 0.005)\ln(2m)+(1.936\pm 0.019), & &n=(3.97\pm 0.07).\\
n&=5, & (-0.502&\pm 0.005)\ln(2m)+(2.18\pm 0.02), & &n=(4.92\pm 0.09).\\
\end{align*}
Approximate values for the (right-hand-side) parameter $n$ were retrieved directly from the  computed  "free" parameters $b=\ln (2n-1)$.
We note that the parameter $a$  has {\it fapp } $a=-1/2$  value  ( e.g. almost $-1/2$,  within  the error bounds).
We recall that the data employed in Fig. 3 - 6    have been computed by means of the spectrum-generating algorithm  which is  not free of a number of error-accumulating factors (like. e.g the
lowest order Strang approximation, the usage of Gram-Schmidt diagonalization procedure, finite bounds for the integration intervals etc.).

Nonetheless, an  affinity  with the nonrelativistic harmonic oscillator spectrum is clearly seen in
 the large mass    regime. In our derivations,  $m= 10$  has been found to set a "sufficiently large"
 threshold value such that for $m\geq 10$  the quasirelativistic
 harmonic oscillator spectrum effectively  displays   (approximates, becomes  very close)
  the nonrelativistic oscillator spectral regularity  $\Delta E = E_{i+1} - E_i  \sim 1/\sqrt{2m}$, for all $i=1,2,...$.

\section{Quasirelativistic finite well.}

Let us consider the eigenvalue problem for $H=T+V$, where $T=T_m= \sqrt{-  \Delta + m^2} -m $ is
 the  quasirelativistic generator and
\be
V(x)=\left\{
       \begin{array}{ll}
         0, & \hbox{$|x|<1$;} \\
         V_0, & \hbox{$|x|\geqslant 1$,}
       \end{array}
     \right.\label{l12}
\ee
with  $V_0>0$.   We use the natural system of units  $\hbar =1=c$  from the start, see the Appendix for a description of
involved  scalings.

 We shall discuss  both shallow and very deep wells   of the size $[-1,1]$.
  In the previous paper \cite{GZ} we have demonstrated that a  sufficiently deep finite  Cauchy  well
is "spectrally close"  to the infinite Cauchy well.  A number of
eigenvalues and eigenfunctions  has been computed for the low-lying
part of the spectrum.

We mention  that for   sufficiently large $n$ the  infinite quasirelativistic  well
 ($[-b,b]$) eigenvalues can be
approximated as follows, \cite{KKM}
 (for a while we reintroduce  dimensional constants, see also the Appendix):
\be E_n  - mc^2 = \left({\frac{n\pi }{2}} - {\frac{\pi }{8}}\right)
{\frac{\hbar c}{b}}  + O\left( {\frac{1}{n}}\right).  \label{appr}
\ee
Interestingly, the right-hand-side of Eq. (\ref{appr})  has been proved  in
Ref. \cite{K} to provide the large $n$ approximation of the
  $[-b,b]$ infinite Cauchy ($m=0$) well eigenvalues.  Subsequently we shall
restore the previously employed $b=1$ and $\hbar =1=c$ notation.

  In  the present Section  we  shall demonstrate that
 the finite quasirelativistic well,  in the small mass regime, becomes   "spectrally close" to  the finite  Cauchy well
   (compare e.g.   Cauchy  versus   quasirelativistic oscillator discussion of Section III).
For another extreme of large masses, we shall demonstrate that  the quasirelativistic well  becomes "spectrally close" to
the familiar   nonrelativistic finite well.
Analytic arguments provided in the the Appendix give support to the conjecture that those extremal behaviors might be
   generic  for  a  wider class of  confining    quasirelativistic problems.

Our numerical procedures  are based on the spectrum-generating
algorithm of Section II, including  all mentioned there cutoff
choices and  the algorithm - related error accumulation
reservations. We use $a=50$ for the integration interval bound.
  The set of trial functions is chosen to be the same  as  that in the discussion of Section III.

\begin{figure}[h]
\begin{center}
\centering
\includegraphics[width=80mm,height=80mm]{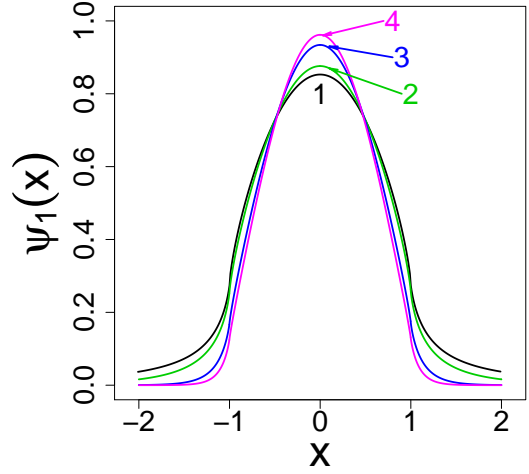}
\caption{Quasirelativistic finite well ground state for  $V_0=5$. Labels $1,2,3,4$  correspond to $m=0.01, 1, 5, 10$, respectively.}
\end{center}
\end{figure}

\begin{figure}[h]
\begin{center}
\centering
\includegraphics[width=50mm,height=50mm]{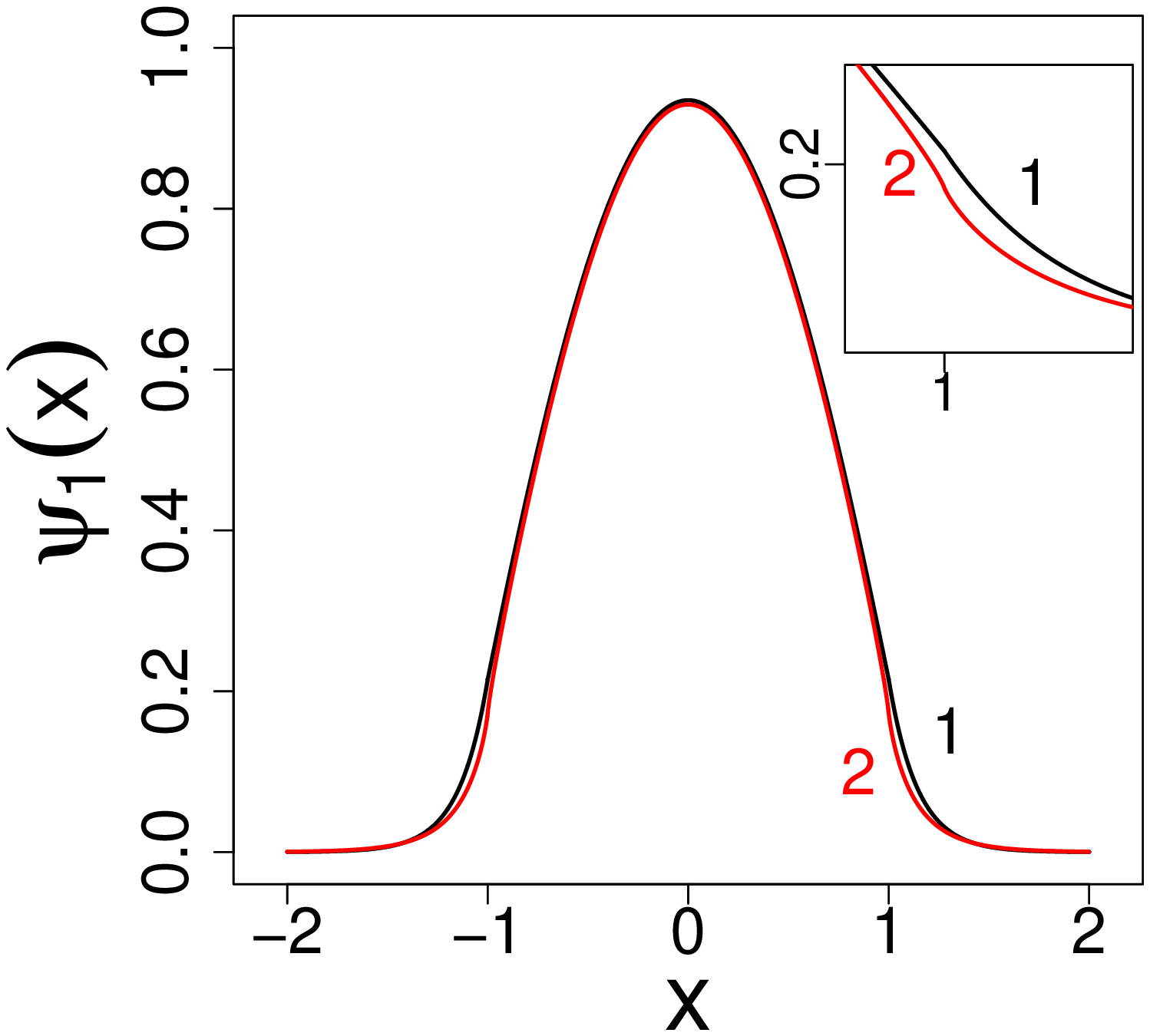}
\includegraphics[width=50mm,height=50mm]{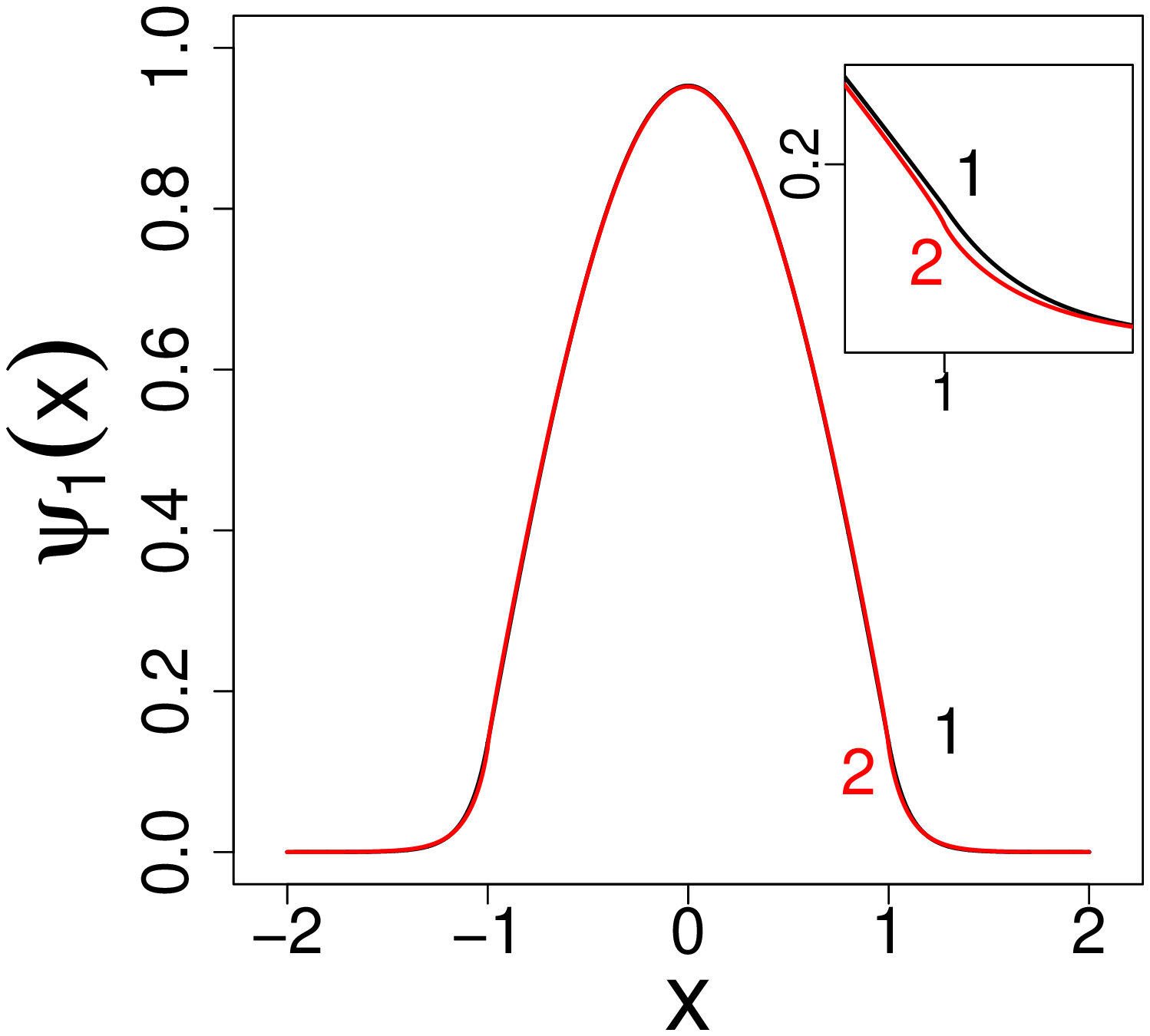}
\includegraphics[width=50mm,height=50mm]{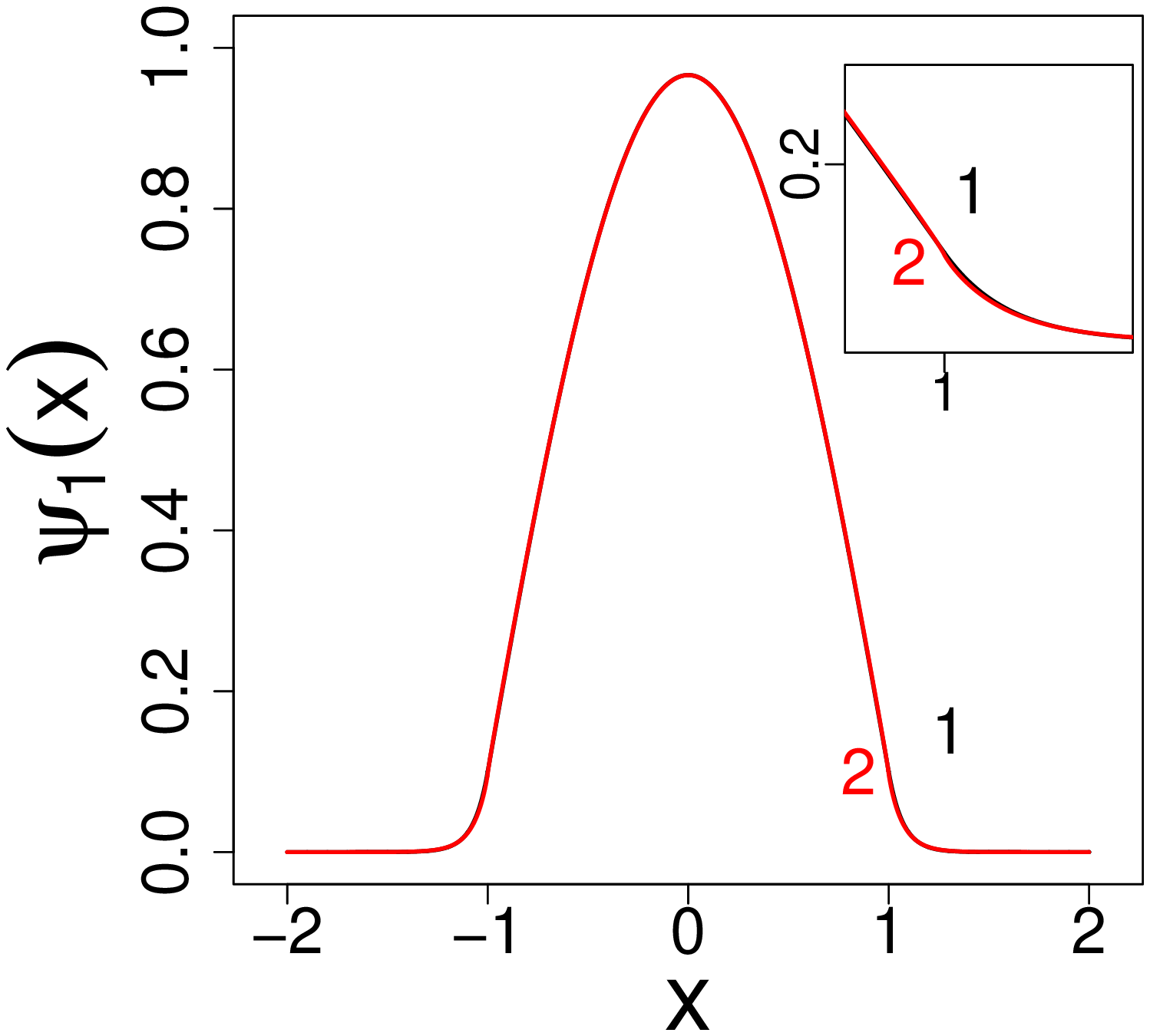}
\caption{A comparison of ground states in case of $V_0=5$ for the nonrelativistic  (label 1) and quasirelativistic well (label 2):
 $m=5$ (left  panel), $m=10$ (middle  panel), $m=20$ (right  panel).}
\end{center}
\end{figure}

\subsection{Shallow well.}

In a finite $1D$ (and $2D$) nonrelativistic  well one normally expects at least one bound state to exist. The well known exception is  the 3D case, when for too shallow wells  (irrespective of their width) bound states
may not exist at all. No  general  statements  of that kind are  known  for  nonlocal  finite  well problems.

We know  the Cauchy well whose depth  is set by  $V_0=5$  has three bound states \cite{GZ}.  However, we  have not explored  before how low $V_0$  need to  be to admit one bound state only.
In the present paper this issue  will be addressed on the level of a quasirelativistic finite well. An extension to  finite Cauchy well will actually come out in the regime of small masses.

 For concreteness and   a direct comparison with results of Ref. \cite{GZ}, let us begin our discussion from the finite  $V_0=5$   quasirelativistic well.
We have extended the stabilization time up to $5000$ small time steps.  (Anticipating further discussion of the large $m$ regime when the Bessel functions become strongly localized, having very narrow
 peaks about their   maxima and minima, a spatial partitioning has been  made finer  $\Delta x= 0.001  \rightarrow  \Delta x= 0.0005$.)

 If  $m$ drops down to a close vicinity of $0$,   quasirelativistic eigenvalues and eigenfunctions   appear to  converge to those  of the finite    Cauchy   well.
To exemplify this  observation  on the level of eigenvalues let us provide explicit   quasirelativistic  ground   state  energy  values in the $V_0=5$ well
  and set them   against  the  respective  $m=0$ value.

  We have   $E_1 =0.9501$ for  $m=0.01$,  $0.9522$  for $m=0.001$ and $0.9538$   for the finite Cauchy well ($m=0$).
Respective eigenfunctions are graphically indistinguishable in the adopted scale.

 In Fig. 7 we  depict quasirelativistic   $V_0=5$   well eigenfunctions
for    graphically distinguishable cases of    $m = 0.01,\,1,\,5,\,10$.
With the growth of $m$ the ground state maximum  increases. Clearly, the eigenfunctions have tails extending beyond the well
 boundaries (e.g. the interval $[-1,1]$), but they decay rapidly  with the growth of $|x|$.
For large $m$ we  detect  a fairly close affinity with the standard (text-book)  nonrelativistic finite well quantum
problem (c.f. the Appendix for relevant data).

In accordance with  the folk lore  wisdom about the nonrelativistic  finite well, in $1D$ at least one bound state is always
in existence.  However,  the  maximal number $N$  of bound states  in the well of a
fixed depth $V_0$   is correlated with the mass $m$ value (we bypass the well width impact, in view of our $[-1,1]$ choice).
Indeed,  the number  of bound states  $N\in\mathbb{N}$  is  constrained by inequalities
\be
\frac{\pi^2}{8V_0}(N-1)^2\leqslant m \leqslant \frac{\pi^2}{8V_0}N^2. \label{number}
\ee
Physically more familiar inequalities in addition to dimensional  constants  explicitly involve the
  width parameter $b$ (the well interior
 is enclosed by $[-b,b]$).  We display for reference the pertinent formula: $\pi ^2 \hbar ^2(N-1)^2
 \leq  8m V_0 b^2 \leq \pi ^2 \hbar ^2N^2$. Our considerations employ $b=1$ and $\hbar =c=1$.
(In passing  we note that in $1D$ and $2D$ well at least one bound state always exists. The
bound  state may not  be granted  to exist  in $3D$ for too shallow wells.)

The above formula allows  to  deduce the number of bound states for a fixed well depth $V_0$ but
different mass values. Thus  e.g. for   all  $m\leq \pi ^2/8V_0$
only one bound state is in existence. Accordingly  the bound  $m < 1.23 /V_0$ tells that for $V_0=5$  one bound state
 only is secured  for masses $m< 0.246$.

For comparison,  maximal  numbers of bound states of the $V_0=5$  well  for various mass values  are displayed in a compact Table VI.
In  the quasirelativistic  case  those were deduced
 by means of the spectrum-generating algorithm. In the  nonrelativistic case
(denoted "standard") these numbers were   deduced from the formula (\ref{number}).
With  the  mass parameter increase,  maximal numbers of bound states show a definite tendency to equalize for both
 local and nonlocal  cases.\
\begin{table}[t]
\begin{tabular}{|c||c|c|}
\hline
mass & quasirelativistic $N$ &  standard  $N$ \\
\hline
\hline
  0.1 & 3 & 1 \\
  0.5 & 4 & 2 \\
  1 & 4 & 3 \\
  3 & 5 & 4 \\
  5 & 6 & 5 \\
  10 & 7 & 7 \\
  \hline
\end{tabular}
\caption{$V_0=5$ well:    maximal number $N$ of bound states for various masses  in quasirelativistic and nonrelativistic cases.}
\end{table}

With the growth of $m$ both  eigenvalues and eigenfunctions  for the nonlocal and local finite well models  "become close" to
each other. To see this spectral affinity, let us  compare respective  eigenvalues in the well $V_0=5$,   for  various
masses  (for m=10, only $7$ eigenvalues
are in existence):
\begin{table}[t]
\begin{tabular}{|c|c||c|c|c|c|c|c|c|c|}
  \hline
  mass & finite well & n=1 & n=2 & n=3 & n=4 & n=5 & n=6 & n=7 & n=8\\
  \hline\hline
  \multirow{2}{1.1cm}{m=10}& quasi & 0.09951 & 0.39217 & 0.86271 & 1.48933 & 2.24605 & 3.10483 & 4.03221 & -\\
  & standard  & 0.10190 & 0.40679 & 0.91211 & 1.61267 & 2.49846 & 3.54752 & 4.68404 & - \\
  \hline\hline
  \multirow{2}{1.1cm}{m=20}& quasi  & 0.05312 & 0.21154 & 0.47264 & 0.83227 & 1.28482 & 1.82341 & 2.43999 & 3.12481\\
  & standard & 0.05379 & 0.21502 & 0.48318 & 0.85739 & 1.33616 & 1.91714 & 2.59636 & 3.36634 \\
  \hline\hline
  \multirow{2}{1.1cm}{m=50}& quasi & 0.02227 & 0.08892 & 0.19968 & 0.35423 & 0.55213 & 0.79272 & 1.07522 & 1.39867\\
  & standard  & 0.02261 & 0.09040 & 0.20334 & 0.36132 & 0.56421 & 0.81181 & 1.10385 & 1.43998\\
  \hline\hline
  \multirow{2}{1.1cm}{m=100}& quasi & 0.01126 & 0.04499 & 0.10113 & 0.17961 & 0.28037 & 0.40334 & 0.54842 & 0.71546\\
  & standard  & 0.01159 & 0.04636 & 0.10431 & 0.18540 & 0.28964 & 0.41695 & 0.56733 & 0.74070\\
  \hline
\end{tabular}
\caption{Quasirelativistic (quasi)  versus nonrelativistic (standard)  $V_0=5$ well: $m$-dependence of eigenvalues}
\end{table}

The resultant eigenvalues  in case of $n>5$, even  for large masses   still differ by few percent.
However we recall that our spectrum-generating algorithm  accuracy
has not been fined tuned to the available extent. A proper  balance between cutoffs, partition units  and the   computations  time   was more important
 for us than  the highest possible accuracy level (diminishing an accumulation of systematic errors)  and that  hampers computation results  for $n>5$. C.f. also
  our comments concluding Sections II and III.B.

The eigenfunction computation  is less sensitive to algorithm generated systematic  errors.   In Fig. 8 the   ground state function  of the  quasirelativistic finite well    is displayed (label 2) and compared with that
for the  nonrelativistic  well (label 1)  for masses  $m=5, m=10, m=20$.  We clearly see that $m=20$, albeit still too small,  may be  tentatively    considered as the   mass threshold above which  the
a concept of  "spectral closeness" of the quasirelativistic and nonrelativistic finite wells  receives quantitative support.

A collection of  excited  eigenfunctions that are parametrized by the mass parametr $m$ is displayed  in Fig. 9. The mass range
 $m=0.01, 1, 5, 10$ is the same as in the ground state Fig. 7.

\begin{figure}[h]
\begin{center}
\centering
\includegraphics[width=50mm,height=50mm]{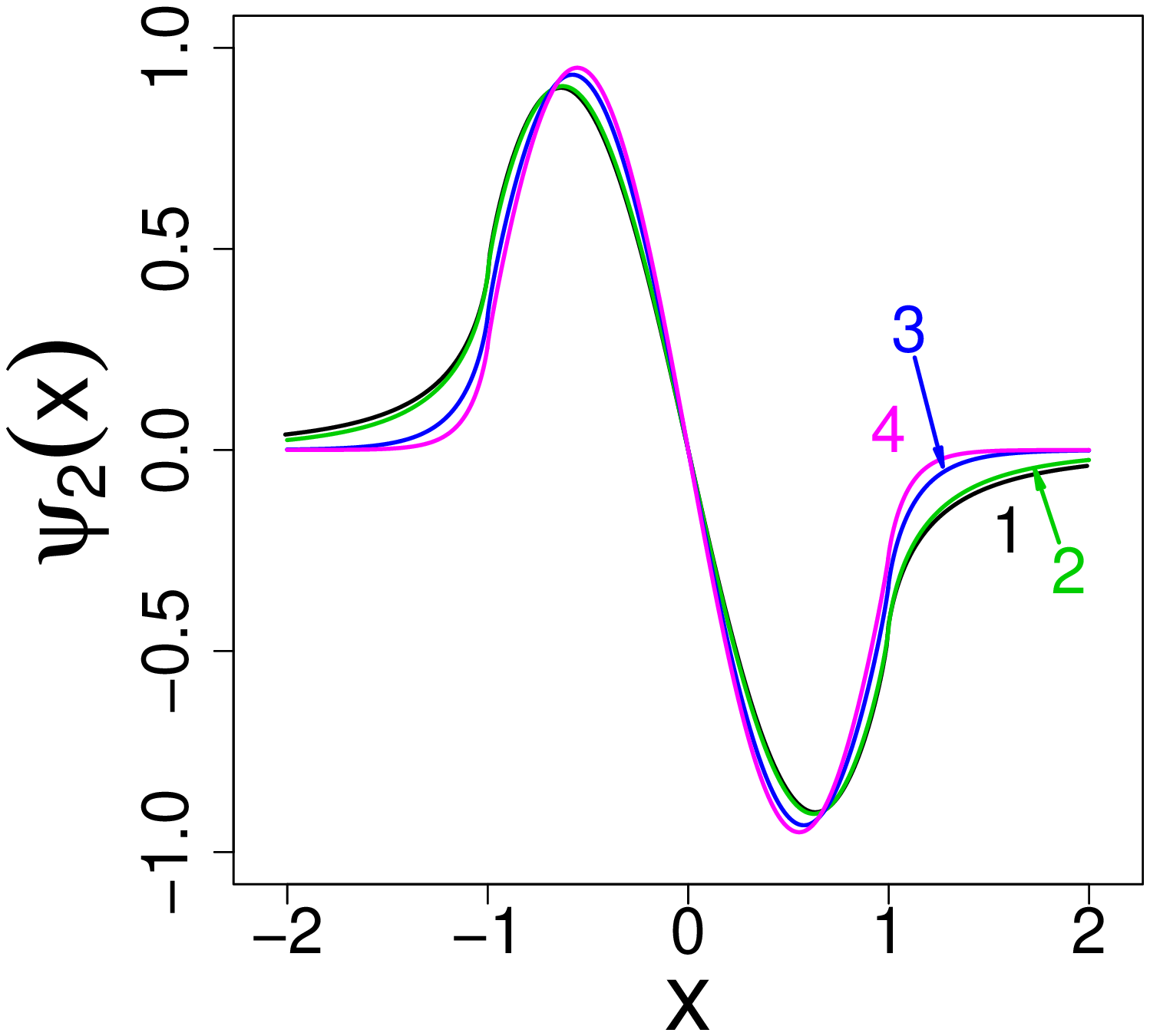}
\includegraphics[width=50mm,height=50mm]{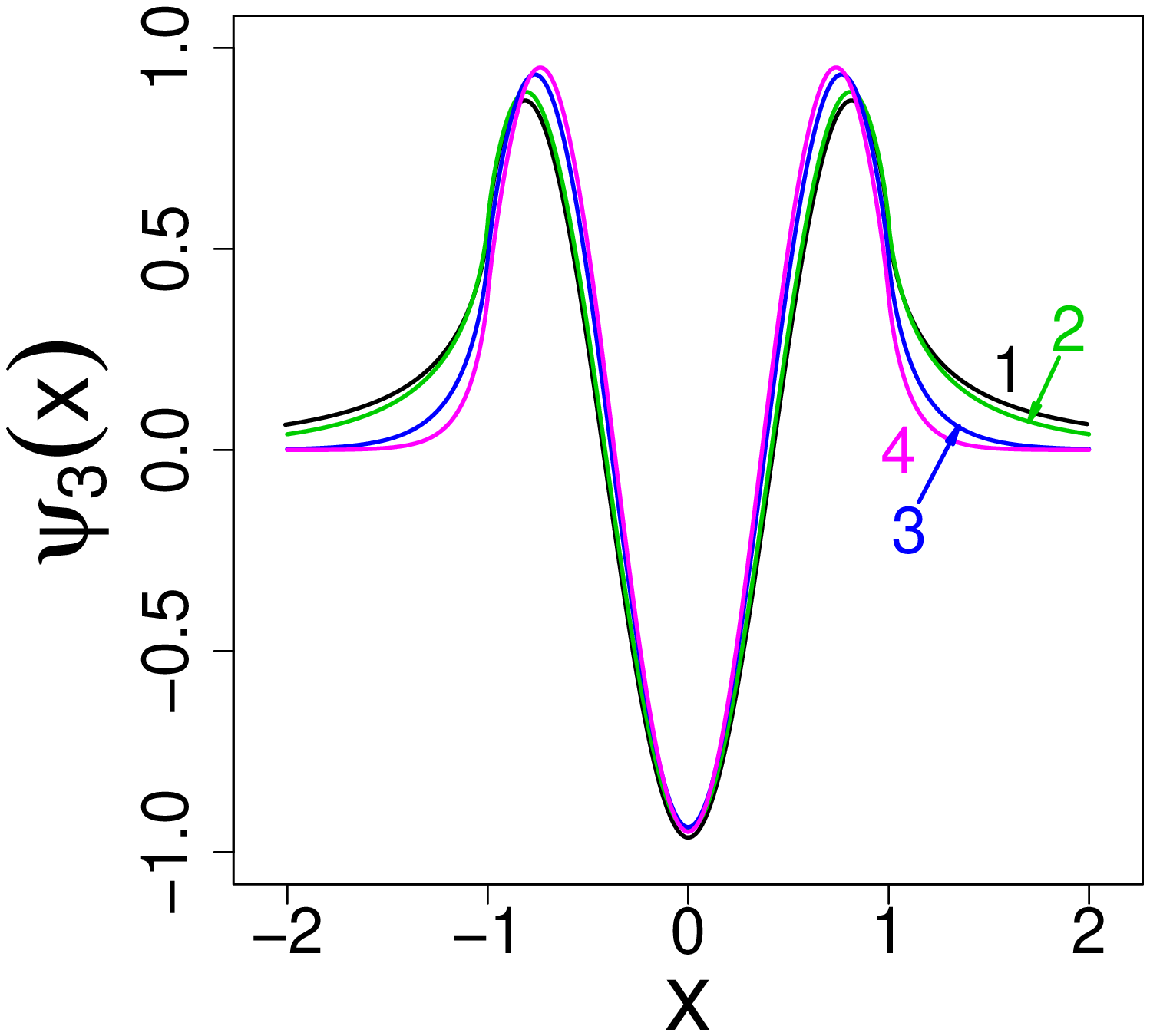}
\includegraphics[width=50mm,height=50mm]{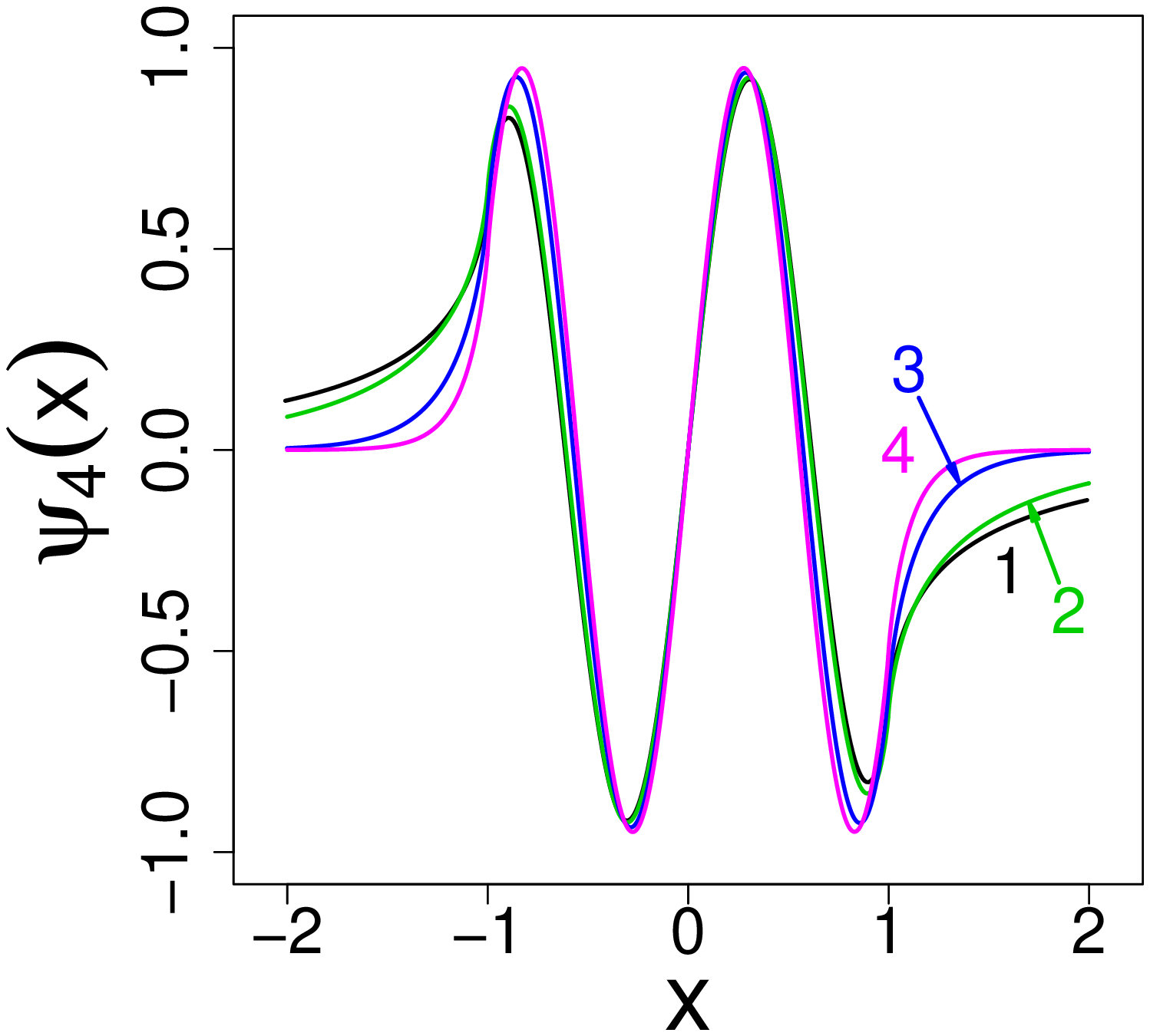}
\caption{Second, third and fourth quasirelativistic  $V_0=5$  well  eigenfunctions.  Masses  $m=0.01, 1, 5, 10$ are labeled
respectively by $1,2,3,4$.}
\end{center}
\end{figure}
\begin{figure}[h]
\begin{center}
\centering
\includegraphics[width=50mm,height=50mm]{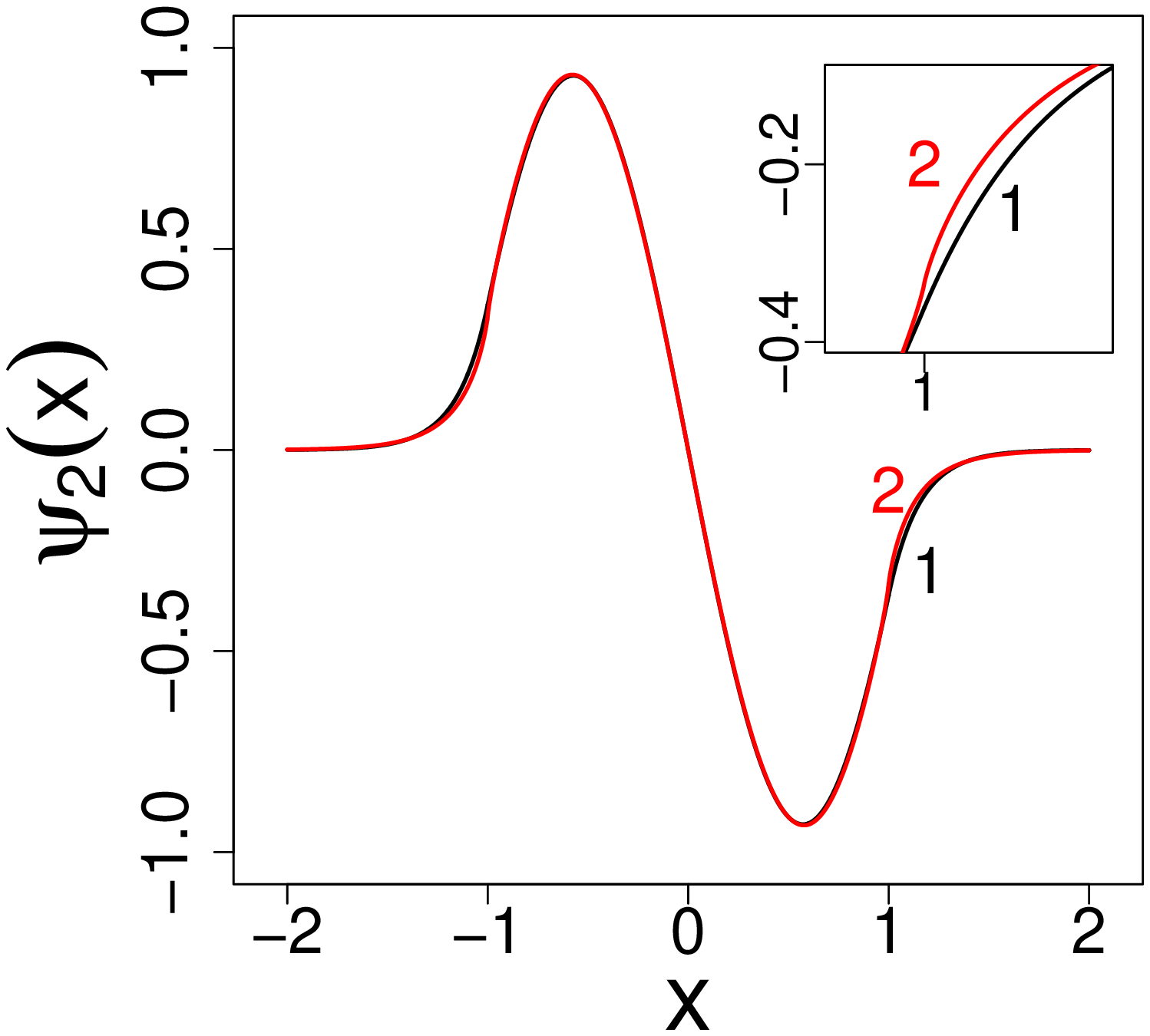}
\includegraphics[width=50mm,height=50mm]{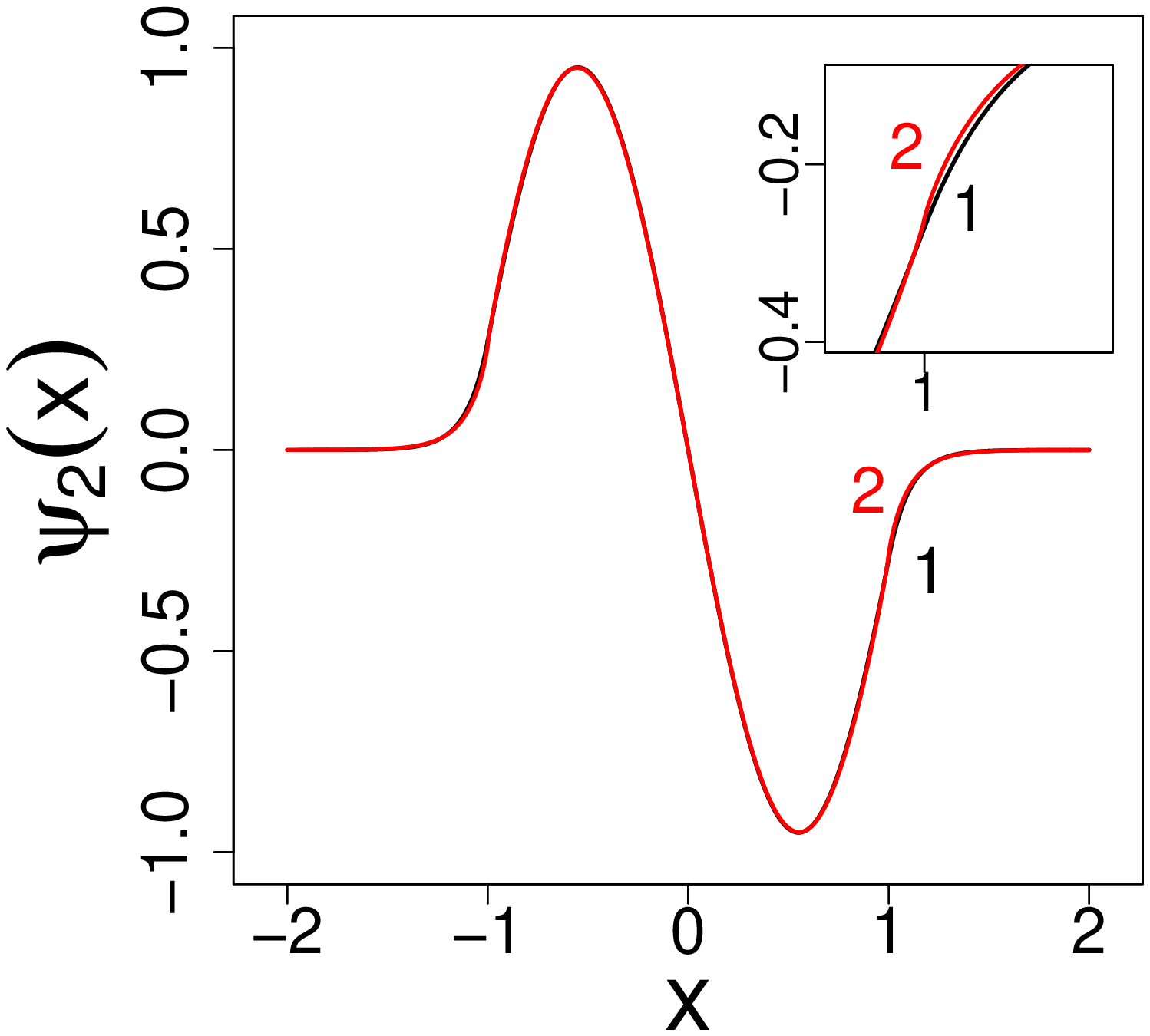}
\includegraphics[width=50mm,height=50mm]{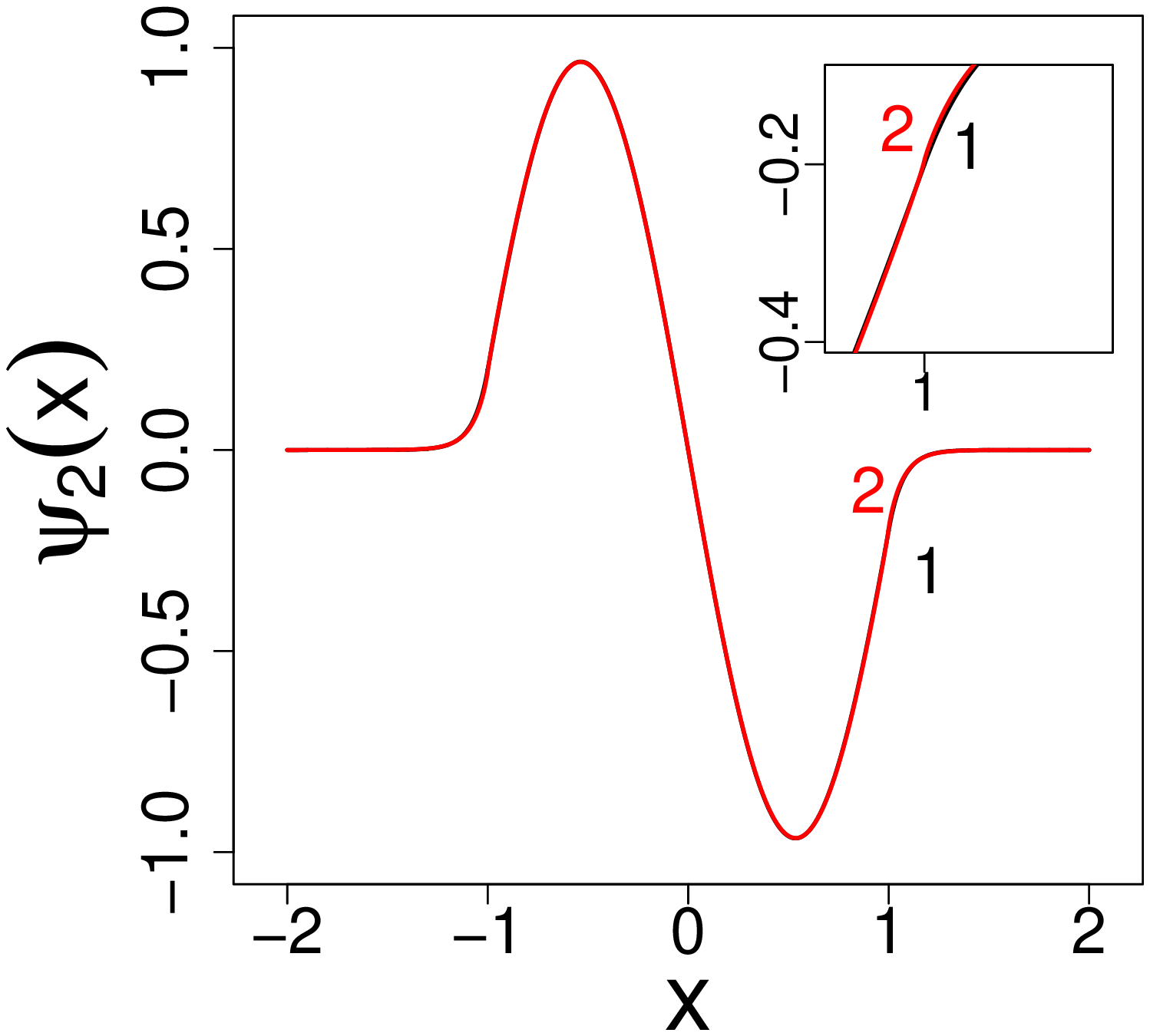}
\caption{ A comparison of the second  eigenfunction in the $V_0=5$  well for nonrelativistic (label 1) and quasirelativistic
(label 2) cases.  Here, $m=5$ (left  panel), $m=10$ (middle  panel), $m=20$ (right panel).}
\end{center}
\end{figure}
\begin{figure}
\begin{center}
\centering
\includegraphics[width=50mm,height=50mm]{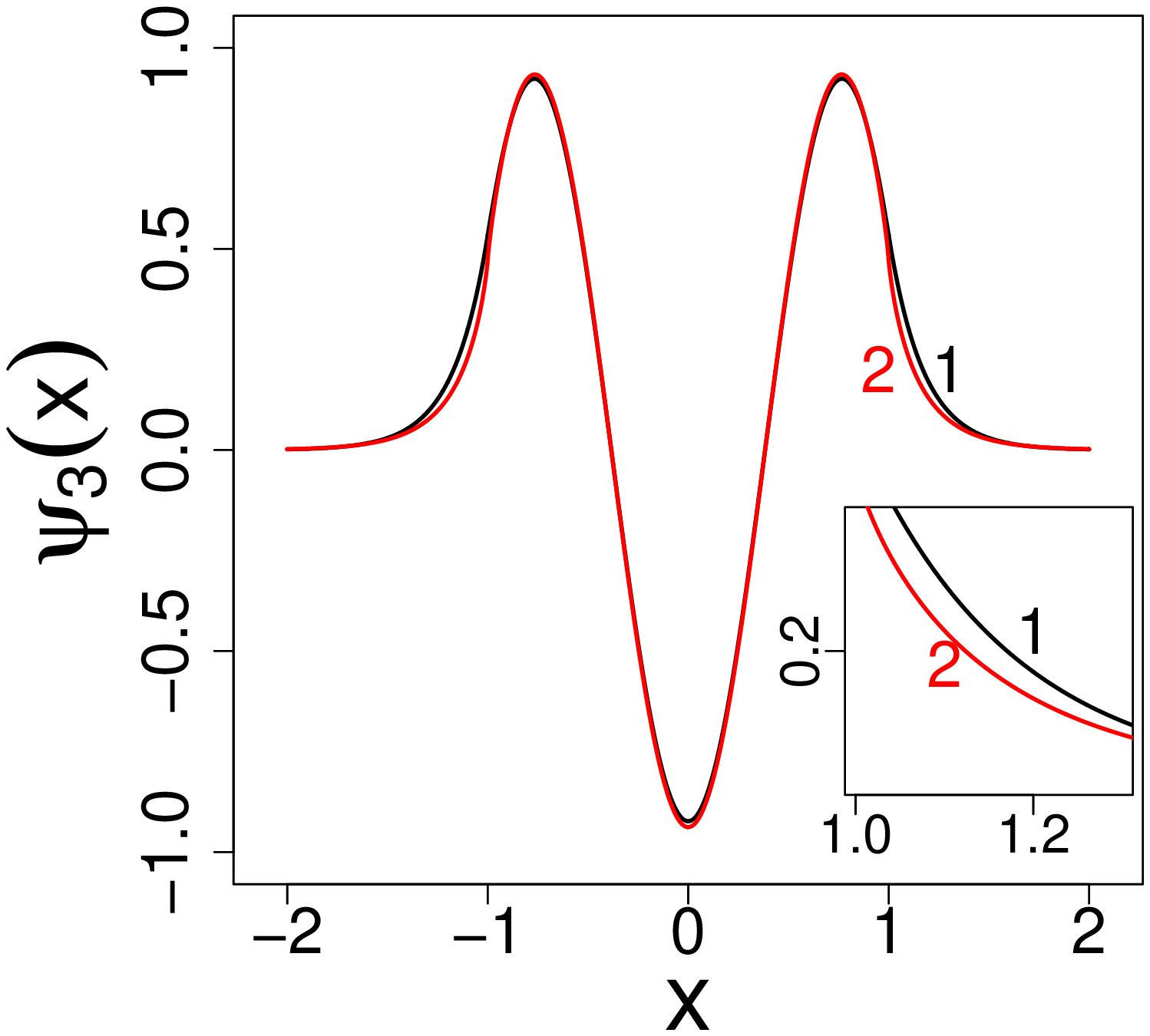}
\includegraphics[width=50mm,height=50mm]{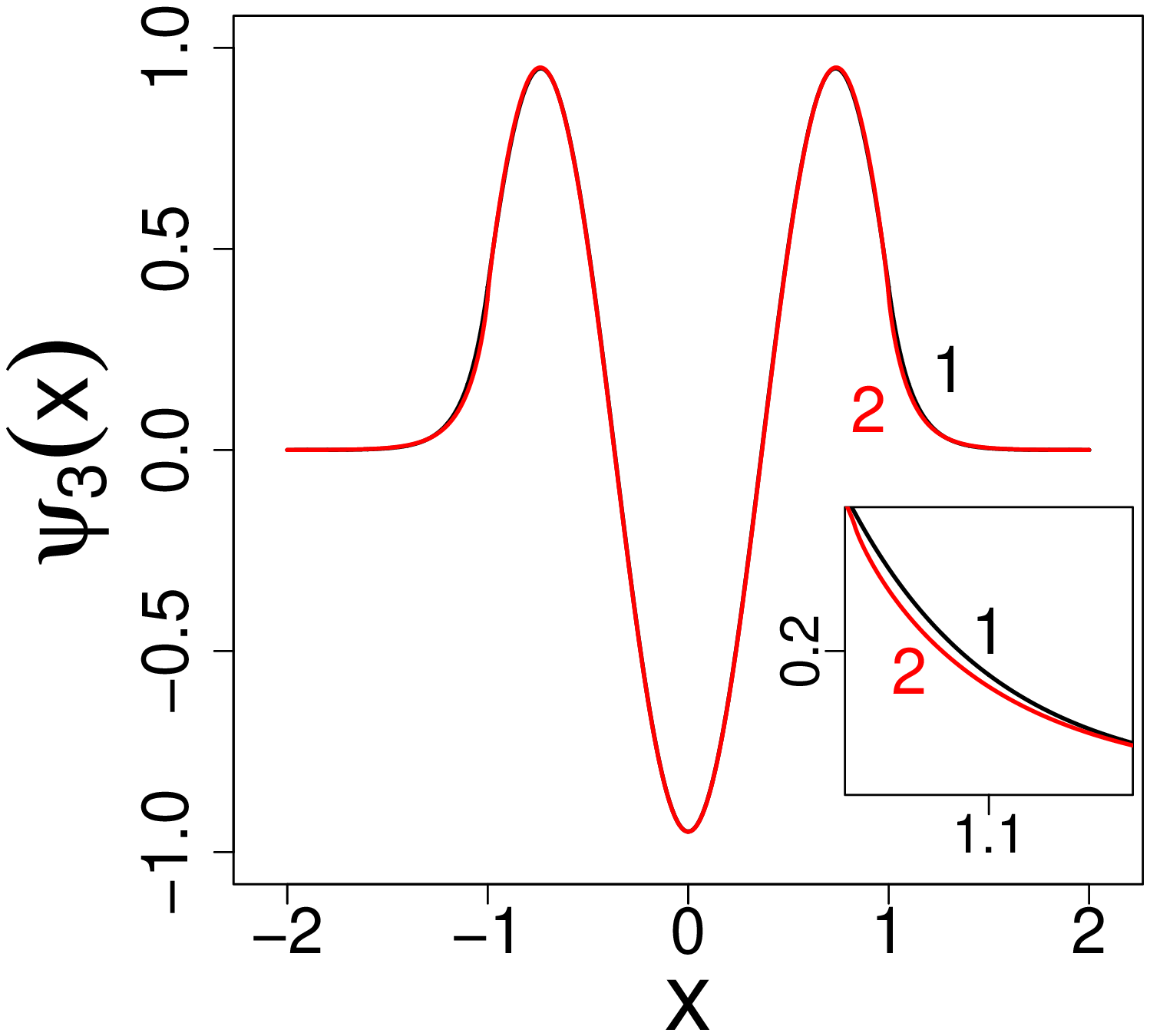}
\includegraphics[width=50mm,height=50mm]{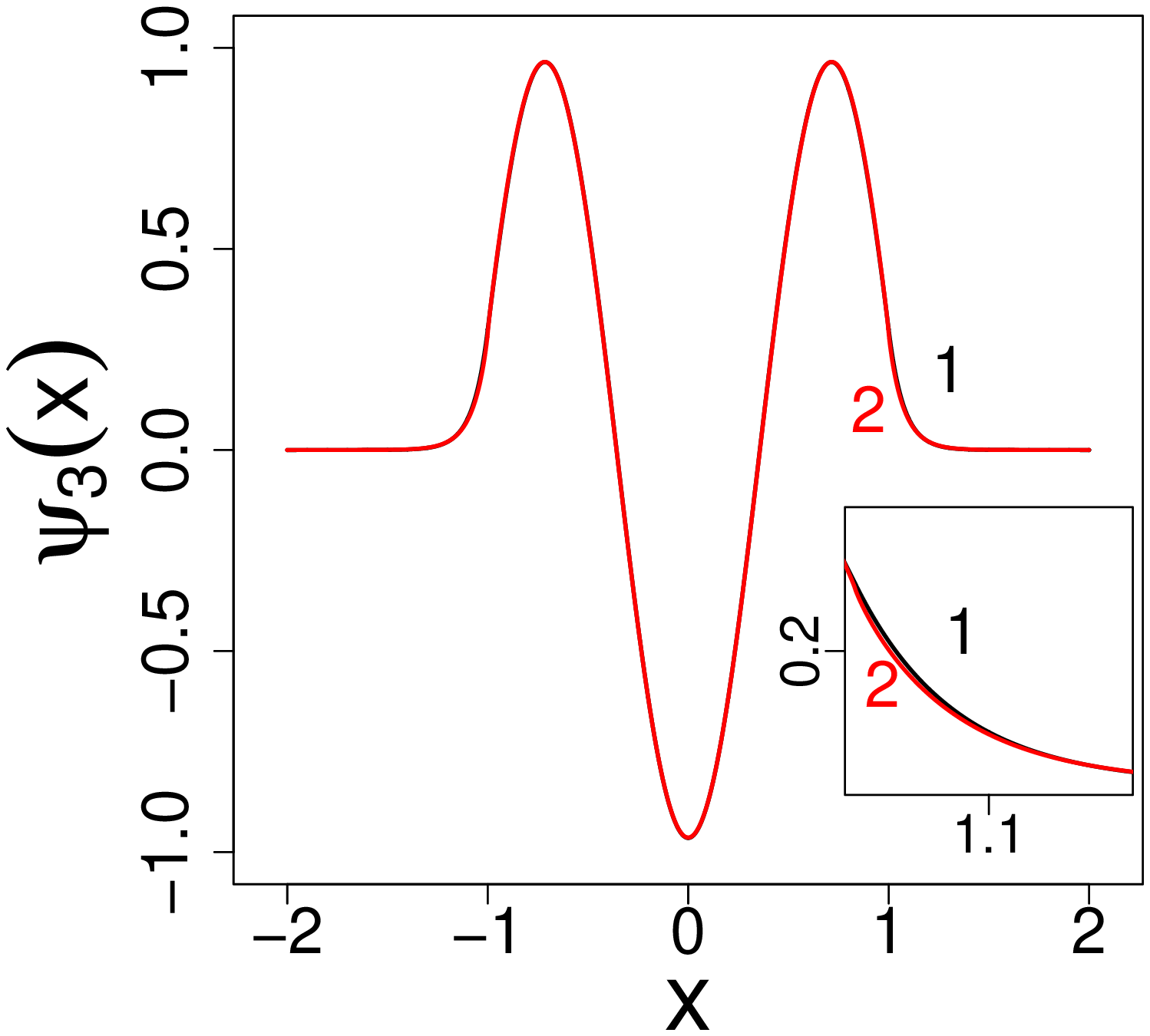}
\caption{A comparison of the third eigenfunction in the $V_0=5$ well for nonrelativistic (label 1) and quasirelativistic
(label 2) cases.  Here, $m=5$ (left  panel), $m=10$ (middle  panel), $m=20$ (right panel). }
\end{center}
\end{figure}

In Figs.  10  and   11   we compare nonrelativistic and quasirelativistic finite  ($V_0=5$) well eigenfunctions for $n=2, 3$  ($n=4,5$ follow the same pattern)
and masses $m=5, 10, 20 $. We get   there a  convincing  support to the previous tentative statement that $m=20$ is
 an optimal   threshold  value. For $m>20$, with a good fidelity,  we can state that
 quasirelativistic  and  nonrelativistic  finite wells are "spectrally close".

\subsection{Deep well versus  infinite well.}

Let us consider a relatively deep well $V_0=500$  (in  Ref. \cite{GZ}   we  have investigated the  well  as deep as $V_0=5000$).
Like in the Cauchy finite well case, a quasirelativistic  deep well is expected to stay  in  spectral affinity with its   infinitely deep partner.
That at least in  relation to the  low part of the spectrum.

For small values of the mass parameter $m$ convergence symptoms towards  $m=0$ spectral solution are clearly seen in a sequence
of ground state  energies  for the  finite  $V_0=500$ well: $E_1=1.1373$ for $m=0.01$, $1.1391$ for $m=0.001$, while
 $E_1=1.1408$  in the  $m=0$  Cauchy case.

Eigenfunctions for small mass values are {\it fapp } graphically  indistinguishable  form their Cauchy relatives \cite{GZ}.
In Fig. 12 we have displayed {\it quasirelativistic}  $V_0=500$ well ground state  for masses  $m=0.01, 1, 5, 10$, where $m=5, 10$
definitely stay beyond the "smallness" range.   For comparison the
{\it nonrelativistic} infinite well ground state  $\cos(\pi x/2)$ has been depicted.  It is clear that all curves stay
in a close vicinity of $\cos(\pi x/2)$, albeit  upon enlargement they  show subtle differences.
 \begin{figure}[h]
\begin{center}
\centering
\includegraphics[width=70mm,height=70mm]{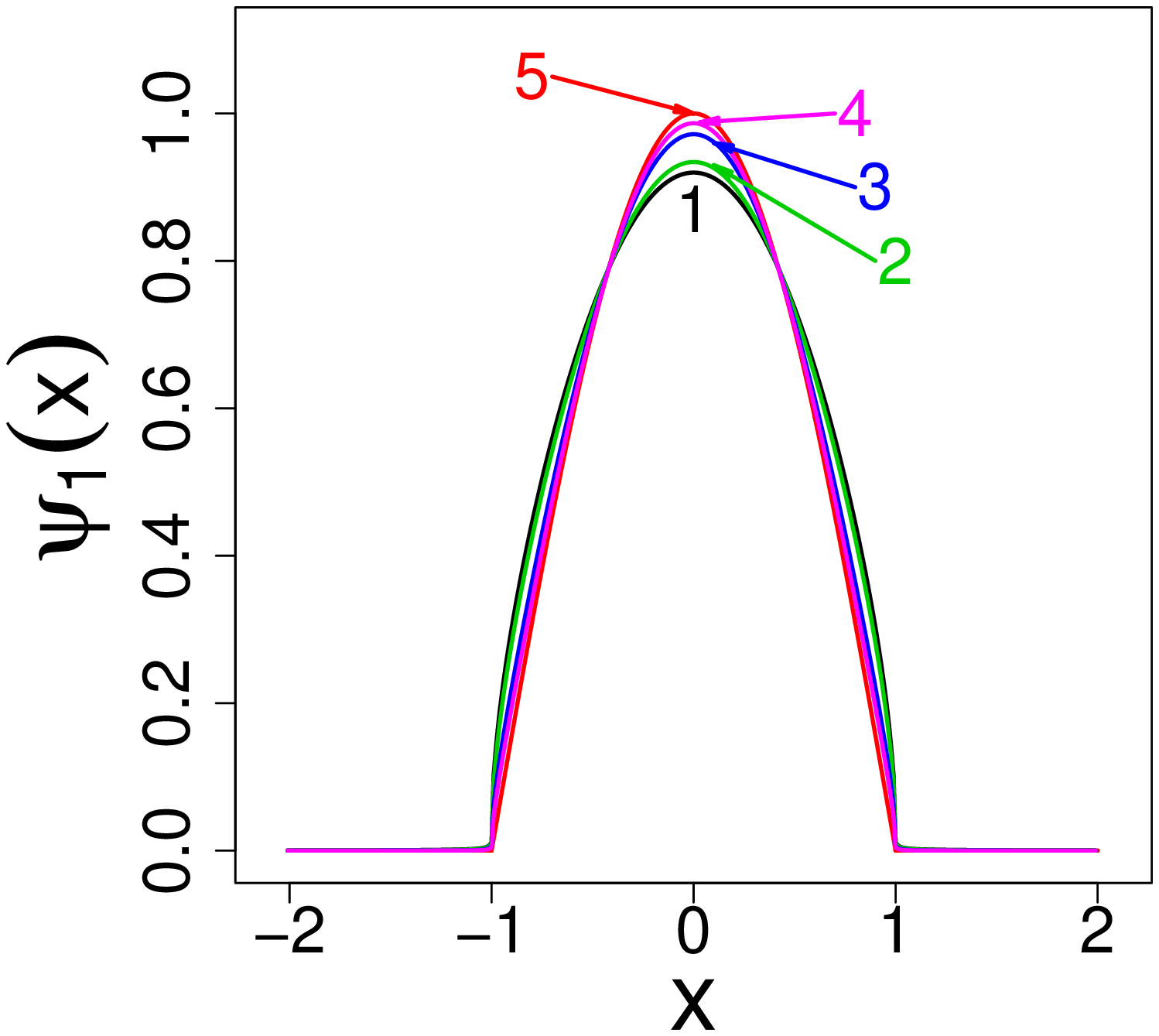}
\includegraphics[width=70mm,height=70mm]{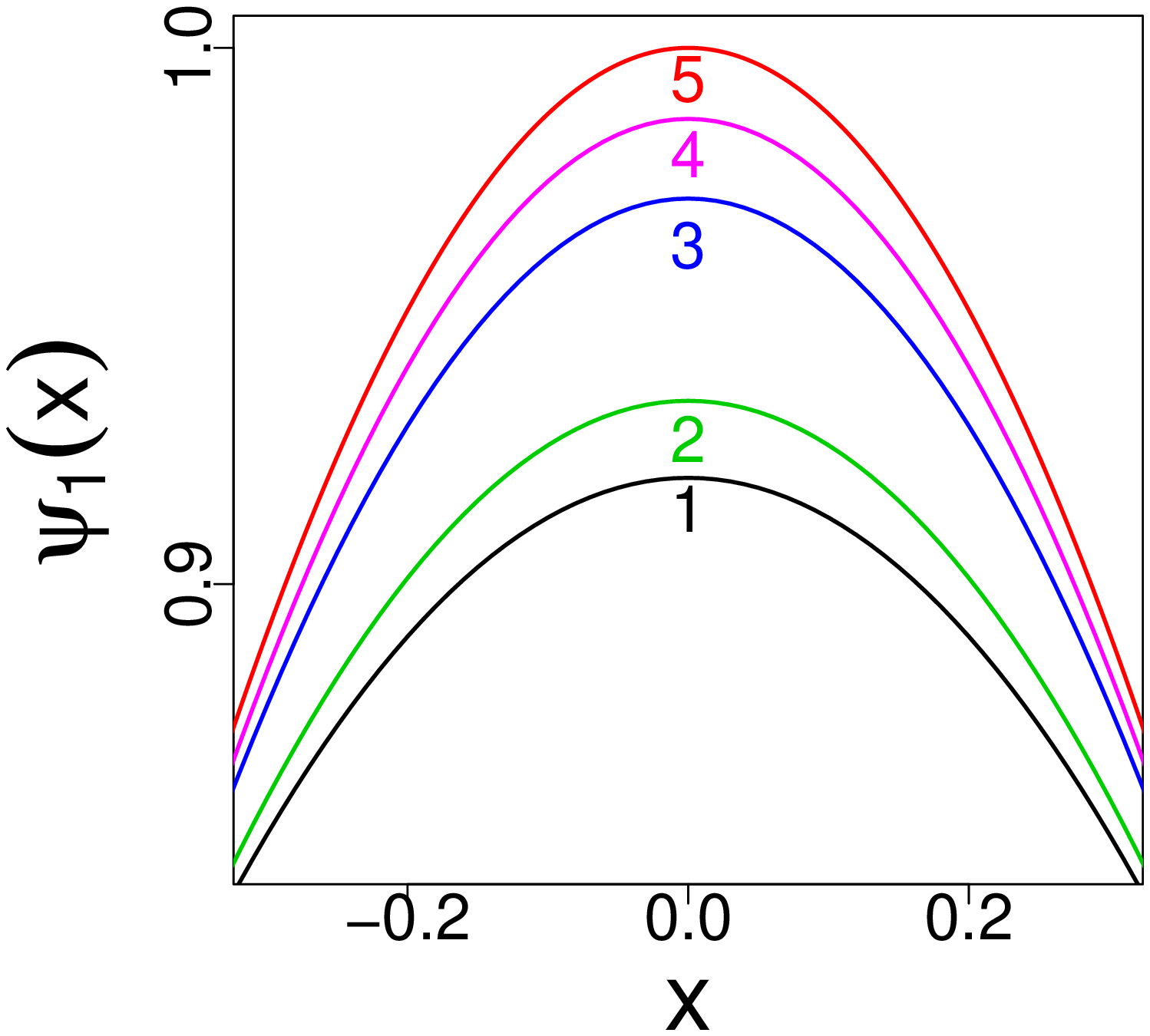}
\caption{Quasirelativistic  $V_0=500$ ground state. Labels  $1,2,3,4$ refer to masses  $m=0.01, 1, 5, 10$. Label $5$ refers to
the nonrelativistic  infinite well  ground state $\cos(\pi x/2)$.
Right panel: an enlargment of the vicinity of the maxium.}
\end{center}
\end{figure}

\begin{figure}[h]
\begin{center}
\centering
\includegraphics[width=70mm,height=70mm]{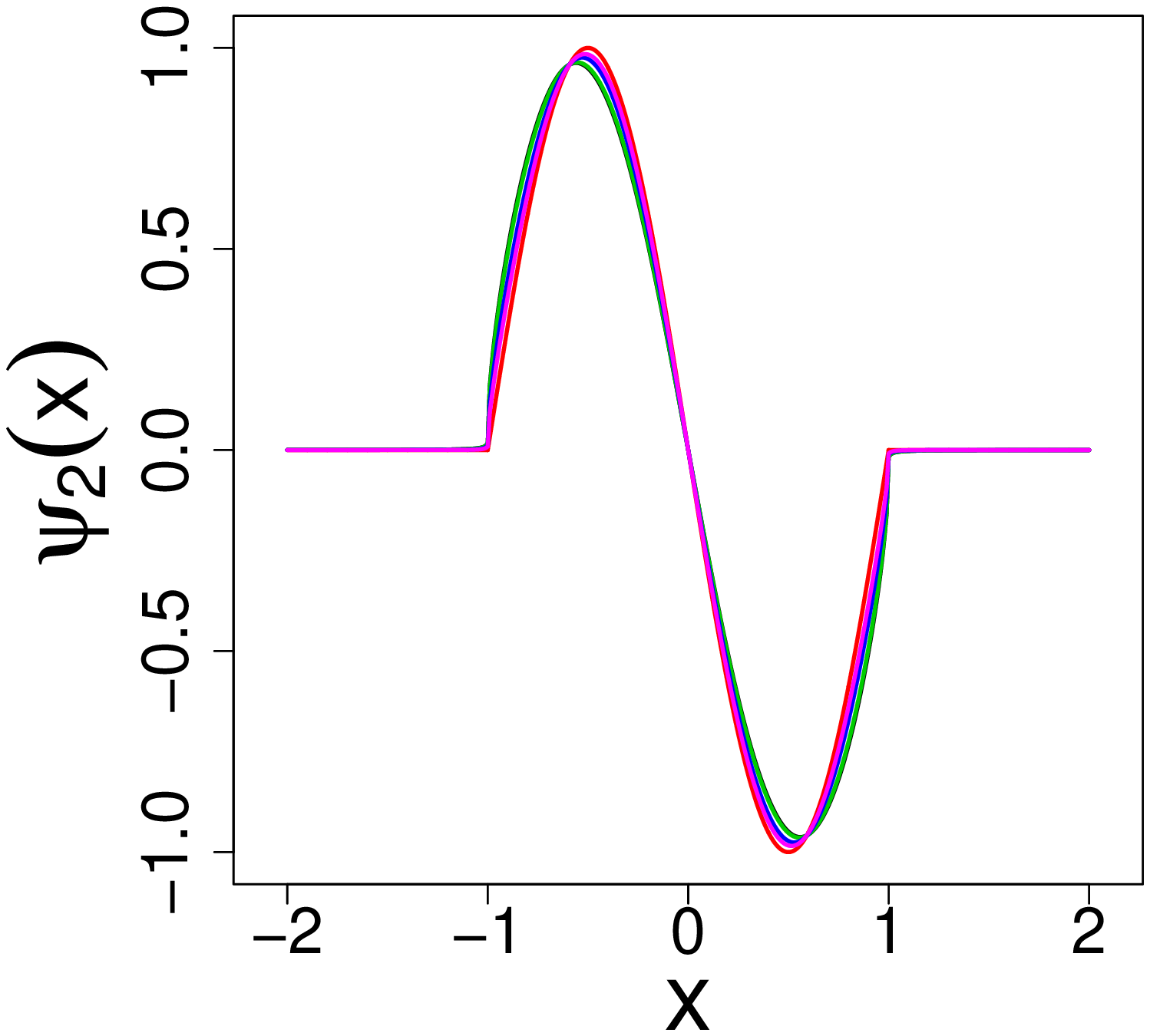}
\includegraphics[width=70mm,height=70mm]{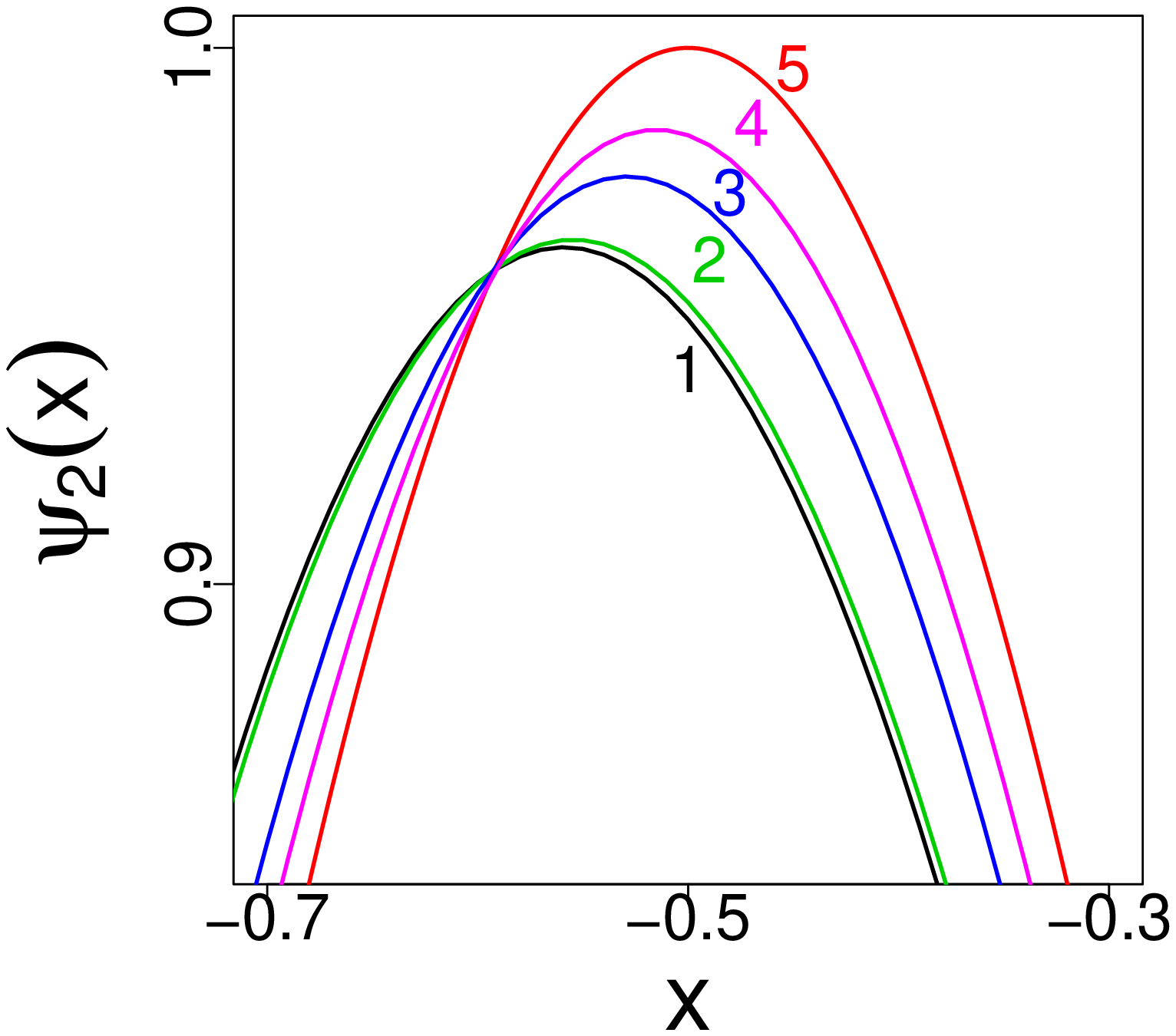}
\caption{First excited state of the  $V_0=500$ well. Labels  $1,2,3,4$ refer to $m=0.01, 1, 5, 10$, label 5  to the curve  $-\sin(\pi x)$.
Right panel: enlargement of  the vicinity of maximum.}
\end{center}
\end{figure}

\begin{figure}[h]
\begin{center}
\centering
\includegraphics[width=70mm,height=70mm]{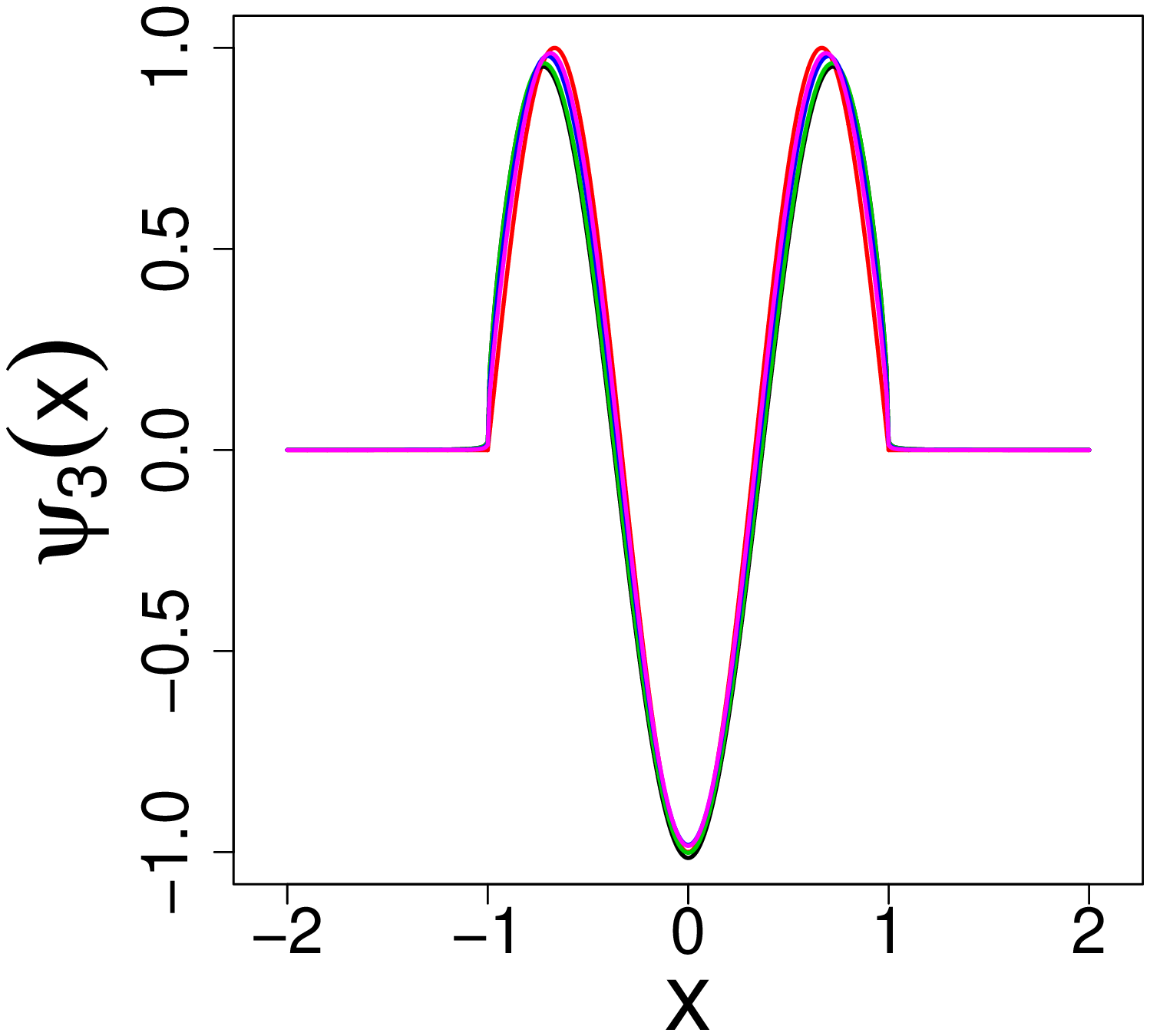}
\includegraphics[width=70mm,height=70mm]{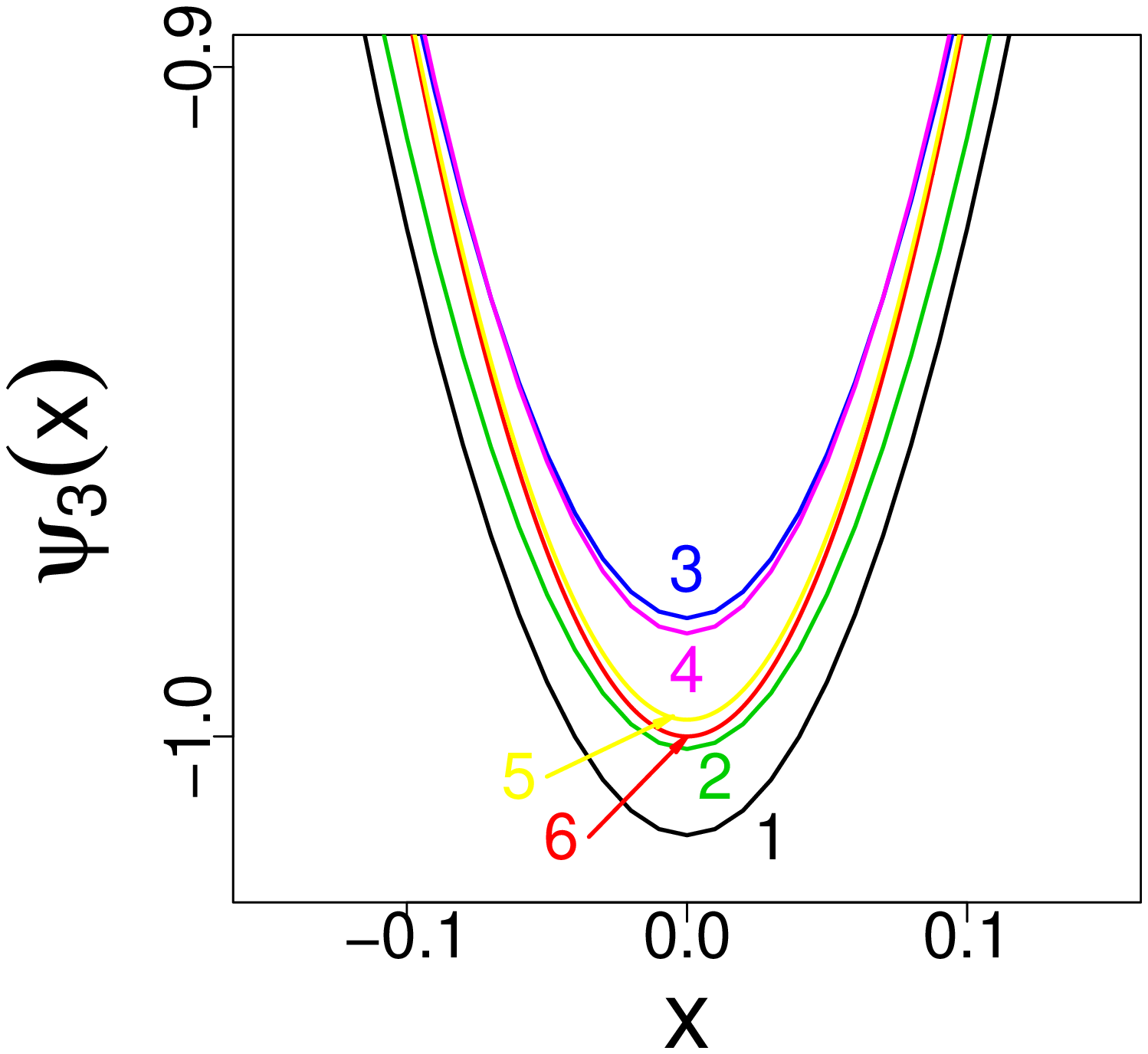}
\caption{Third eigenfunction for  $V_0=500$. Labels now refer to  $m=0.01, 1, 5, 10, 50$, label  6   to the   curve  $-\cos(3\pi x/2)$.
Right panel depicts an enlargement of  the vicinity of the minimum.}
\end{center}
\end{figure}

\begin{figure}[h]
\begin{center}
\centering
\includegraphics[width=70mm,height=70mm]{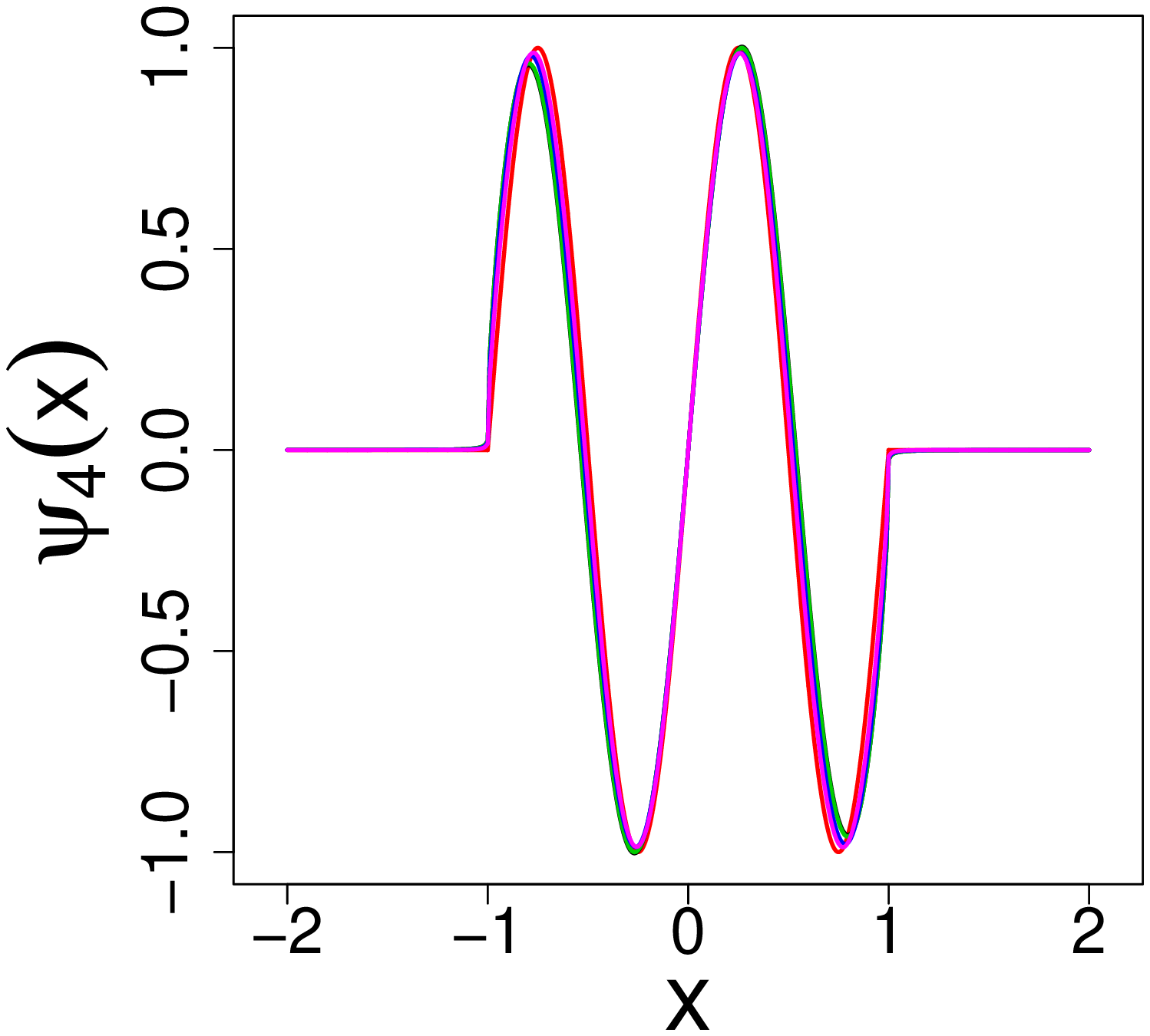}
\includegraphics[width=70mm,height=70mm]{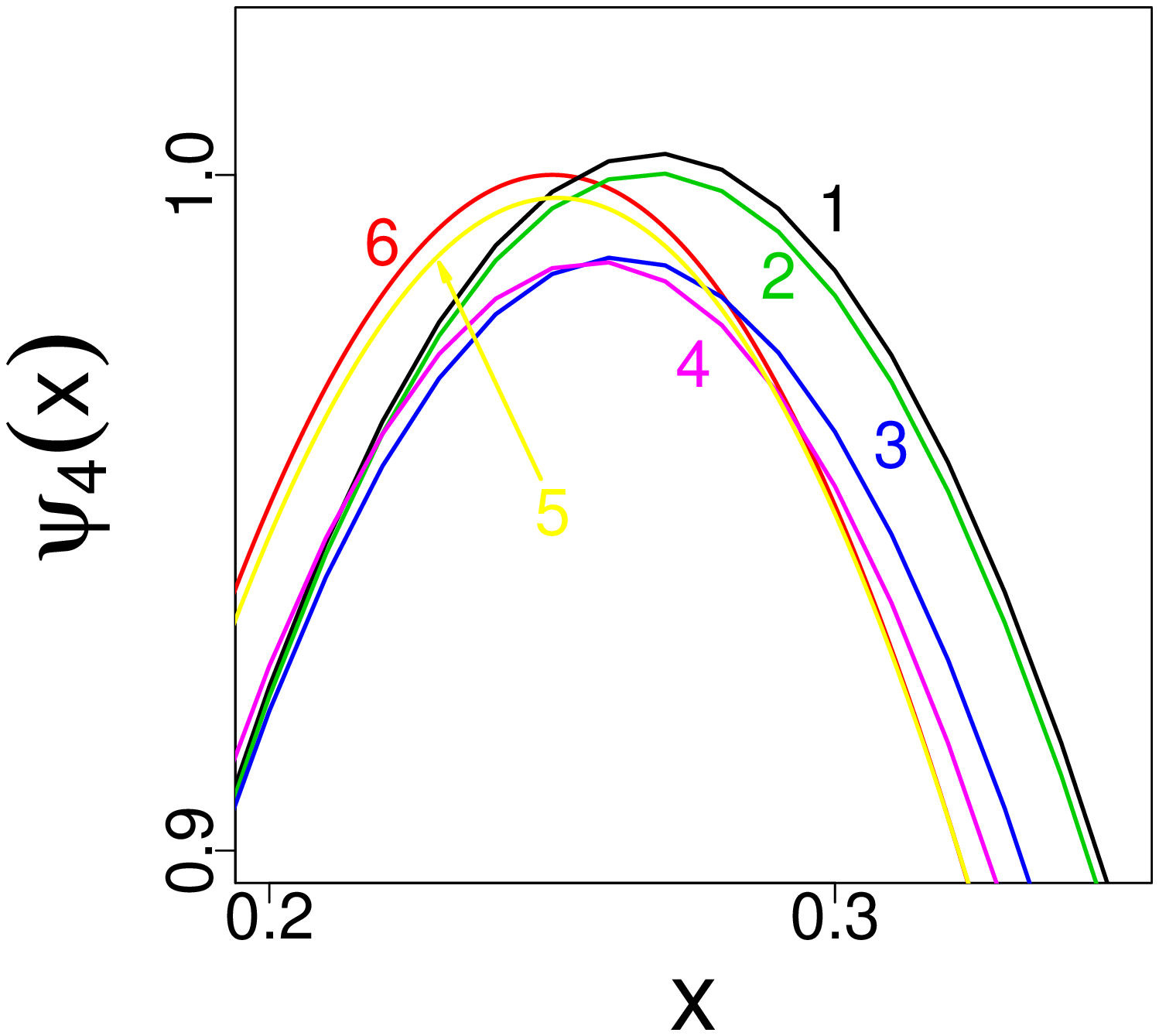}
\caption{Fourth eigenfunctions for $V_0=500$. Labels  $1,2,3,4,5$ refer to  $m=0.01, 1, 5, 10, 50$, label 6 to the curve  $\sin(2\pi x)$.
Right panel: enlargement of the vicinity of the maximum.}
\end{center}
\end{figure}

We note that in case of $V_0=500$, for small $m$ respective ground states stay in a close vicinity of the infinite Cauchy well \cite{GZ}.
To the contrary, if $m$ is sufficiently large, respective ground states converge to $\cos(\pi x/2)$ which is a nonrelativistic ground state for an infinite well.
The same pattern of behavior is detectable for excited states displayed in Figs. 13, 14 and  15.

The large $m$ regime  locates  excited states respectively   in the vicinity of at  $-\sin(\pi x)$ (Fig. 13),
  $-\cos(3\pi x/2)$ (Fig. 14) and   $\sin(2\pi x)$ (Fig. 15).
Due to the presence of $m=0.01$ curves we in fact have a transparent interpolation between the  infinite  Cauchy and nonrelativistic infinite well
approximations of  the   deep quasirelativistic well. The convergence may not be uniform, see  Fig. (15).

In the left panel of Fig. 16 the m-dependence of first five  deep  well ($V_0=500$)  eigenvalues   has been displayed.
For a direct comparison the corresponding
Cauchy well  ($V_0=500,\,  m=0$) eigenvalues were depicted as well.  In the right panel the $n$-dependence of computed eigenvalues  is
displayed for  masses $m= 10, \, 20, \, 50,\, 100$.
 The  convergence towards Cauchy well data while  $m$ drops down to $0$ is clearly
seen. To the contrary,  for  large  $m$  an approach towards  the  corresponding nonrelativistic well
spectral data  can be directly   read out from the figures. We shall validate the latter statement by  a  more detailed  data analysis.

\begin{figure}[h]
\begin{center}
\centering
\includegraphics[width=70mm,height=70mm]{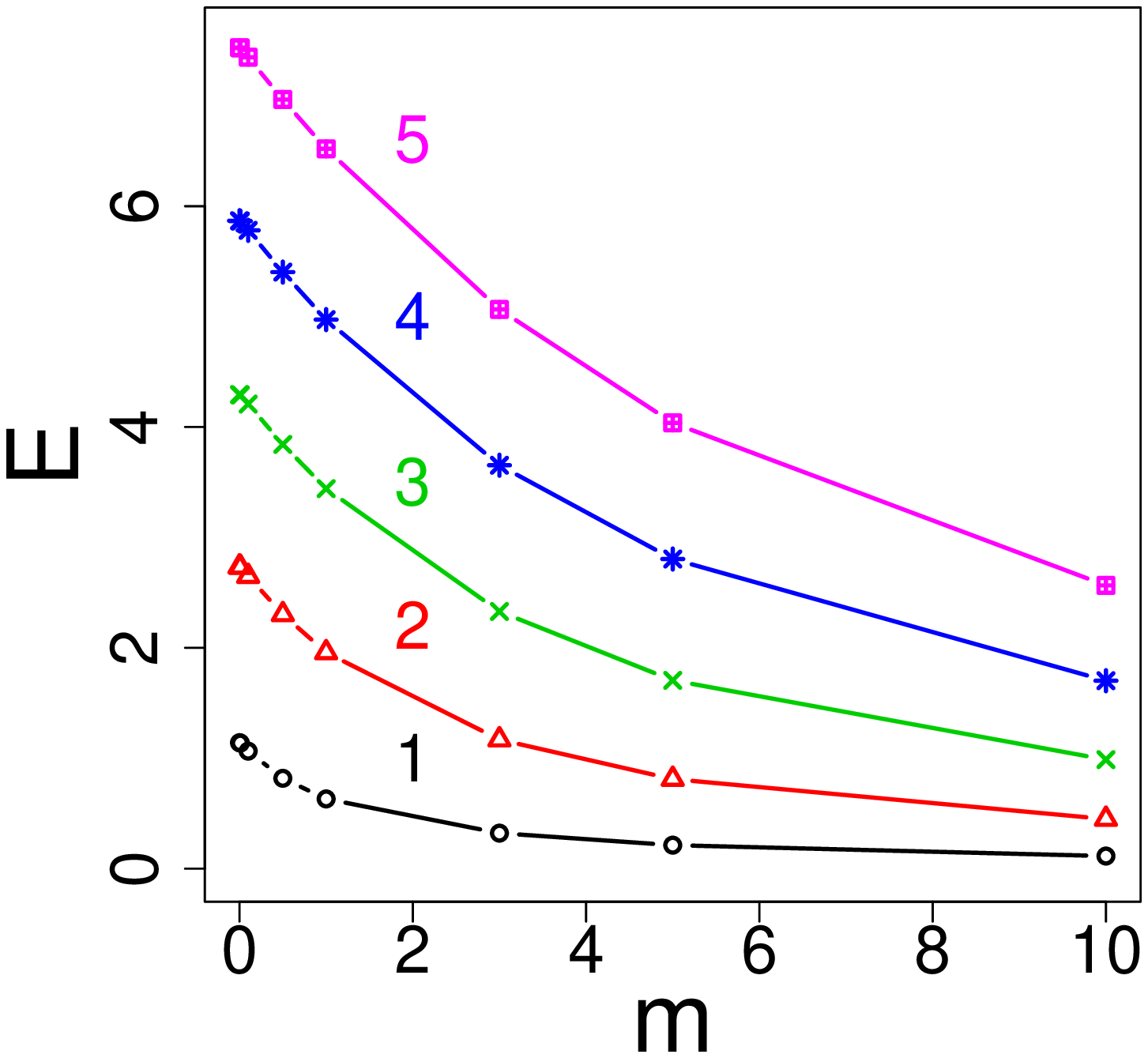}
\includegraphics[width=70mm,height=70mm]{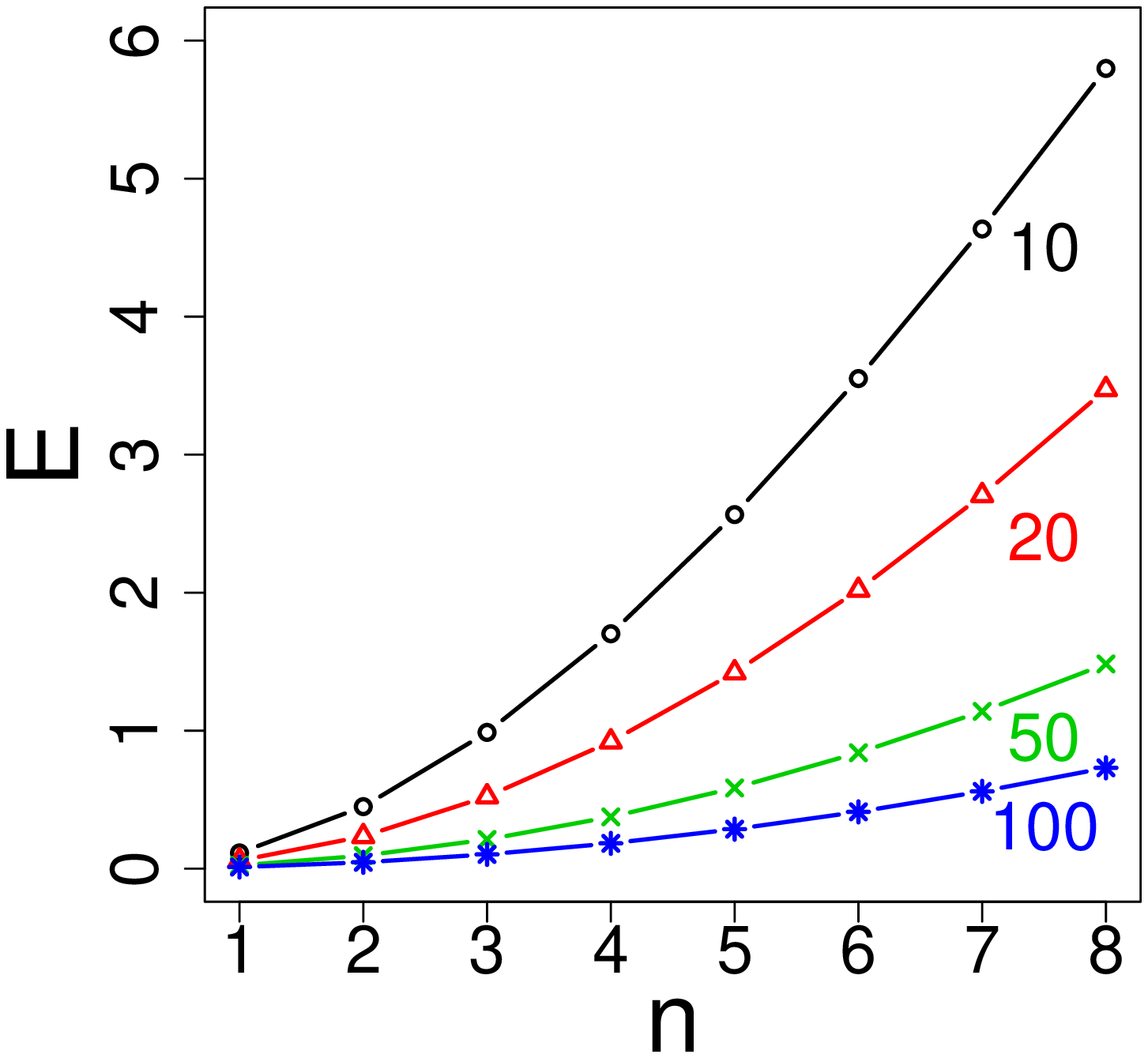}
\caption{$V_0=500$ quasirelativistic well.  Left panel:  $E_n$ dependence on $m$, $n=1,2,3,4,5$.
 Right panel: computed eigenvalues   are  depicted against $n=1,2,3,4,5$.  For each  mass  value    ($m=10,20, 50, 100$)
  we depict  a  curve which is an optimal fit to
 the data.}
\end{center}
\end{figure}

On the basis  of simulation data, we may fairly accurately deduce best fitting analytic forms for  curves associated with masses
 $m=10, 20, 50, 100$ (depicted in Fig. 16) and $m= 200$ (not depicted so far.
Since we expect a convergence (with the growth fo $m$)  to nonrelativistic well spectral data,  let us  consider as a useful  reference
an approximate formula for the nonrelativistic  deep well spectra \cite{GK, BFV}:
\be
E_n^{V_0}\approx E_n^\infty\left(1-\frac{4}{\pi\sqrt{V_0}}\right)=\frac{\pi^2 n^2}{8m}\left(1-\frac{4}{\pi\sqrt{V_0}}\right)\label{approx}
\ee
We note that $\frac{4}{\pi\sqrt{V_0}}<0.06 $  for  $V_0=500$.

For each mass parameter in the right panel of Fig. 16,  the fitted  fitting curve actually can be described by means of  of an
 approximate  analytic formula (derived directly from the data).  For direct  comparison,  the ground state energy $E_1^{st}$
has been evaluated by means of a nonrelativistic formula (\ref{approx}).
The  spectral affinity  of the quasirelativistic  well  with the nonrelativistic well for large mass values
appears to be validated with no trace of doubt.

\begin{align*}
m&=10, & (0.1191&\pm 0.0049)n^{1.8929}, & & E_1^{st} \sim 0.1163, \\
m&=20, & (0.0596&\pm 0.0025)n^{1.9643}, & &E_1^{st} \sim   0.0582,\\
m&=50, & (0.0236&\pm 0.0013)n^{1.9937}, & & E_1^{st}  \sim  0.0233,\\
m&=100,& (0.0117&\pm 0.0007)n^{1.9983}, & &E_1^{st}  \sim   0.0116,\\
m&=200,& (0.0058&\pm 0.0004)n^{1.9996}, & &E_1^{st}  \sim  0.0058.
\end{align*}

We note that an approximate  formula (\ref{approx})  has the form  $E_n^{st} \sim {\frac{\alpha }{m}}\, n^2$.
 It is a  convergence   $\beta \rightarrow 2 $  of the exponents
  $\beta  = 1.8928,\, 1.9643, \, 1.9937, \, 1.9996 $ in  the above    $n^{\beta }$ entries,    which is most indicative.

\section{Outlook}

We have investigated in minute detail spectral properties (eigenvalues and eigenfunctions   shapes) of   nonlocal confining
 quantum models  associated with  the quasirelativistic generator. Harmonic and finite well potentials were considered.
Computation accuracy is very high in the low part of the spectrum and specifically eigenfunctions  shapes can be reproduced with
a  fidelity level that was never reached before in the nonlocal context, c.f. also Ref. \cite{SG}.

 For example it was known that both the infinite Cauchy well and    the infinite
quasirelativistic  well  have eigenfunctions whose shapes are  {\it similiar} to those of trigonometric functions (e.g. eigenfunctions of
 the corresponding  infinite  nonrelativistic well). This similarity, albeit appealing,  is  merely elusive,  Our computer-assisted
  results,  both in the present paper and in Ref. \cite{SG},
confirm  that  true shapes  considerably  differ from  nonrelativistic ones.

Obviously,  one may set a  suitable acceptance (robustness)  level  within  which these  differences become immaterial.
  However the modern view on  quantum phenomena proves that even  extremely subtle discrepancies might be observable,
   ultimately  acquiring   a  profound meaning, with an  impact upon the development or refinements of the existing theory
   and experiment as well.

   The mass range $m \in (0,\infty )$
  has been explored and  the   spectral affinity ("closeness")  with  (i) $m=0$ ultrarelativistic (Cauchy) case  for $m\ll 1$  and
  (ii)  standard  nonrelativistic quantum eigenvalue problem for $m\gg 1$, has been established.   This spectral affinity might be
   a generic property of all  confining quasirelativistic models, irrespective of the number of space dimensions.
 We have given analytic hits towards this conjecture in the  Appendix D.

 Translating  our observations  to typical elementary  particle masses (spin  being disregarded), we realize that
    e.g. neutrino  (and light quark)  masses  would   well fit to the approximation in terms of the Cauchy models,
    while  proton mass  (perhaps surprisingly)  would   rather  fit to the nonrelativistic
    approximation  of the quasirelativistic Hamiltonian.

\section{Appendices}

\subsection{Lowest eigenvalues of the Cauchy oscillator and their approximate  values.}

Approximate formulas (10)   and (11) for Cauchy oscillator eigenvalues  reflect the fact that these eigenvalues
 are normally divided into two subclasses.  The  approximate  eigenvalues $ E_{n=2k-1}^{appr}=  (3\pi/2)^{2/3} (n + 3/4)^{2/3}$,
that are numbered by  $k=1,2,3,...$ and thence refer to odd $n$   labels,  $n=1, 3,5,...$, actually  correspond to  even
eigenfunctions. The eigenvalue   stands for  the minus zero  of the Airy function derivative.
A complementary formula $ E_{n=2k}^{appr}= (3\pi /2)^{2/3} (n +  1/4)^{2/3}$
refers to even label  $n=2,4,6...$ and  odd eigenfunctions.  The eigenvalue stands  for the minus zero  of the Airy function.
See e.g. \cite{SG,LM}.

 We  note that  the  formulas (10), (11) can be written in a compact form encompassing all consecutive $n$-labels:
$$
E^{appr}_n=\left({\frac{3\pi}{8}}\right)^{2/3} \left[8n  + (-1)^n \right]^{2/3}.
$$

Our robustness threshold will be the fourth or fifth decimal  digit in presented results. We point out that  while evaluating
 Airy function zeroes (we term them "exact") one can  use an
arbitrarily large   number of decimal digits, like   $14$  or more, see e.g. \cite{LM,A}.

It turns out that approximate formulas (10) and (11) give a fairly good approximation for
Cauchy oscillator eigenvalues   not necessarily for large $n$ only, but actually beginning from the bottom one $n=1$.
Indeed  for $E_{n=2k}$ eigenvalues we have:

\begin{eqnarray}
&& \quad E_2^{exact}=2.3381,   \quad E_2^{appr}= 2.32025,\nonumber \\
&& \quad E_4^{exact}= 4.0879,  \quad E_4^{appr}=4.08181,\nonumber \\
&& \quad E_6^{exact}=5.5206,   \quad  E_6^{appr}=5.51716,\nonumber \\
&& \quad E_8^{exact} =6.7867,   \quad E_8^{appr} = 6.78445, \nonumber \\
&& \quad E_{10}^{exact} = 7.9440, \quad E_{10}^{appr} =7.94248, \nonumber \\
&& \quad E_{12}^{exact} = 9.0226, \quad E_{12}^{appr} =9.02137, \nonumber \\
&& \quad E_{14}^{exact} =10.0402, \quad E_{14}^{appr} = 10.03914, \nonumber \\
&& \quad E_{16}^{exact}=11.0085, \quad E_{16}^{appr}=11.00776,\nonumber \\
&& \quad E_{18}^{exact}=11.9360, \quad E_{18}^{appr}=  11.93532. \nonumber \\
\end{eqnarray}

For  $E_{n=2k-1}$  eigenvalues, a  comparison of exact and approximate outcomes  goes as follows:
\begin{eqnarray}
&& \quad E_1^{exact}=1.0188, \quad E_1^{appr}=1.11546,\nonumber \\
&& \quad E_3^{exact}=3.2482, \quad E_3^{appr}=3.26163,\nonumber \\
&& \quad E_5^{exact}=4.8201, \quad E_5^{appr}=4.82632,\nonumber \\
&& \quad E_7^{exact}= 6.1633, \quad  E_7^{appr} = 6.16712, \nonumber \\
&& \quad E_9^{exact}= 7.3721, \quad  E_9^{appr} = 7.37485, \nonumber \\
&& \quad E_{11}^{exact}= 8.4884, \quad  E_{11}^{appr} = 8.49050, \nonumber \\
&& \quad E_{13}^{exact}= 9.5354, \quad  E_{13}^{appr} = 9.53705, \nonumber \\
&& \quad E_{15}^{exact}=  10.5276, \quad  E_{15}^{appr} = 10.52897,\nonumber \\
&& \quad E_{17}^{exact}=11.4751, \quad E_{17}^{appr}=11.4762,\nonumber \\
&& \quad E_{19}^{exact}=12.3848, \quad E_ {19}^{appr}=12.3857. \nonumber \\
\end{eqnarray}

\subsection{Quasirelativistic well:  $m$-dependence of lowest five eigenvalues
for wells depths $V_0=10,\, 20,\, 50,\, 100$. Tables VIII-XI.}

\begin{table}[h]
\begin{tabular}{|c||c|c|c|c|}
  \hline
  $V_0=10$      & $m=5$ & $m=10$ & $m=20$ & $m=50$ \\
        \hline\hline
  $E_1$ & 0.19087  & 0.10444 & 0.05480 & 0.02244  \\
  $E_2$ & 0.73186  & 0.41180 & 0.21829 & 0.08969  \\
  $E_3$ & 1.54865  & 0.90659 & 0.48793 & 0.20153  \\
  $E_4$ & 2.56136  & 1.56631 & 0.85967 & 0.35764  \\
  $E_5$ & 3.70609  & 2.36600 & 1.32820 & 0.55761  \\
  \hline
\end{tabular}
\caption{$V_0=10$, $m$-dependence of eigenvalues. }
\end{table}
\begin{table}[h]
\begin{tabular}{|c||c|c|c|c|}
  \hline
  $V_0=20$      & $m=5$ & $m=10$ & $m=20$ & $m=50$ \\
        \hline\hline
  $E_1$ & 0.19911  & 0.10788 & 0.05617 & 0.02283  \\
  $E_2$ & 0.76223  & 0.42582 & 0.22401 & 0.09125  \\
  $E_3$ & 1.60971  & 0.93699 & 0.50062 & 0.20502  \\
  $E_4$ & 2.65763  & 1.61809 & 0.88199 & 0.36381  \\
  $E_5$ & 3.84054  & 2.44327 & 1.36276 & 0.56728  \\
  \hline
\end{tabular}
\caption{$V_0=20$, $m$-dependence of eigenvalues. }
\end{table}
\begin{table}[h]
\begin{tabular}{|c||c|c|c|c|}
  \hline
  $V_0=50$      & $m=5$ & $m=10$ & $m=20$ & $m=50$ \\
        \hline\hline
  $E_1$ & 0.20605  & 0.11104 & 0.05743 & 0.02316  \\
  $E_2$ & 0.78747  & 0.43773 & 0.22885 & 0.09256  \\
  $E_3$ & 1.65949  & 0.96255 & 0.51149 & 0.20802  \\
  $E_4$ & 2.73417  & 1.66097 & 0.90106 & 0.36921  \\
  $E_5$ & 3.94414  & 2.50604 & 1.39200 & 0.57572  \\
  \hline
\end{tabular}
\caption{$V_0=50$, $m$-dependence of eigenvalues. }
\end{table}
\begin{table}[t]
\begin{tabular}{|c||c|c|c|c|}
  \hline
  $V_0=100$      & $m=5$ & $m=10$ & $m=20$ & $m=50$ \\
        \hline\hline
  $E_1$ & 0.20919  & 0.11245 & 0.05803 & 0.02332  \\
  $E_2$ & 0.79883  & 0.44321 & 0.23116 & 0.09321  \\
  $E_3$ & 1.68170  & 0.97429 & 0.51661 & 0.20948  \\
  $E_4$ & 2.76799  & 1.68054 & 0.91000 & 0.37182  \\
  $E_5$ & 3.98941  & 2.53443 & 1.40563 & 0.57977  \\
  \hline
\end{tabular}
\caption{$V_0=100$, $m$-dependence of eigenvalues. }
\end{table}

\subsection{Eliminating and reintroducing dimensional constants}
\subsubsection{Oscillators.}

{\it (i) Quasirelativistic oscillator.}

The dimensional version of  the  Hamiltonian reads   $H^{dim} = \sqrt{-\hbar ^2c^2 \Delta + m^2c^4} - mc^2   + kx^2/2$, while we have been computing
 the  spectral solution  for $H = \sqrt{- \Delta + m^2} - m  + x^2$.  The relationship  between  $E_n^{dim}$ and $E_n$ needs to
  be settled. The scaling procedure is entirely equivalent to the choice of natural units  accompanied by  getting rid of $k/2$.

Let us  consider scaling transformations inspired by the following form  of  $H^{dim}$:
$$
H^{dim}=    c^2[ \sqrt{- {\frac{\hbar ^2}{c^2}} \Delta + m^2 }  -  m  + {\frac{k}{2 c^2}} x^2]  =
 c^2 [\sqrt{- \tilde{ \Delta } + m^2} - m  +  \kappa \tilde{x}^2].
$$
where we denote $\tilde{x} = c x/\hbar  $ and $\kappa = k\hbar ^2/2c^4$.
One more scaling transformation can be executed by means of a substitution:
$\tilde{x}= \check{x} /\kappa ^{1/3}$, followed by  $\tilde {E_n} = \kappa ^{1/3} \check{E}_n$, $m= \kappa ^{1/3} \check{m}$.
Clearly, we arrive at
$$
H^{dim} = c^2 \kappa ^{1/3} [\sqrt{ - \check{\Delta } + \check{m}^2}  - \check{m}   + \check{x}^2] = c^2 \kappa ^{1/3} \check{H} =\hbar c
 \left({\frac{k}{2\hbar c}}\right)^{1/3}\check{H}
 $$
where $\check{H}$ has a  canonical    form employed in  computational routines of Section III, compare e.g. Eq. (9).

If  we denote  $f(x)= \check{f}(\check{x})$, then  there holds
 $$
 H^{dim}f(x)= c^2 \kappa ^{1/3} \check{H} \check{f}(\check{x})
$$
where $\check{x}= (\kappa ^{1/3} c/\hbar )\, x = (k/2\hbar c)^{1/3} x$, $m= \kappa ^{1/3} \check{m}$    and $E^{dim}_n =  c^2 \kappa ^{1/3}  \check{E}_n$.
 Eigenfunctions of $\check{H}$ are by construction normalized (c.f. Section III),  hence  to extend this property to eigenfunctions
of   $H^{dim}$ we need to compensate  the  change of integration variable from $\check{x}$ back to $x$
(we recall  that $f(x)= \check{f}(\check{x})$).

  Since $d\check{x}=(\kappa ^{1/3} c/\hbar )\, dx$, the  $L^2(R)$-normalized eigenfunction
   $\check{f}(\check{x})$ of $\check{H}$
 gives rise to the $L^2(R)$-normalized eigenfunction  $\psi (x)$  of $H^{dim}$,   according to
$$
\check{f}(\check{x}) \rightarrow  (\kappa ^{1/3} c/\hbar )^{1/2} f(x)= (k/2\hbar c)^{1/3} f(x)= \psi (x).
$$
All that modifies an  integration   interval   from $[-\check{a},\check{a}]$ on the $\check{H}$ level  to
$[-a,a]$,   with
 $a=  (2\hbar c/k)^{1/3} \check{a}$   on the level of  $H^{dim}$.\\

{\it (ii)  Cauchy oscillator.}

In the derivation of the spectral solution \cite{SG} we have used a scaling transformation
which connects the eigenvalues  $E^{dim}_n$  of $H^{dim}=  \hbar c |\nabla | +  k x^2/2$ with those  (e.g. $E_n$)
 for $\check{H} = |\nabla | + x^2$. Obviously, it is a special  $m=0$ version of the previous $m \neq 0$ derivation. Namely, we have
   $E^{dim}_n=   (k \hbar ^2 c^2 /2)^{1/3} \check{E}_n$.  Accordingly, we have  $[-a,a]$  with
  $a=  (2\hbar c/k)^{1/3} \check{a}$.

\subsubsection{Wells.}

{\it (i) Infinite Cauchy well.}

The dimensional energy operator reads $H^{dim} = \hbar c |\nabla |$, while   Dirichlet boundary conditions impose
the "infinite well constraint" at  boundaries $[-b,b]$ of the well. By setting  $x= b\check{x}$ we introduce a dimensionless
"space" label $\check{x}$.  Hence $H^{dim} = (\hbar c/b) \check{H}$ where $\check{H}= |\check{\nabla }|$.
The   Dirichlet boundary conditions for $\check{H}$  now  refer to another (dimensionless)
 interval $[-1,1]$,  that in view of  $\check{b} =  1$.  We  note that  the   dimensionless "energy" unit for
   $\check{E}$  equals $1$, which translates to an energy unit  $(\hbar c/b)$ in case of $E^{dim}$.
   The integration interval  $[-\check{a},\check{a}]$ is mapped into  $[-a,a]$  with $a= b \check{a}$.\\

{\it (ii)  Finite Cauchy well.}

We have $H^{dim} = \hbar c |\nabla |  + V_0^b(x)$,  where $V_0^b(x)= V_0>0$  for $|x|\geq b$ and
 vanishes  in the interval $(-b,b)$. By setting  $x= b\check{x}$ we get  $H^{dim}= (\hbar c/b)  \check{H}$ where
  $\check{H}= |\check{\nabla }| + \check{V}_0^{\check{b}}$  and  $\check{V}_0^{\check{b}} = (b/\hbar c) V_0$
  for $|x|\geq 1$, while  being equal $0$ in $[-1,1]$.  Obviously $\check{V}_0^{\check{b}}$ is a dimensionless quantity,
   "measured" in units $1,2,3...$, while  $V_0^b$    in    units $(\hbar c/b)$. Like before,
    $[-\check{a},\check{a}]$  goes over to $[-a,a]$  with $a= b \check{a}$. \\

{\it (iii)  Quasirelativistic finite well.}

As before we take   $\epsilon =(\hbar c/b)$ to set an energy scale. Accordingly
$ H^{dim}= \left( {\frac{\hbar c}{b}}   \right)  \check{H}$, where
$$
\check{H} = \sqrt{ - \check{\Delta } + \check{m}^2} - \check{m} + \check{V}_0^{\check{b}}
$$
where $\check{V}_0^{\check{b}} = (b/\hbar c) V_0$  for $|x|\geq 1$, while  being equal $0$ in $[-1,1]$.
The "mass" parameter  $\check{m}= b/\lambdabar _C$  is dimensionless. Here  $\lambdabar _C= \hbar  /mc $
  is the  familiar reduced  Compton wavelength associated with a quantum particle of mass $m$.
   Again  $[-\check{a},\check{a}]$  gets replaced by  $[-a,a]$ with
$a= b \check{a}$.\\

\subsubsection{Length and energy scales.}

 It seems useful to comment on the role of the omnipresent   factor  $\hbar c$  which contributes to ultimate (dimensional) energy scales.
  In conjunction with $b$ it appears as  an energy scaling factor $\epsilon = \hbar c/b$.
  Since $\hbar c =1.975 \, [GeV]\cdot [fm] = 1.975 \cdot 10^{-6} \,[eV]\cdot [m]$, then  e.g.
  $b= 1 [nm] =10^{-9}\,  [m]$  results
   in $\epsilon = 1.975\,  [keV]$,
   $b=  10^{-8}\,   [m]$  gives rise to  $\epsilon = 197.5\,  [eV]$, while $b=1 [\mu m]$ to  $1.975\,  [eV]$.\\

In the previous subsection $\check{m}= b/\lambdabar _C$ with $\lambdabar _C= \hbar  /mc $ has been dimensionless. Thus, given concrete
$\check{m} \in (0,\infty ) $ value,  the related  $\lambdabar _C$ sets the length scale  and in reverse (given $b$).
  For example, if $m$ is the electron rest mass, we have $\lambdabar _C= 386 [fm] =0.00386 [\AA] $.  Then $b= 10^{-10} [m]$
  implies  $\check{m} \sim 2.6 $. On the other hand, presuming e.g. $\check{m} =26$   and the electronic $\lambdabar _C$  we end up with
 $b=10^{-9} [m]$.

 Concerning the dimensional  mass $m$ choice we have a number of other physical  options. Thus  e.g. accepting that  the electron
  mass  $m_e\sim 0.511 [MeV]/c^2$, we can easily recompute $\lambdabar _C$  to refer to  other  elementary  particles.
   Thus e.g. for the proton $m_p \sim 938 [MeV]/c^2$ we have $m_p/m_e \sim  1836$. Analogous proportionality
    factors can be introduced  e.g.  for the electron neutrino  $m_{\nu }\sim 2.2 [eV]/c^2$, muon neutrino $m_{\mu } \sim  170   [keV]/c^2$,
    neutral pion $m_{\pi } \sim 140 [MeV]/c^2$, kaon $m_K \sim 494 [Mev]/c^2$.
Since  for  the exemplary case of the   electron neutrino we  have  $m_e/m_{\nu } \sim 232.3 \cdot 10^3$,  the corresponding reduced
Compton wavelength reads  $\lambdabar _C^{\nu } =232.3 \cdot 10^3
  \lambdabar _C \sim    896.7  [\AA] $.

\subsection{Ultrarelativistic ($m\ll 1$)  and
nonrelativiastic ($m\gg 1$) mass  extremes   of  the quasirelativistic  kinetic energy operator $T_m$.}

 An analytic approach to $m\ll 1$ and $m\gg 1$ regimes of $H=T_m+V$  is best exemplified by resorting to the quasirelativistic operator
 $T_m = \sqrt{\Delta + m^2}   - m$. The standard reasoning employs  the Fourier representation \cite{GS,L}.

Reintroducing  the  physical constants (on may keep $\hbar =1=c$ intact as well),  the quasirelativistic  operator  $T_m$
is presumed to   act  upon  functions in the domain of $H=T_m +V$
\begin{equation}\label{rfse1}
 (T_m + mc^2) \phi (x)=\sqrt{m^2c^4-\hbar^2c^2\frac{\partial^2}{\partial x^2}}\  \phi(x).
\end{equation}
Denoting $\tilde{f}(k)= (2\pi )^{-1/2} \int_{-\infty}^\infty f(x)e^{-ikx}dx$, $f(x)=  (2\pi )^{-1/2} \int_{-\infty}^\infty \tilde{f}(k)e^{ikx}dk$ and
interpreting  the action of the square root operator  in
terms of the series expansion, we  readily arrive at  the  following formal  Fourier representation:
\begin{eqnarray} \label{rfse3}
 (T_m +mc^2)\phi (x)   =
\frac{mc^2}{\sqrt{2\pi}}\int_{-\infty}^\infty \tilde{\phi }(k) dk\ \sqrt{1-\frac{\hbar^2}{m^2c^2}\frac{\partial^2}{\partial x^2}}\
e^{ikx}=\nonumber \\
\frac{mc^2}{\sqrt{2\pi}}\int_{-\infty}^\infty \tilde{\phi }(k) dk\ \left[1- \frac{\hbar^2}{m^2c^2} \frac 12 \frac{\partial^2}{\partial x^2}-
\left(\frac{\hbar^2}{m^2c^2}\right)^2\frac 18 \frac{\partial^2}{\partial x^2}-...\right]\ e^{ikx}= \label{rfse4} \\
 \frac{mc^2}{\sqrt{2\pi}}\int_{-\infty}^\infty \tilde{\phi }(k) dk\ \sqrt{1+\frac{p^2}{m^2c^2}}\ e^{ikx}  =
 \frac{1}{\hbar\sqrt{2\pi}}\int_{-\infty}^\infty dp\ \sqrt{m^2c^4+p^2c^2}\, \,  e^{ipx/\hbar}
\tilde{\phi }(p). \nonumber
\end{eqnarray}
We note an  explicit presence of  $\hbar /mc=\lambdabar _C$  and  $p=\hbar k$.

All our derivations and the previous discussion of the "spectral affinity" (closeness) of various systems (like e.g. this of the quasirelativistic
 and nonrelativistic  oscillators in the large $m$ regime) crucially rely of  the presence of confining potentials.
Then only, we can expect that  the Taylor series  with respect  to $ p^2/m^2c^2=  k^2 \lambdabar _C^2$ may be terminated after the first order term:
\be
\frac{mc^2}{\sqrt{2\pi}}\int_{-\infty}^\infty \tilde{\phi }(k) dk\ \sqrt{1+ k^2\lambdabar ^2_C}\ e^{ikx} \sim
\frac{mc^2}{\sqrt{2\pi}}\int_{-\infty}^\infty \tilde{\phi }(k) dk\ [1 +  (1/2)  k^2\lambdabar ^2_C]\ e^{ikx} =\\
mc^2 \phi (x) - {\frac{\hbar ^2}{2m}} \Delta \phi (x).
\ee
This property can be granted only if the function $\tilde{\phi }(k)$  gives substantial contributions only from $k$
obeying $k^2\lambdabar ^2_C \ll 1$, vanishing rapidly otherwise. That is inseparably linked with the
previously considered   nonrelativistic  ($m\gg 1$)  regimes, where physical constants $\hbar $ and $c$ are kept fixed  while  $m$
is being varied.

In passing we note that  for the  electron  $\lambdabar _C =0.00386 \AA $  is a fairly small
proportionality factor, while for the  electron neutrino we  have   $\lambdabar _C^{\nu } =
  \lambdabar _C \sim    896.7  [\AA] $ which  is  considerably larger ($\sim 2\cdot 10^5$ times).

Although we have anticipated the existence of the mass $m=0$ limit in the quasirelativistic  confining  contexts,  our  tacit  presumption of
 the  nonrelativistic  regime  $p^2 \ll m^2c^2$   has directly  led to an   expansion of  $mc^2 \sqrt{1+ (p^2/m^2c^2)}$
  into Taylor series with respect to $\sim p^2/m^2c^2$ and  evidently we are left with    no   room  for $m \rightarrow 0$ therein.

 Nonetheless,    we can safely  put  $m=0$, after the series resummation - in the last entry of the formula
   \eqref{rfse4} - so arriving at the correct  form of the  Fourier image of $|\nabla |$.
To justify  the latter option  we should  consider the ultrarelativistic
  regime  with $p^2 \gg m^2c^2$ which is granted only if  the function $\tilde{\phi }(k)$  gives substantial contributions only from $k$
obeying    $k^2\lambdabar ^2_C \gg 1$, vanishing rapidly otherwise.    Then, we  may expand
  $|p|c \, \sqrt{1 + (m^2c^2/p^2)}$   with respect to  $ m^2c^2/p^2= (k^2\lambdabar _C^2)^{-1}$.
 Keeping the leading term of the series only, we arrive at  the  required $m \ll 1$ outcome:
\be
{\frac{mc^2}{\sqrt{2\pi}}} \int_{-\infty}^{\infty} \tilde{\phi }(k) dk\, \sqrt{1+ k^2\lambdabar ^2_C}\ e^{ikx} =
 {\frac{\hbar c}{\sqrt{2\pi}}} \int_{-\infty}^{\infty} \tilde{\phi }(k) dk\, |k|\,  \sqrt{1+  (k^2\lambdabar ^2_C)^{-1}}\, e^{ikx}
\sim \hbar c |\nabla |\phi (x).
 \ee

\end{document}